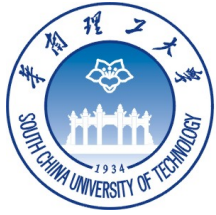

# 华南理工大学
## South China University of Technology

# 博士学位论文

## 基于多源数据融合的城市公交系统乘客出行模式挖掘及其应用研究

| 作 者 姓 名 | 刘永鑫 |
|---|---|
| 学 科 专 业 | 交通信息工程与控制 |
| 指 导 教 师 | 翁小雄 教授 |
| 所 在 学 院 | 土木与交通学院 |
| 论文提交日期 | 2018 年 6 月 |

# Study on Key Technologies of Transit Passengers' Travel Pattern Mining and Applications based on Multiple Sources of Data

A Dissertation Submitted for the Degree of Doctor of Philosophy

**Candidate**：Liu Yong Xin

**Supervisor**：Prof. Weng Xiaoxiong

South China University of Technology

Guangzhou, China

分类号：　　　　　　　　　　　　　　　　　　　学校代号：**10561**

学　号：　**201410101381**

华南理工大学博士学位论文

# 基于多源数据融合的城市公交系统乘客出行模式挖掘及其应用研究

| | |
|---|---|
| 作者姓名：刘永鑫 | 指导教师姓名、职称：翁小雄 教授 |
| 申请学位级别：工学博士 | 学科专业名称：交通信息工程与控制 |
| 研究方向：智能交通数据挖掘 | |
| 论文提交日期：2018 年 6 月 22 日 | 论文答辩日期：2018 年 6 月 2 日 |
| 学位授予单位：华南理工大学 | 学位授予日期：2018 年 6 月 25 日 |

答辩委员会成员：

主席：　　徐建闽

委员：　　邓兴栋　翁小雄　卢凯　林培群

# 华南理工大学
# 学位论文原创性声明

本人郑重声明：所呈交的论文是本人在导师的指导下独立进行研究所取得的研究成果。除了文中特别加以标注引用的内容外，本论文不包含任何其他个人或集体已经发表或撰写的成果作品。对本文的研究做出重要贡献的个人和集体，均已在文中以明确方式标明。本人完全意识到本声明的法律后果由本人承担。

作者签名：刘永鑫　　　　日期：2018 年 6 月 22 日

# 学位论文版权使用授权书

本学位论文作者完全了解学校有关保留、使用学位论文的规定，即：研究生在校攻读学位期间论文工作的知识产权单位属华南理工大学。学校有权保存并向国家有关部门或机构送交论文的复印件和电子版，允许学位论文被查阅（除在保密期内的保密论文外）；学校可以公布学位论文的全部或部分内容，可以允许采用影印、缩印或其它复制手段保存、汇编学位论文。本人电子文档的内容和纸质论文的内容相一致。

本学位论文属于：

□保密（校保密委员会审定为涉密学位论文时间：___年__月__日），于___年__月__日解密后适用本授权书。

☑不保密，同意在校园网上发布，供校内师生和与学校有共享协议的单位浏览；同意将本人学位论文提交中国学术期刊(光盘版)电子杂志社全文出版和编入 CNKI《中国知识资源总库》，传播学位论文的全部或部分内容。

(请在以上相应方框内打"√")

作者签名：刘永鑫　　　　　　日期：2018·6·22
指导教师签名：翁少雄　　　　日期：2018.6.22.
作者联系电话：15918706589　　电子邮箱：229568069@qq.com
联系地址(含邮编)：

# 摘 要

随着城市化进程的不断加快，城市公交系统作为城市重要的交通网络之一，面临着线网规模不断增长、客流动态变化不断增强、乘客出行需求日趋多样等诸多挑战。但一直以来公交行业缺乏有效了解乘客的出行特征以及个体线路选择偏好的手段，难以将乘客出行规律以及偏好特性运用到线网规划调整中。因此，根据公交运营单位采集的乘客出行刷卡记录挖掘其行为模式，分析乘客出行偏好，并随之调整线网及运力配置，对提升公交网络运行效率和服务质量，改善城市居民便利度具有重要的意义。本研究旨在通过多源数据融合修复原始公交运营数据的缺失信息，复现乘客 O、T、D 时空轨迹，然后从乘客的轨迹片段中挖掘出行模式，最后，用数据挖掘成果结合乘客个体出行轨迹和行为特征对公交线网进行优化。具体研究内容如下：

（1）在数据源缺陷修复层面，针对目前我国城市公交数据系统普遍存在的数据缺陷，提出一套在有缺陷数据环境下对一票制公交信息系统的数据进行预处理的方案。主要内容包括了数据采集系统时间误差的校正方法和 GPS 报站缺失数据的推断方法，一方面，所提供的方法将数据事件转化为离散时间信号序列，然后借助数字信号同步方法找到了数据源间的时间误差，并加以校正；另一方面，结合历史数据和无时间误差的乘客刷卡时间戳，修复缺失的报站数据，最终，得到全体乘客准确的上车刷卡时空轨迹数据集。有效的提升了一票制公交系统的数据质量。

（2）在乘客 O、T、D 出行轨迹复现层面，针对典型下车推断算法无法容忍信息缺失的缺陷，提出了公交乘客完整公交出行信息提取方法。首先在数据存在部分缺失的条件下，引入概率推断、居住地预估计等方法还原乘客的出行时空轨迹。其次，针对传统出行阶段识别算法错误识别乘客短暂活动的缺陷，在时空阈值的基础上提出了改进的出行阶段识别方法。最后，借助完整的 O、T、D 出行信息数据分析了目标城市公交客流的时变特征与满载率的空间分布特性。

（3）在乘客个体出行模式分析层面，针对城市公交乘客出行轨迹碎片化难以分析乘客个体出行模式的现状，提出了基于出行拓扑关系图的新型数据融合方法，从轨迹片段中提取公交乘客的闭合出行链，并分析了乘客的活动特征。首先将个体乘客多天的出行轨迹进行聚类、时空叠加，形成 OTD 出行拓扑关系图；其次在拓扑关系图的基础上搜索乘客的闭合出行链；再次，建立乘客不闭合出行轨迹集合与闭合出行链之间的内在关联；最后，基于全体乘客的闭合出行链集，从不同层面分析了目标城市的公交乘客出

行特征。提出了从公交乘客个体行为分布角度挖掘城市公交客流集散通道的策略。

（4）在乘客线路选择偏好挖掘层面，借助乘客出行链数据以及 OTD 时空轨迹信息构建了乘客的公交出行线路选择偏好模型，并对不同乘客的活动规律进行了研究。首先，借助前序数据挖掘乘客出行链数据以及 OTD 时空轨迹信息，提取了影响乘客出行线路选择的多个因素，建立了不同服务差异化场景下的乘客出行线路选择偏好模型并进行验证。最后，将乘客按不同的决策敏感因素进行分类，分析了其在工作日与周末的出行规律。

（5）最后，在利用乘客行为特征进行线路优化层面，本研究针对目前线路优化方法的不足，提出了一种结合乘客个体出行偏好与公共投入转化率的公交线路优化方法。借助粒子群优化算法与客流分布数据推演得到了不同优化目标下新线的所有可行解；然后，以目标城市典型公交线路为案例，从客流量、乘客出行效率、运力配置等方面讨论并选出最优线路方案；再次，对比了新线加入前后的客流分布以及出行乘客时间效率变化，将乘客个体出行行为数据挖掘以及知识发现的成果应用到公交线路优化中，为城市公交系统的运营调度管理提供决策依据和理论支撑。

**关键词**：公共交通系统；数据挖掘；出行链；线路优化；出行特征

# Abstract

With the rapid progress of urbanization, public transit systems have been regarded as the most critical mode of transportation. Nowadays, urban transit networks are facing several challenges: expanding network scales, changing variability of ridership along with variously growing reachability demands of residents. Unfortunately, the transportation agencies lack of efficient methods to understand and utilize passengers' travel pattern and service preferences so as to benefit the design and adjust of bus network. Therefore, it's of great significance to leverage the huge quantity of passengers' transaction records to mine their travel demands along with behavioral patterns so as to adjust bus headways and dwelling stations thereby contributing the performance enhancement and quality of service in urban transportation systems. In this research, we propose a series of methodologies to mine transit riders' travel pattern and behavioral preferences, and then we use these knowledge to adjust and optimize the transit system of collaborative city. The contributions of this research are:

Firstly, We propose a series of methodologies for data preprocessing in order to increase the data validity of our collaborator. Our major contribution of this part are: a) we propose a novel approach to rectify the time discrepancy of data between the AFC (Automated Fare Collection) systems and AVL (Automated Vehicle Location) system, our approach transforms data events into discrete signals and then applies time domain correlation the detect and rectify their relative synchronization discrepancies. b) By combining historical data and passengers' synchronized ticketing time stamps, we induct and compensate missing information in AVL datasets. Our approach can greatly enhance data validity of urban transit systems of China.

Secondly, In order to leverage the drawbacks of state-of-art algorithm for passengers' alighting point estimation, we first introduce maximum probabilistic inference and passengers' home place estimation to recover their complete transit trajectory from semi-complete boarding records. Then we propose an enhance transfer activity identification algorithm which is capable of specifying passengers' short-term activity from ordinary transfer procedures. Finally, we provide our analysis on the whole city's temporal-spatial distribution of ridership using recovered passenger trajectory.

Thirdly, To discover passengers' rigid travel demand from fragmented passengers' transit trajectories, we propose a novel graph based data fusion mechanism. We first cluster each passenger's trajectory data in multiple days and construct a Hybrid Trip Graph (HTG). We then use a depth search algorithm to derive the spatially closed transit trip chains from HTG;

Finally, we use closed transit trip chains of passengers' in the whole city to study passengers' travel pattern from various aspects. Finally, we proposed another option to discover transit corridors of the target city by aggregating the passengers' critical transit chains.

Fourthly, to obtain deeper understanding on transit passengers' route choice preference, we first derive eight factors that may have influence on passengers' transit route choice, and then construct each passengers' transit route choice model in regard of different service scenarios they have encountered. Next, we verify and assure our model by using ridership re-distribute simulation experiments. Finally, we conduct a comprehensive analysis on the temporal activity patterns of passengers with different route choice preference.

Finally, to make use of the rich information of passengers travel pattern thereby providing better transit services. We first provide a novel transit route optimization method by integrating passengers' transit route choice models and particle swarm optimization algorithm. Then, we then use a real-world case to study the result of various optimization target from the following perspective: passenger flow, passengers' time efficiency, vehicles' headways. Finally, we compare the ridership distribution and passengers' travel time efficiency before and after the establishment of new routes. Our research can provide useful insights in leveraging the power of Big Data to enhance the performance of public transit systems.

# 目 录

# 第一章 绪 论

## 1.1 研究背景

城市公共交通作为承担城市客流运输的重要交通方式，是倡导城市"宜居"环境的关键性环节，通过大力发展综合公交体系改善城市交通拥堵状况已成为各级地方政府的共识。随着城市交通基础设施建设的不断推进，多模公共交通（常规公交、BRT、有轨电车、地铁及其中转换乘枢纽）的出行模式变得愈加普遍。它给人们带来了快捷、安全和环保的出行方式，吸引着越来越多的居民从私家车转向公共交通出行。但是，公交乘车环境不尽人意，高峰期的客流拥挤和延误又将一部分乘客反弹回私人机动车出行。而公交企业却处于高峰期车辆调度紧张、平峰期车辆人员闲置、资源时空匹配无法满足客运需求的两难境地。究其根源，**运营单位无法从存在着结构性缺陷的运营数据中获得客流分布特征以及乘客的出行行为模式**，以至于公交运营单位无法准确获取有效信息并优化线网以及运力分配。

首先，虽然我国绝大部分城市的公共交通 IC 卡票务系统已投入使用多年，刷卡出行占比已超过 70%，信息系统中已积累大量的居民出行记录数据，但该系统主要是为了票务管理，不记录完整的出行信息。例如，一票制公交系统不记录乘客下车站点，地铁数据库有上下车站点，但不记录换乘站点；部分城市 BRT 系统甚至同时缺少换乘信息与乘客下车信息。**因此，城市公共交通运营数据虽然信息量极大，但存在器质性缺陷，缺失关键信息，不能清晰的反映乘客的客观出行需求，难以直接运用于公交行业管理与交通规划。**

其次，我国常规的公交系统评价体系多侧重于公交线网密度、公交站点覆盖率等静态指标的度量，评价内容侧重于公交线网布局和规划的合理性。而公交系统是一个动态系统，**现有的基于静态指标的公交系统评价体系，无法客观评价公交系统调整前后性能也无法刻画公共投入的转化率，不能为政府制定有效的综合交通政策提供决策支持。**

再次，现有公共交通系统规划多是运用传统的四阶段法模型，基于基于小样本交通调查以及不同交通方式的客流分配模型，对初始网络方案进行客流服务评价，并不断调整后形成公交网络方案，**本行业公共交通网络、线路以及节点规划与设计方法大多仍停留在定性和经验层面，缺乏定量的理论支持，对公交网络优化调整方法、模型与算法的研究尚存不足。**

为解决以上问题，构建高效的城市综合交通系统，需要借助科学的规划、调度、管

理手段，要求管理部门及时、准确、全面地掌握居民出行尤其是公共交通出行客流的时空分布规律、乘客出行行为模式和线网动态服务质量。大数据技术作为目前信息处理的前沿技术，具有强化多维数据之间的关联关系，弱化数据完整性及精确性的特点，特别适合目前形式多样、结构各异、信息完整度不一致的跨部门，跨行业海量异构化数据处理，为政府与行业部门及时、准确的了解出行特征，为预测出行客流态势与进行公共交通线网优化调整提供了有力的技术保障。

## 1.2 研究目的和意义

论文旨在基于已有的城市公交系统数据，如刷卡数据、车辆进站报站数据、线网地理信息数据、线路基本信息数据等，通过**数据挖掘手段复现乘客的出行轨迹，包括上下车地点和时间、换乘地点和时间，并基于乘客的出行信息进一步挖掘乘客的闭合出行链以及出行线路选择偏好的为代表的出行模式，最后提供一套基于个体乘客偏好的公交线路优化方案。**

本文在建立多部门参与的综合交通信息整合机制条件下，从跨部门、跨区域的多源大数据分析入手，对不完全信息下公交乘客出行模式分析方法进行了研究。首先，全样本地抽取城市公交刷卡数据、车辆 GPS 数据以及系统运行数据，在数据源存在缺陷的情况下提供一套具备普适性的数据预处理方案（时间误差软校准方法与缺失数据推断方法）。其次，利用乘客行为规律特点以及概率推断方法，完成多源数据时空关联融合，从不完全信息的原始数据中挖掘出城市居民公共交通出行轨迹（起点 Origination-换乘 Transfer-终点 Destination，简称为 OTD）。其三，通过提取乘客出行特征与出行模式，进行多层次的数据融合，以获得市域、行政区镇的公交客流分布与时空特征。其四，进行公交客流分布的数据推演与实证分析，从海量数据中挖掘乘客路径选择分析其时变规律，并推演线网调整后的客流重分配趋势。其五，依据动态客流数据及乘客偏好模型，建立公交系统优化调整模型，提出在保障乘客出行效率基础上优化运力配置和公交站点布置的快速公交线路设计方法。

因此，利用数据挖掘手段对公交乘客出行模式（含出行分布与选择偏好）进行研究和分析，对于现阶段"政府购买公共服务"背景下，进一步改善当前城市公交系统的规划、设计、运营、管理都具有重要意义。

## 1.3 研究内容和技术路线

通过海量数据挖掘，从不完全信息中挖掘还原公交乘客的出行模式，并提出快速公

交线路优化方案,以期从规划、设计、运营、管理等角度全面提高现有城市公交系统的服务能力和水平,本文的主要研究内容包括三部分:

**第一部分:公交运营数据缺陷处理与乘客完整出行信息提取,主要包含以下研究内容:**

(1)缺陷数据源背景下的数据预处理方法:针对目前"一票制"公交系统的数据缺陷,提出了乘客刷卡数据与车辆进站报站数据间的时间误差校正方法,该方法无需对现有公交数据采集系统进行硬件更新,还能借助本文提供的信息压缩方案高效的完成时间误差校准;此外,融合历史数据与乘客刷卡记录提出了缺失的报站数据信息还原方案。

(2)乘客完整出行信息提取方法。改进传统的下车站点推断方法,提升其对缺失数据的容忍度;然后,改进了传统的出行阶段(起点、换乘、终点)识别算法,使其能够识别乘客短时活动的,通过跟车调查验证了算法的有效性,并分析目标城市的客流时空分布以及线网的满载率。

**第二部分:公交乘客出行模式(闭合出行链与线路选择偏好两部分)分析及线路选择偏好模型,主要包含以下内容:**

(1)乘客闭合出行链挖掘方法。在乘客出行时空轨迹的基础上,提出了一种能够融合多天碎片化出行轨迹的乘客出行拓扑关系图构建准则,以及在此基础上的乘客闭合出行链搜索、关联算法;此外,对目标城市乘客的闭合出行链进行了分析,从个体乘客出行分布的基础上,提出了一种城市的公交客流集散通道挖掘方法。

(2)乘客线路选择偏好的分析。在乘客个体出行链以及OTD数据的基础上,提取了每个乘客多个关键特征,构建了乘客的线路选择偏好模型,以客流重分配推演验证了模型的有效性。还分析了具有不同特征的乘客在工作日以及周末的出行规律。

**第三部分:基于公交乘客个体出行偏好的公交线路优化方法:**

针对目前线路优化方法的不足,提出了一种基于乘客个体出行偏好的公交线优化路方法。借助粒子群优化算法与客流分布数据推演得到了不同优化目标下新线的所有可行解;然后,以目标城市典型公交线路为案例,从客流量、乘客出行效率、运力配置等方面讨论并选出最优线路方案并对比了新线加入前后的客流分布以及出行乘客时间效率变化,将乘客个体出行行为数据挖掘以及知识发现的成果应用到公交线路优化中,为城市公交系统的运营调度管理提供决策依据和理论支撑。

根据研究内容本研究的技术路线如图1-1所示,**总体思路为:从有缺陷、碎片化的数据中复现乘客的O、T、D出行轨迹,挖掘闭合出行链和行为特征,最后利用乘客行**

为特征指导公交线网优化。

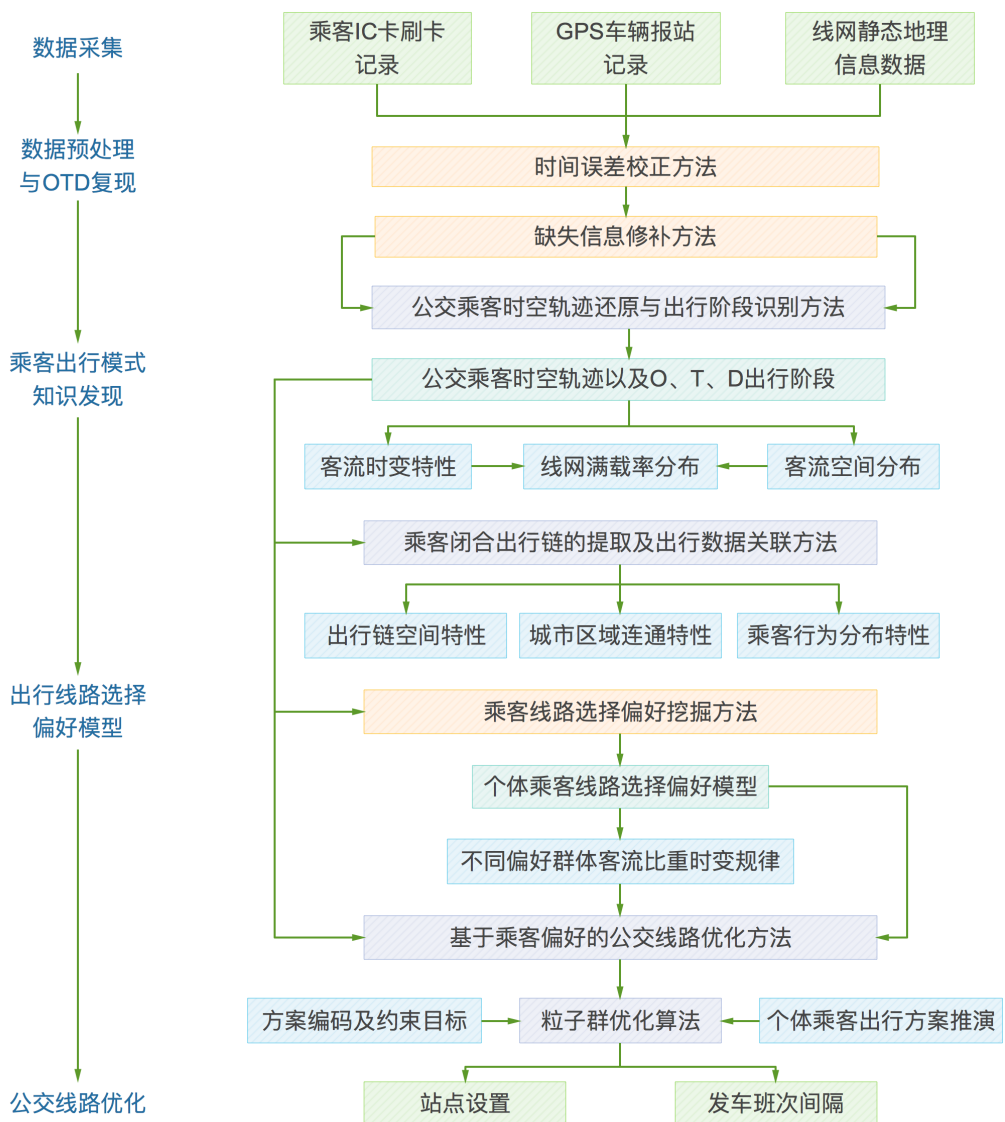

图 1-1 本研究总体技术路线

Fig. 1-1 Overall workflow of this research

## 1.4 创新点

依据本文的 5 个主要研究内容，本文主要的创新点分为以下三个方面阐述：

一、公交运营数据缺陷处理与乘客完整出行信息提取方法：

（1）提出了一种能自动校正数据源间时间误差与修复缺失信息的多源数据预处理方法，得到可靠的乘客上车刷卡轨迹数据。

（2）改进了传统的乘客下车站点推断与出行阶段（起点、换乘、终点）识别方法，增强其在数据存在缺失时的鲁棒性，并利用跟车调查数实证据进行验证，最后得到客流的时空分布规律以及线网满载率的分布特性。

二、公交乘客个体出行模式分析及线路选择偏好模型：

（1）提出了一种在碎片化出行轨迹数据的数据场景下，融合乘客个体多天的出行OTD数据，还原其闭合出行链的方法；借助该方法挖掘、分析了目标城市的乘客深层次的出行分布规律。

（2）提出了一种从乘客出行链以及出行轨迹中挖掘其出行线路选择偏好的方法；对目标城市乘客线路选择偏好进行了分析，发现了乘客面对不同的服务差异化场景会表现出不同的偏好特性，依据偏好特性划分了乘客群体，分析了不同群体占客流比重时变规律。

**三、基于公交乘客个体出行偏好的公交线路优化方法：**

提出了一种基于乘客个体出行偏好的公交线优化路方法。借助粒子群优化算法与客流分布数据推演得到了不同优化目标下新线的所有可行解；然后，以目标城市典型公交线路为案例，从客流量、乘客出行效率、运力配置等方面讨论并选出最优线路方案并对比了新线加入前后的客流分布以及出行乘客时间效率变化，为城市公交系统的运营调度管理提供决策依据和理论支撑。

## 1.5 论文结构

本文总共分为七大章节，主要介绍内容如下:

第一章，绪论，主要介绍了本文的研究背景和研究背景、研究目的及意义、研究内容及创新点、论文结构与技术路线。

第二章，国内外研究综述，主要介绍了本文研究过程中涉及到的重要理论基础，包括城市公交系统数据挖掘的基本方法、城市公交系统中乘客行为分析方法以及公交线网优化方法。

第三章，多源数据预处理方法。本章旨在针对我国目前各大城市公交数据系统存在的缺陷，提出一套能用于在有缺陷的数据源下对一票制公交信息系统具有普适性的通用化数据预处理方案，以期得到准确、可靠的乘客刷卡时空轨迹数据集。包括多源数据间的时间误差校正方法，和报站系统缺失数据推断方法。

第四章，公交乘客完整出行信息提取方法。改进了经典的下车站点推断与换乘识别方法，引入概率推断、居住地预估计等方法用以在数据缺失的条件下最大程度还原乘客的出行时空轨迹。借助OTD数据分析了该城市公交客流的时变特征与满载率的空间分布特性。

第五章，基于时空轨迹片段的乘客闭合出行链挖掘方法。本章提供了一套通过多天

数据叠加在轨迹片段中还原公交乘客闭合出行链的方法，并给出了将乘客不闭合出行轨迹集合与闭合出行链建立关联的方法。基于全体乘客的闭合出行链集，从不同层面分析了珠海市公交乘客的特性，特别是空间分布特性。

第六章，乘客线路选择偏好挖掘。本章借助乘客出行链数据以及 OTD 时空轨迹数据挖掘乘客的公交出行线路选择偏好并建立数学模型，并分析了具有不同特征的乘客群体的出行时变规律。

第七章，基于乘客线路选择偏好的公交线路优化方法。本章借助乘客出行偏好分布模型，设计一套基于粒子群优化与客流分布推演的公交线路优化方法与案例。

**本文所讨论的公交系统乘客出行模式在宏观上表现为客流的时空分布及其衍生特性；微观上表现为每一名乘客的闭合出行链及其线路选择偏好。**

文中涉及的全部地理信息数据可视化图均依托课题研究过程中自主研发的数据分析平台生成，其技术架构详见附录 **1**。

# 第二章 国内外研究现状

根据本文研究内容和技术路线，本章，国内外研究现状文献综述分别从运营数据中提取客流分布、乘客出行模式挖掘以及公交线网优化三方面对具有代表性较强的文献进行讨论。

## 2.1 公交线网客流分布提取方法研究现状

目前，公交系统数据挖掘应用中，最主要的目的就是获得乘客的下车站点进而求得客流的时空分布。现阶段最主要的做法为通过乘客的连续刷卡记录推断其下车站点，完成换乘识别的基础上，通过叠加得到乘客出行的时空分布规律；或根据数学模型以概率推断的方式，按站点吸引法推算客流 OD 分布矩阵。其中，利用乘客出行轨迹以及出行阶段获得 OD 矩阵是目前的研究热点。

我国乃至国外大多数城市的公共交通系统采用 IC 卡一票制的交易方式收费，一票制刷卡方式简化了乘客的乘车付费流程，但为乘客出行轨迹数据提取、分析等带来了严峻挑战。从一票制刷卡交易数据（主要包含乘客标识、车次、线路、交易时间戳）中，提取乘客出行轨迹的难点在于识别乘客的下车站点以及标记出行阶段。对此有两种方案：启发式规则推断法与概率推断法，其中启发式规则推断法从个体乘客出发还原其出行轨迹，而概率推断法则由具有相似规律的群体出发完成客流量在线网上的分配。

### 2.1.1 OD 矩阵概率分配法

OD 矩阵概率分配法主要思想为：首先，根据各车站的上下车客流量构建分布矩阵，并寻找每个 OD 对相应的乘客；然后，对于只有上车站点的乘客，根据站点引力模型或者相似度匹配按概率高低排序分配到各候选站点上，其关键点在于构建客流分配概率向量。该方法的代表案例为：

马晓磊等[1]，利用公交 IC 卡及 GPS 数据，对公交 IC 卡乘客上车站点推算进行研究。针对安装车载 GPS 设备的车辆，运用 GPS 数据与 IC 卡数据融合算法进行推算；对于无车载 GPS 设备的情况，为适应一票制 IC 卡数据挖掘，对贝叶斯决策树算法进行改进，允许节点跳跃，推算上车站点，并且利用 Markov 链特性降低算法的运算复杂度。

胡继华等[2]，提出一种对刷卡乘客分类推断上、下车站点并扩样叠加轨迹的方法。首先通过融合公交车 GPS 数据和 IC 卡数据来推断不同类别刷卡乘客的上、下车站点。然后，对全部乘客的出行轨迹进行叠加。

胡继华等[3]，分析了影响乘客选择下车站点的因素，提出了基于乘客个体特征的站

点吸引权概率模型，最后提供了验证单条公交线路上下车客流量的方法。

徐璐等[4]，使用了马尔科夫模型，建立了基于客流对称性的模型。利用乘客平均出行时间对模型进行校验和修正。

王超等[5]，在己知各站点上车人数的基础上，利用多种分布模型对推断公交线路 OD 矩阵，多种模型得到的结果基本相似，而改进的结构化模型可以得到更完整的 OD 矩阵；

高联雄等[6]，从智能公交卡付费信息和调度信息中挖掘公交车辆行程时间信息、进而利用行程时间和站点上下车人数信息估计公交动态 OD 矩阵的方法。

概率分配法具有能从较为宏观上反映客流分布规律且计算量较小，能够用于线网规划，但其忽略了大量有价值的乘客轨迹信息。

### 2.1.2 乘客出行轨迹推断法

"两票制"公交刷卡交易系统（如地铁系统、分段计价公交）中，无需对乘客出行轨迹进行特殊处理。然而对于"一票制"公交系统的原始数据，需要运用乘客出行轨迹推断算法根据乘客的出行行为规律，制定一系列规则，来判定乘客的下车站点参考点，其中最著名的准则为最短路径换乘或到达目的地原则，代表研究案例如：

章玉等[7]，基于 GPS 数据时间层次聚类（阈值 60s），获取乘客刷卡上车站点信息，然后基于乘车站距分布的下车站点获取算法（Tsygalnitzky 算法）获取下车站点；设定换车时间阈值、对乘客换乘进行判断。

王玥月等[8]，提出了公共交通通勤出行链提取的"四阶段"法，即出行链结构的提取、通勤出行行为的判别、出行阶段起讫点时空信息匹配、以及出行阶段行程距离和时间匹配计算。并利用站点空间位置最短距离法判断换乘下车站点。

陈君[9]等，WANG-Wei 等[10]，基于 IC 卡数据与 GPS 车辆定位数据，基于连续出行法，挖掘每个出行者的 OD 信息，并利用交通调查数据进行验证，结果表明，上车站点的挖掘成功率达到 90%，下车站点挖掘成功率仅为 57%，但从各个站点下车人数比例上看，交通调查数据与推断得到的数据相近。

**相比于概率推断法，基于启发式规则的方法可以跟踪每一个出行者，因而提供了更加详细的信息，有助于后续进一步分析乘客行为特征**，因此，有研究提供了对其进行改进的案例并尝结合"两票制"公交数据以验证其准确性，典型案例如下：

Munizaga M.A 等[11]，以圣地亚哥市为例，基于改进下车站点预判（同时考虑最小时间步行与最短步行距离 1000 m）法对两周的公交、地铁刷卡数据进行预处理。然后利用 30 min 为阈值区分换成点与目的地，利用 2 h 分割两次出行，其中，上车站点匹配成

功率高达 99 %，同时，用改进的下车站点预判法，利用均一化时间系数法，更合理的估计换乘站点，避免了最近距离换乘判决法的盲目性。

Munizaga M 等[12]，同样以该圣地亚哥为例，提出了一系列改进经典出行链方法和提高模型精度的策略：a)基于客流时间分布规律的城市交通出行起始时间分割方法。b)考虑乘车距离与欧几里德距离比值的出行阶段分割方法以识别乘客的短时目的地。c)对于每天只出现一次的刷卡记录，取第二天的第一次刷卡记录作为该乘客的下车站点。还借助志愿者进行验证，测试结果表明下车站点匹配准确率达到 84.2%，出行阶段划分准确率达到 90%。

Gaudette P [13]，采集蒙特利尔市公交刷卡数据与 GTFS 线路调度数据，并以地铁站为时空锚点，结合出连续出行链中最短距离换乘假设，还提出每天最后一次出行下车站点推断依据为当天首次出行上车站点的假设，提取出个体乘客 O-T-D 记录，得到客流时空分布。同时，作者将数据挖掘得到的 OD 结果导入 TRANSIMS 进行分析，分析结果表明所提出的换乘分配算法，可以得到的换乘客流与通过数据挖掘得到的换乘客流间存在较小偏差。

Alsger A 等[14]，以布里斯班市公交两票制公交系统的数据，假定无下车站点信息的条件下，对经典的 O-T-D 提取方法进行检验。检验结果表明，不同的换乘距离与换乘时间阈值设定都会对结果产生影响，且通过经典假设的得到的 OD 准确度仅为 65%，通过数据分析，作者提出乘客每天最后一次出行的下车站点为候选集中与当天第一次出行上车站点距离最近的站点。同时，准确度提升为 75%。

从以上案例观察，经过改进的直接推断法，其客流推断精度虽然宏观数值上仍低于概率推断法，但也已具备较高的使用价值。另一方面，对直接推断法与概率分配法带来较大影响的因素在于数据源的质量与完整性。目前大部分城市的公交数据系统并不能保证数据完整性，因此客流数据挖掘过程常常受到干扰，短时期内还需要寻找一套能够兼容缺陷数据源的同时，准确合理还原乘客出行轨迹以及出行分布的方法。

### 2.1.3 客流分析中的数据缺陷应对方法

公交数据挖掘以及应用各个环节，能获得可靠结论的前提是获得完整准确的 AFC 与 AVL 数据，但现实情况下，数据缺失与信息不一致无法避免，部分城市公交系统数据源存在严重的数据不一致甚至信息错误，主要表现在以下两方面：首先，公交车载刷卡计费系统（AFC）与车辆定位系统（AVL）独立，时间基准不同步（部分地市超过 45%的公交运营车辆中，系统时间误差超过 35 分钟），导致无法通过乘客的刷卡时刻与车辆

的时空轨迹匹配，推算出乘客准确的上、下车站点；其次，受城市建筑阴影效应影响，车载定位设备常常出现定位失败或通信丢包的现象，无法提供到完整的车辆时空轨迹。影响乘客出行信息推断与还原，3.75%的缺失轨迹数据即可造成超过 25%的乘客刷卡上车匹配失败[15]。

为克服以上缺陷，部分文献提出了基于启发式判定规则的数据处理方案：Steve Robinson 等[16]，首先针对公交智能卡系统存在的缺陷，分析了产生错误数据的原因。其次，提出了一种通过数据挖掘定性判断车载系统错误与针对乘客个体出行特征，纠正错误下车打卡数据的方法。并建议综合采用最大换乘次数、最长换乘时间、不重复线路、最长出行时间、直线系数等多种因素判断乘客的换乘行为。

Ma Xiao-lei 等[17]，基于马科夫模型与贝叶斯决策树，构建了一套不需要借助其它外部数据源，即可推断刷卡乘客上车站点的方法。该方法假设站点间的平均行程时间服从高斯分布，并假设后一站的概率只与当前站点及平均行程时间有关，进而构建马科夫决策树模型。并对该模型进行验证，发现仅有不到 20%的站点能够完全判断正确，剩下 80%的站点存在 2 到 3 站误差，该方法为 GPS 数据无法得到时判断乘客下车站点提供了一定的思路。

Zhang F 等[18]，基于乘客智能卡刷卡数据、车辆定位数据、IC 卡交易金额与余额、收费费率以及乘客换乘特性，得出一系列判断乘客下车站点的约束条件。再利用条件随机场模型与协同滤波算法对不完全信息下的公交乘客 O-T-D 进行提取，试验结果表明，该方法可以达到较高的准确度。

Kusakabe T 等[19]，以大坂为案例，基于智能卡 OD 数据与交通调查数据构建的先验概率模型，提取到达时间、车站内停留时间、起伏点，结合朴素贝叶斯分类法，提供了一种基于数据融合技术的出行目的判断方法。

以上案例虽然为从有缺陷的数据源中还原公交乘客的出行信息提供了一定思路，但未能从根本提升现有公交数据系统的数据质量，**从不完整甚至有缺陷的信息中还原公交乘客 OTD 时空轨迹仍是目前本领域的难点**。

## 2.2 公交乘客出行模式分析方法

公交乘客的出行模式分析可以按分析目的分为宏观与微观两个层面，宏观层面上，公交乘客出行特征挖掘主要的研究目的是为公交运营管理提供宏观层面上的决策支持，而在微观层面上，出行链则更有利于对乘客群体行为特征的进行分析。

## 2.2.1 公交乘客出行特性挖掘

宏观层面上的数据挖掘被用于了解客流时空变化规律或线网资源的使用率，以期为公交系统规划以及整体政策调整提供依据，主要案例如下：

Bagchi M. 等[20]，从韩国实行的分段计价地铁和公交公 IC 数据中提取包括人均刷卡次数、乘客出行的时间和空间分布等指标，并用于公交市场分析以及公交企业的运营管理。

Seaborn C. 等[21]，利用伦敦市公共交通 IC 卡数据，研究了乘客在公交与地铁两大公共交通系统之间相互换乘的时间阈值，将相互独立的轨迹连接成完整的出行链，并在此基础上，对每日公共交通出行链总数、每天人均出行链数、出行阶段数、地铁与公交混合出行数等进行了统计分析。

Catherine M. 等[22]，以加拿大魁北克市为案例，分析了乘客刷卡量在时间上和空间上的分布特征，然后提取了公交站点的被使用频率的时变规律，最后，通过聚类分析了不同种类乘客公交出行的时变化模式。

Wonjae J. 等[23]，通过韩国公交智能卡数据，挖掘了乘客公共交通出行时间和换乘特征，绘制了行程时间空间分布图，分析了换乘时间、换乘需求等参量的空间分布，为线网优化提供依据。

Foell S. 等[24]，以里斯本的公交系统为例，融合 AFC 与 AVL 数据，基于经典假设得到个体乘客 O-T-D 序列，分析城市公交客流的时间分布。然后，从站点到线路两个层面，分析了公交系统的使用概率分布。再次，通过相似性比较，作者发现乘客在工作日站点、线路的使用相似度远高于周末。最后，分析了连续乘车事件在时间上的概率相关性。

Guangxia L. 等[25]，以新加坡两票制度公交为例，用时空阈值法由三个月的出行链中寻找停留时间与访问次数靠前的站点，提取时间序列的傅立叶参数后，再利用 local kenemization 算法进行投票判断居住点。该文献表明，提取出的职住分布与新加坡人口调查的职住分布具有更强的相关性。

龙瀛等[26]利用连续一周的两票制公交刷卡数据，构建了基于乘客乘车过程的通勤出行识别模型，和基于乘客在停驻点到达时间和逗留时间的职住地识别方法；然后，结合用地性质分布，分析了北京 3 大典型居住区和 6 个典办公区的通勤出行特征。研究结果与 2005 年居民出行调查结果相符。

## 2.2.2 基于出行链的乘客行为模式分析方法

出行链研究是乘客交通出行行为研究的重要部分，其描述了出行者从起始点出发，经过若干个目的地，再返回起始点的出行全过程。它摒弃了传统交通行为模型中表征每一次出行的孤立静态性，真实揭示了城市交通出行全过程的连续性特征，体现了乘客交通出行的连续动态性[27]，可以认出行链是刻画乘客出行模式的重要手段。国外基于出行链的研究起步较早，研究范围也较广，主要从出行链空间拓扑、出行链中各阶段交通方式选择两方面进行。主要案例综述如下：

Thomas F.Golob 等[28]，在 2000 年时研究了家庭活动参与和出行链生成的联立模型，作者基于基本的出行生成模型，加入了时间利用的观点，结合了出行活动和出行时间将模型进行改进。该模型共同预测了活动、出行时间和出行生成作为居民特征和可达性指数。同时表明基本模型将活动分为工作和非工作两类，可以拓展三种活动：生存、娱乐以及义务活动。该模型同时分析了在家工作的出行链和活动参与的特征。

Valiquette 等[29]，利用蒙特利尔交通出行调查数据，基于环路（开环、闭环、单环、多环）、锚点（居住地、工作地、学校）、出行目的对居民出行链进行分类讨论；数据统计结果表明，儿童跟随出行、拥有私家车，家庭组成等因素都会影响居民出行链的复杂度、直径、频率产生重大影响，同时，以工作为目的出行链成为了城市居民最主要的出行链。

Liu Zhenru 等[30]，在以出行链为基础的居民出行行为研究综述中认为，对基于出行链的分析方法应该基于出行链活动，深入分析居民对出行链的选择，进而对出行行为进行研究，为制定交通规划和交通管理政策提供一个更好的决策支持。

Yuhwa Lee 等[32]，基于家庭在户外的出行链的活动相互联系假设，运用模型估计来解释家庭类型和机构、时间分配以及出行链模式之间的关系。户外的生存性活动、维持性活动以及自由活动的维持时间被当做独立的变量，同时来检测家庭出行链中的时间分配策略。研究结果表明家庭内部关系、家庭类型和结构以及家庭户主的属性都和出行链行为之间存在内在的联系。

Thomas F.Golob 等[33]，使用悉尼从 1997 年开始的持续的居民出行调查数据，研究五年一组中个人出行活动链，建立了出行行为、年龄、收入和生活环境的模型。研究发现，在复杂的出行链中，老年人的出行需求由开车转变到了乘坐小汽车，最后转到了公共交通上，尤其是单人出行和几乎所有的女性。最后针对老年人群体出行不便而提出对公共交通、交通基础建设的建议。

Xin Ye 等[34]，检验了乘客活动出行模式和模式选择的复杂性之间的内在关系。通过对工作出行链和非工作出行链的模型进行探讨，统计模型检验结果表明对于工作出行和非工作出行，出行链的复杂程度决定了模式选择。

Stefaan V. Walle 等[35]，使用比利时的调查数据进行分析研究后发现在公交出行中，步行时间、候车时间、换乘对公交出行有负面影响，居民公交出行能够接受的最长等待时间和候车时间由总的出行时间决定。另外在郊区地区，人们的出行行为比更复杂，小汽车、自行车等步行以外的方式被作为公交的接驳方式，以填补出行方式链中缺失的部分。

MA-Xiaolei 等[36]，以闭合出行链法建立每个乘客的出行轨迹，对城市居民的出行链时空序列进行聚类分析，提取潜在的规律，同时基于历史数据，分析出行模式的稳定性（或支持度）。其中，分析出行模式稳定性使用的输入属性包括：出行天数，相近的第一次乘车时间数。相似的线路序列数量，相似站点数列数量。出行稳定度用粗糙集理论对通过聚类算法发现的 5 种行为模式构造了快速分类规则。

国内基于出行链的乘客出行行为研究起步较晚，主要研究方向集中在出行活动分析以及交通方式选择分析方面。其中，基于交通方式选择分析方面，主要借助交通出行调查获得居民的出行日志[37]，并用出行链对活动特征进行建模，最后分析不同活动对于交通方式选择的偏好特性。主要案例包括：

Xianyu 等[31]，借助 MNL 模型，分析出行交通方式、职-住出行链与出行者社会经济属性间的联系，该研究结果表明，工作出行为出行者的首要出行链组织因素，而出行方式决定于活动类型；

李妲等[38]，对北京市节假日出行特征进行分析，以出行链为基本分析单元，找出了出行链的时间特征、活动模式以及出行的交通方式选择上的不同，建立了基于活动的出行选择模型。

褚浩然等[39]，提出了出行链的特征指标，并运用多项 Logit 模型建立的交通方式选择模型，验证了平均出行链长度对交通方式选择有显著的影响的假说，选择乘坐私人交通方式的可能性与出行链长度成正比。

宗芳等[40,41]，运用活动链和中途停驻的概念，建立了基于活动的出行时间和方式选择模型，并用公交优先策略进行了模拟分析。

李萌、王迎等[42,43]，对活动链的各阶段交通方式与居民的基本属性的关系进行分析，通过居民活动链和出行方式链建立了出行方式选择模型。

基于出行活动的分析是从个体行为出发,将居民的各种出行以链的形式进行跟踪分析,目的是剖析交通出行的微观机理,研究人们的出行行为特征和选择特征。主要对出行时间、出行方式、以及出行目的地等方面进行分析,进而构建出行方式选择模型。主要研究案 例包括:

杨柳等[44],对城市居民的出行前选择行为进行了分析,运用出行链的分析方法,总结了我国大城市通勤出行链和非通勤出行链的基本特征。从微观的角度构建了基于出行链的出行时间、出行方式以及出行目的地的选择模型,其基于活动的交通需求预测方法为规划人员提供了一种新的出行需求预测方法。

石心怡等[45],研究了出行链的时间分布特征以及交通方式组成,通过对土地利用性质与出行链的关系分析,得出了出行链的空间分布特征规律:不同的用地性质吸引了不同类型的出行链,基于工作、基于生活以及基于休闲的出行链分别分布在城市不同的交通走廊上。

马飞等[46],分析复杂通勤出行链脆弱性的影响因,采用预抽样数据和主成分分析法对主要影响因素进行识别,提取交通网络拥堵、公共交通衔接不畅、机动车停车困难、交通诱导信息误差和主观因素等 5 个公因子,设计了复杂通勤出行链脆弱性感知的测度 指标体系;运用因子分析方法得出各影响因素之间的关系。研究结果表明交通网络拥 堵和公共交通衔接不畅对复杂通勤出行链脆弱性感知水平的贡献度最大。

何流等[47],对居民日出行情况进行建模与仿真。将出行者分为3类,分别建立学生出行次数 logistic 回归模型、老年人出行次数泊松分布模型及就业者出行选择模型。结合城市宏观社会经济数据,对模型进拟合,得到不同类别出行者的出行次数。以中国中部城市郑州市为例,预测了居民的平均出行次数。

吕婷婷等[48],探索南京市市郊居民的通勤出行规律,并利用 MNL 模型,对南京市市郊通勤者出行方式及出行链选择进行研究及实证建模,表明实际就业-居住关系的均衡有助于减少市郊居民的通勤时间和通勤距离,职住一体化程度越高的区域职住分离度越低;

**综上所述,国内外针对出行链的研究范围较广,涉及到多个相关的研究主题。国内目前基于出行链的研究主要集中于出行链属性与居民社会经济属性关联关系、出行链模式选择以及方式选择模型等方面;且大部分关于出行链的研究都是基于居民出行问卷调查数据进行的**,费时费力,同时数据不能及时更新,并不能满足当今快速变化的交通的要求。随着新型数据采集方法的兴起,部分案例开始采用时空定位数据,并借助数据挖

掘手段挖掘出行者的出行链及其行为特性，代表案例如下：

Zhao J 等[49]，将共享单车用户的出行链按拓扑特征分类并进行分析，结果表明，闭合自行车出行链的截止时间为 1 小时，而非闭合自行车出行链的截止时间为 30 分钟；研究还表明，男性骑行者在周末相较女性更容易单环形成闭合出行链，女性骑行者则有更大概率形成多环路出行链；

Ying Hu 等[50]，利用共享汽车 GPS 轨迹数据，将共享汽车形成的闭合出行链分类，并按照半径、停驻时间比、停驻点数量对共享汽车的出行链进行聚类分析，该研究表明绝大多数共享汽车用户所产生的出行链具有的形式为多点、短时停留、中等半径出行链。

王俊兵等[51]，利用北京公交刷卡数据完成了不同时段不同出行模式的出行链提取，还原了乘客出行选择，进行宏观出行特征的统计分析，包括乘客全日刷卡量的时间分布特征、乘客出行的空间分布特征、出行时间和距离的计算及换乘特征。

**在公共交通领域，不依赖交通调查，仅采用数据挖掘手段对乘客的闭合出行链进行还原与挖掘分析仍是一个难点**，现阶段公共交通数据挖掘及应用虽在一定程度上揭示了乘客的活动规律，但大部分研究案例均未采用出行链来对乘客行为建模；另一方面，公共交通领域闭合出行链的研究国内多停留在轨迹数据还原阶段[52]。

## 2.3 公交线路优化方法

公交线路优化方法是目前公共交通领域的热点，与常规优化问题类似，公交线路优化的目的在于寻找合适的站点数量与运力配载，使得线网规划或调整阶段的目标函数在给定 OD 矩阵和约束条件限制取得极大值。换而言之，将线路优化以及快速公交线路设计设问题转化为数值优化问题后用智能优化或启发式算法求解，目前国内的研究中，公交线网优化常采用多参数加权组合优化，代表案例为：

聂瑶等[53]，提出公交线网的分层优化模型：快线层以直达客流密度最大为目标函数；普线层以乘客总出行阻抗最小为目标函数；支线层则以线网覆盖率最大为目标函数。采用蚁群算法求最优解。

罗孝羚等[54]，构建了以乘客出行时间最小化为目标的公交线网及发车频率的混合模型，并设计了相应的改进遗传算法求解该模型。

胡继华等[55]，针对公交线网规划问题, 提出一种多目标的公交线网规划模型，以最小化乘客总出行时间和总换乘次数，最大化线网的需求密度为目标函数。利用禁忌搜索和模拟退火两种算法对模型进行求解, 并利用 Sioux-falls network 数据集进行验证。

蒋阳升等[56]，结合现有的可达性相关研究现状，提出了公交线网广义可达性的相关概念。在此基础上，建立了公交线网空间可达性的优化模型，并设计了遗传算法对模型进行求解。

徐茜等[57]，以系统总成本为目标函数，以线路长度、频率等为约束条件的公交线网设计模型。对该模型采用模拟退火算法求解，并对线网解进行调整。

齐振涛等[58]，从直达客流量的角度对公交最优路径进行求解。采用公交线路必经中间站点的选取方法，将 k-最短路径算法进行改进，用于求解最大直达客流量路径；以最大直达客流量、人均最小成本加权作为优化目标，建立了基于网络人均出行时间成本最小的公交线路优化模型。

蒋阳升等[59]，以客流强度与可达强度匹配最佳和可达性值最大为目标，以首末站布局位置要求为约束的公交线网可达性优化模型，用遗传算法求解，通过算例进行验证。研究结果表明新方案可达强度与客流强度匹配程度有所提升。

郭戎格等，宋子航等[60,61]，以出行者出行成本最小、运营商运营成本最小为目标建模，考虑定制公交"直达、零换乘"特点，结合约束条件并采用遗传算法求解。

国外，对公交线路优化的起步较早，但较之于国外大部分案例中不涉及发达的公交线网，因此研究思路与国内研究差异较大。部分国外研究案例目标城市中，公交站点稀少，发车班次间隔超过 30 分钟，导致乘客的换乘时间成本剧增，因而专门出现了以减少换乘时间损失为目标的线网优化案例：

F Cevallos 等[62]，以佛罗利达某公交系统为例，对车辆班次时刻表进行调整，以尽可能降低乘客的换乘候车时间，该案例同样将线网优化问题转换为数值优化问题，并采用了遗传算法进行求解。

Y Hadas 等[63]，利用客流数据结合设计了动态变化的时刻表以加入区间车，以使乘客出行时间达到最小同时减少可能的换乘候车时间以及需要的步行换乘距离，该案例最后使用了动态规划算法寻找最优解。试验表明最高可以减少 17% 的乘客出行时间。

M Schröder 等[64]，通过调整班次时刻表达到减少乘客换乘时间以的方法提升乘客换乘质量。

Y Shafahi 等[65]，通过调整车辆车头时距、时刻表等手段达到了优化换乘的目的，该案例中同样采用了遗传算法寻找最优解。

同时，在公交线网发达的大城市，则是采用了与国内目前研究方案相似的多目标组合优化方法：

CE Cortés 等[66]，提供了一套基于遗传算法的多目标公交线网运营策略优化方法，该方法允许经营者调整发车班次以及使用区间车，由遗传算法寻找一套实时调度策略，让乘客等待时间最少的东西尽可能降低运营成本。

L Fu 等[67]，提供了一种将区间车与全站车进行协调调度的方案其目的是让乘客的等待时间、车辆行驶时间以及乘客的出行时间之和取得极小值。由于涉及站点较少，故使用了穷举法求得全局解。

SUN Chuanjiao 等[68]，提出了一套通过控制 BRT 线路运营车辆车头时距以使得乘客的出行时耗与运营成本达到综合极小值的方法，该案例使用了可变长度编码的遗传算法寻求最优解，数值仿真表明该方法可以有效减少双方的综合损耗损耗。

F Zhao 等[69]，提出了一种用户等价出行时间以及换乘次数最少同时线网最大化覆盖用户出行需求的公交线网优化方案，并对比了禁忌模拟退火法以及梯度下降法来求解最优解，结果表明模拟退火算法能够在绝大多数情况下表现的更好。

M Nikolić 等[70]，则是采用了人工蜂群算法，优化目标同样是最少换乘的前提下使得大部分被服务的乘客出行时间最短。

Y Hadas 等[71]，提出了一种使得空座率以及无法上车的乘客同时最小化的公交线路优化运营方案，该方案根据乘客出行需求同时调整发车频率与所采用的车型。

对比国内外的案例，可以发现以下特点：首先，采集的数学手段与优化过程并无过多差别，但相比之下，**国外研究案例会更多考虑平衡公交公司的运营成本与乘客时间花费，并利用实际的线路对优化方法进行验证**[72]；其次，针对公交线路优化问题，也有研究案例采用了改进的搜索算法[73,74]，但本质上并无特别显著的区别；再次，**现阶段所有的线路优化问题均欠缺对乘客个体线路选择以及行为偏好的考量**（对于线路选择目前多按候车时间按概率分配[75]），**未能将乘客个体出行行为数据挖掘以及知识发现的成果应用到公交线路优化中**。

## 2.4 文献综述结论

本章探讨了公共交通领域的数据挖掘案例以及将数据挖掘成果应用于城市公交线网优化的方法与案例。经总结，得到以下结论：

（1）从城市公交多源数据中还原乘客的出行轨迹是业界重点探讨的课题。

（2）现阶段，公共交通数据源在信息完整性方面尚不理想，有缺陷的数据严重影响数据挖掘结论的有效性。

（3） 采用新型数据挖掘手段，在不依赖人工交通调查的前提下，挖掘乘客的闭合出行链以及行为特征的研究仍有待深入。

（4）现阶段通用的公交线网以及线路优化方案尚未考虑乘客的个体方案选择偏好，多用统一化模型覆盖个体乘客，有待提出更合理的方案。

# 第三章 公交系统多源数据预处理方法

随着公交乘车 IC 卡及 GPS 报站系统的广泛使用，我国目前各大城市均能提供不同种类的公交数据以供分析应用。但是，公交信息系统（尤其是一票制公交的信息系统）设计目的均是为了票务清分，未考虑借助票卡数据以及车辆定位数据研究乘客的出行行为以及区域客流时空分布规律的需求。因此从原始刷卡乘车数据还原乘客的时空轨迹，存在较高的技术难度。另一方面，公交 IC 卡数据与车辆报站数据间往往存在时间同步错误以及数据丢失的现象，导致信息不一致，后续的关联匹配以及一系列的数据挖掘无法继续进行。数据质量控制目前仍为公交系统中的空白，尚未引起足够重视，也没有一套完整的数据质量测控标准体系。**综上所述，公交系统中的原始数据存在器质性缺陷，不能直接用以进行深层次分析。**

**本章以及第四章共同围绕的主题是公交运营数据缺陷处理与乘客完整出行信息提取。**在本章，我们首先针对公交 IC 卡收费系统与 GPS 报站系统间的时间同步误差提出了一种基于事件时空关联与信号关联匹配的时间误差校正方法；其次，针对 GPS 报站数据缺失，提出了一种融合乘客刷卡时间戳与区间行程时间的报站系统缺失数据推断方法；结合这两种方法完成了数据源存在缺陷条件下的公交乘客上车站点匹配。全章内容旨在提出一套能用于在有缺陷数据环境下对一票制公交信息系统具有普适性的数据预处理方案，以期得到准确、可靠的乘客刷卡时空轨迹数据集。同时，本章所述的数据处理方法无需对公交系统的硬件进行升级改动，也能用于处理其他有缺陷的历史数据。

## 3.1 数据源及总体数据流图

本节内容涉及两大数据源，分别为乘客刷卡数据与 GPS 报站数据。我国大部分采用"一票制"策略的城市，所提供的公交运营数据关键字段如表 3-1 所示。由表 3-1 可知，乘客刷卡数据源（即 AFC 数据）与 GPS 报站记录（即 AVL 数据）共同的信息字段为车牌号与线路号。而在大多数情况下，乘客的刷卡时刻均在进站后以及车辆出站后一小段时间内，因此通过时间匹配可以确定乘客的上车站点。但前提是 IC 卡收费系统的时间误差不能大于允许的匹配误差阈值，且对应时刻必须有对应的报站记录。

数据预处理如图 3-1。具体来讲，数据采集初步格式转换完成后，分别进行了乘客刷卡数据时间误差校正与车辆定位数据信息修补。**时间误差校准方法将数据由事件域转为信号域，并借助了数字信号处理方法成功提取出时间误差，并进行校正**；车辆定位数据修补的关键在于建立合理、高效的概率分布模型，准确推断出丢失的数据。本章

所讨论的数据预处理算法最终输出结果为经过了时间误差校正以及缺失站点数据修补的乘客刷卡事件时空轨迹数据集。

表 3-1 数据源关键字段

Table 3-1 Principle information in the related data sources

| 数据源 | 关键信息字段 |
| --- | --- |
| 乘客刷卡数据 | IC 卡卡号、刷卡时刻、车牌号、线路号 |
| 车辆报站数据 | 车牌号、线路号、车站名、进站时刻、出站时刻 |
| 车辆坐标与线路表 | 线路名、方向（上行或下行）、站点列表、站点坐标、线路轨迹坐标 |
| 车辆调度记录 | 车辆调度记录&线路、发车班次、车牌号、发车时刻、经停（或绕行）站点 |

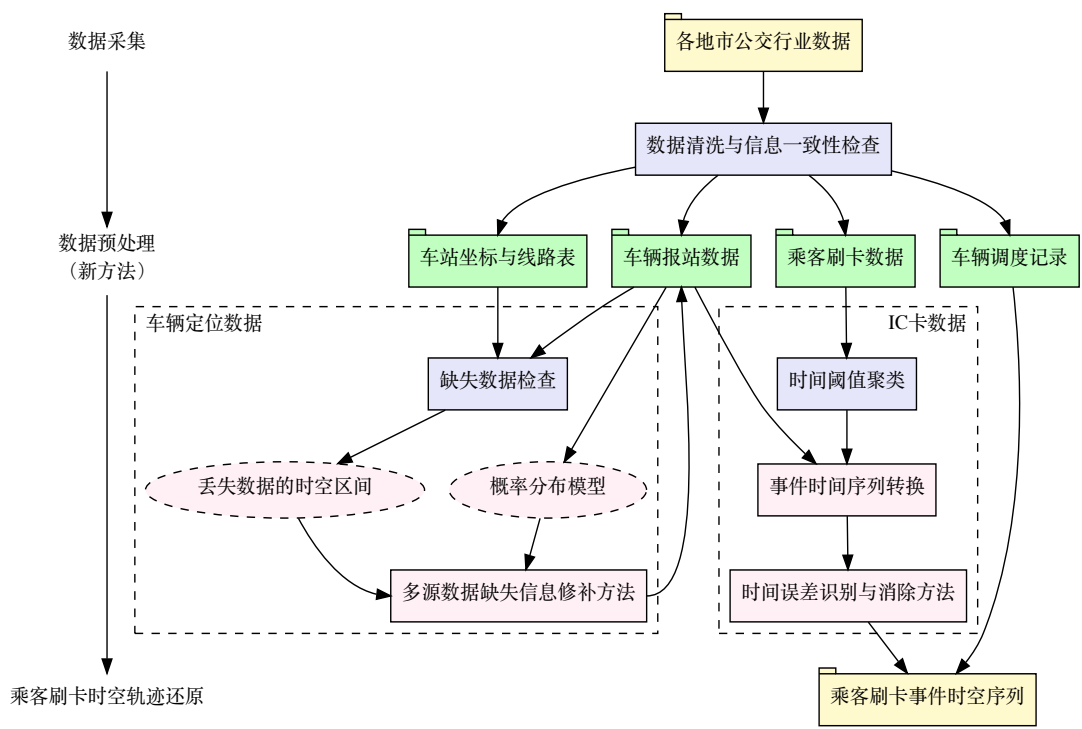

图 3-1 数据预处理总体流程

Fig. 3-1 Overall workflow for data preprocessing

## 3.2 时间误差自动消除的乘客上车站点匹配方法

目前，已有的公交乘客上车站点推断方法按照基本原理可以分为两类：概率推断法[17,76,77]、直接推断法[18,78]。两种方法各有优缺点。

概率推断法通过假设车辆经过每一段连续站点对所构成的路径的时间均服从高斯分布。通过轨迹记录器取得每段路径的行程时间分布，建立马可夫模型或贝叶斯概率推断模型；由每一趟车的起始站点为原点，起始时刻为零时刻，计算乘客刷卡时刻与零时刻间的时间差，并借助概率图模型寻找使后验概率取得极大值的站点。该方法可在没有

AVL 数据的情况下，仅凭已经训练好的模型推断出乘客的刷卡上车站点，但该方法也存在明显的缺陷，公交车辆在两站点对间的行程时间会受到道路条件、车流量、驾驶员状态、能见度等因素的叠加影响，并不总是服从高斯分布，得到的模型存在偏差。

直接推断法假设乘客在车辆进站与到达相邻下一站的时间内完成上车、刷卡，因此，该方法采用一定的时间阈值（通常取 45s[79,80]）将乘客与 AVL 数据源中的车辆报站时刻进行时间关联，将 AVL 报站数据中与刷卡时刻最接近且满足时间阈值限制的站点推断为上车站点。该方法简单、易于实现无需对概率分布进行假设，也无需对概率推断模型进行拟合，能够准确、唯一确定每一名乘客的上车站点。但该方法适用前提是 AFC 自动收费系统与 AVL 自动报站系统间的时间误差已知且为恒定值。若两系统间存在未知时间误差，则无法得到上车站点或直接得到错误的上车站点。为改进该方法的缺陷，本研究在直接推断法的基础上，提出了基于事件关联与时间误差自动校准的乘客上车站点推断算法。

### 3.2.1 算法总体流程

本节提出的时间误差自动校正的公交乘客上车站点匹配方法总体流程如图 3-2 所示，该算法主要步骤如下：

**Step 1** 对每辆车的每个班次的运营数据，提取其所有乘客上车刷卡时刻，并按照时间先后进行排序，其中所述每一趟行驶定义为从起点站到终点站；

**Step 2** 通过跟车调查或者历史数据合理选择乘客刷卡数据时间分割阈值为$\varepsilon$，将相邻时间间隔小于$\varepsilon$的刷卡记录归为同一上车站点，即可将同一辆车同方向各站点的乘客刷卡数据分离；

**Step 3** 提取每个班次每站点第一个刷卡的乘客时间戳形成的时间序列；再提取公交车辆在该班次每站点进站时刻，形成的二元（只有0与1）时间序列；

**Step 4** 将以上两时间序列中不为零的时刻进行时域脉宽延拓，即每个不为零的时刻转换为一个宽度$t_w$的正脉冲，形成两路新的信号序列；

**Step 5** 将两路信号序列加上时间推移量$\tau$后进行时域相关运算，寻找使得时域相关运算结果为极大值的时间推移量。如果时间匹配精度允许大于5s,则可采用数据压缩方案对两时间序列进行预处理后匹配。数据压缩的具体思路是：根据采样定理，在不损失信息的前提下对信号序列进行抽样，减少需要处理的信号序列长度从而压缩搜索范围。

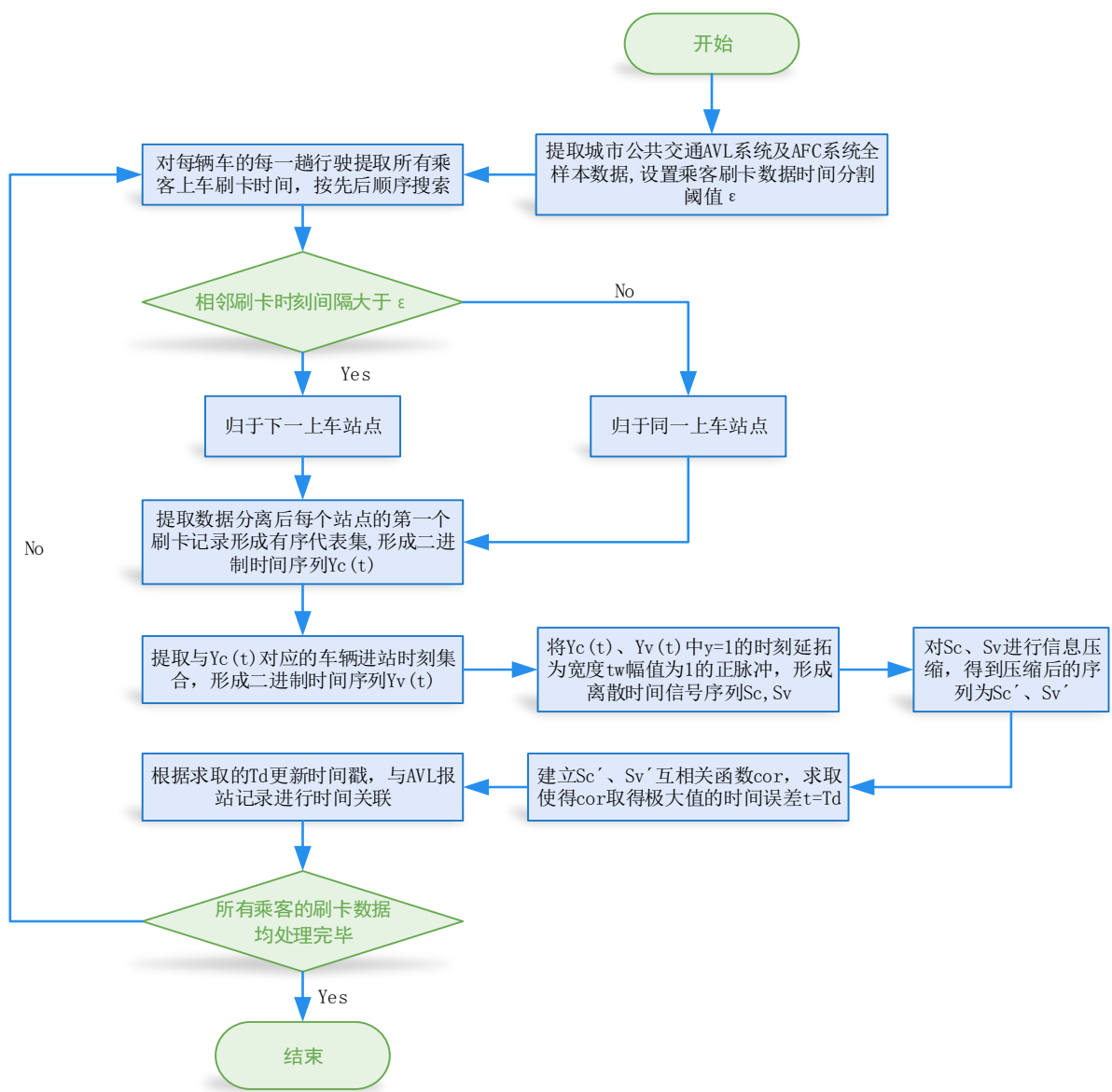

图 3-2 时间误差自动消除的上车站点匹配方法流程

Fig. 3-2  Overall workflow for time error rectification boarding correlation

### 3.2.2 算法关键步骤

#### 3.2.2.1 乘客刷卡数据时间分割

乘客刷卡事件与公交车辆的到站事件间存在确定的关联。假设车辆 $p$ 在某一趟行驶（定义为从起点站到终点站）时，产生的所有 AFC 乘客刷卡数据定义为$C_P$，在第$i$个车站产生的乘客刷卡记录形成的有序集合为$C_i = c_i^1, c_i^2, ..., c_i^m$，产生的报站记录为$V_i$。则：在$C_i$所有元素与$v_i$关联度最高的元素为该站第一条刷卡记录$c_i^1$，本研究定义$c_i^1$为站点 $i$ 的代表元素，车辆在每个车站的乘客刷卡数据中的代表元素可等价于该车在此站的进站事件。为从某车辆全部 AFC 乘客刷卡数据$C_P$中分离出各站点的乘客上车记录$C_i$，本研

究对南方某城市公交车进行跟车调查，记录 1297 名乘客的上车刷卡时刻，得到的同一车辆停站后相邻乘客上车刷卡间隔时间分布如图 3-3 所示：

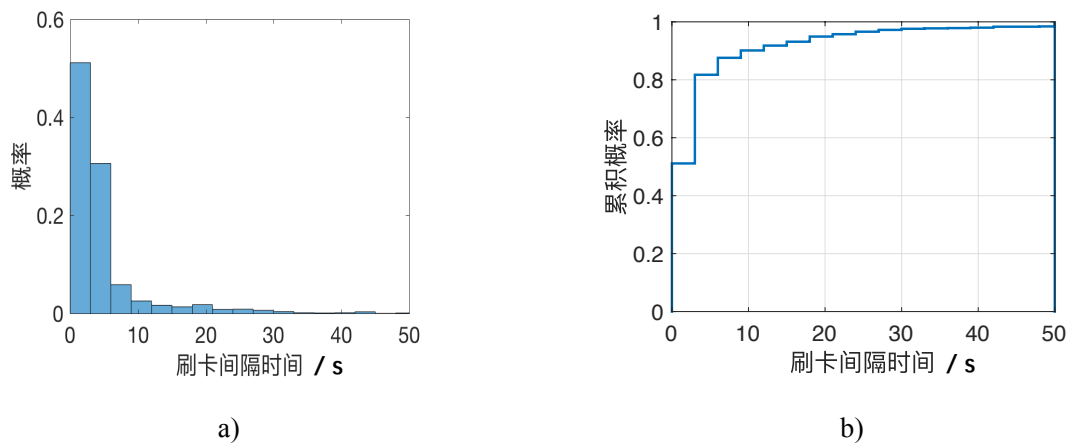

a)                                   b)

图 3-3 相邻乘客刷卡时间间隔分布

Fig. 3-3 Distribution of intervals of smart card events within consecutive boarding passengers

由图 3-3a，大部分相邻乘客的刷卡间隔时间分布在 1~10s 的区间内，从图 3-3b 累计分布上看，20s 已覆盖了超过 90%相邻乘客的刷卡时间间隔。

通过 AVL 车辆报站数据得到该城市公交站点间车辆行程时间分布如图 3-4 所示，

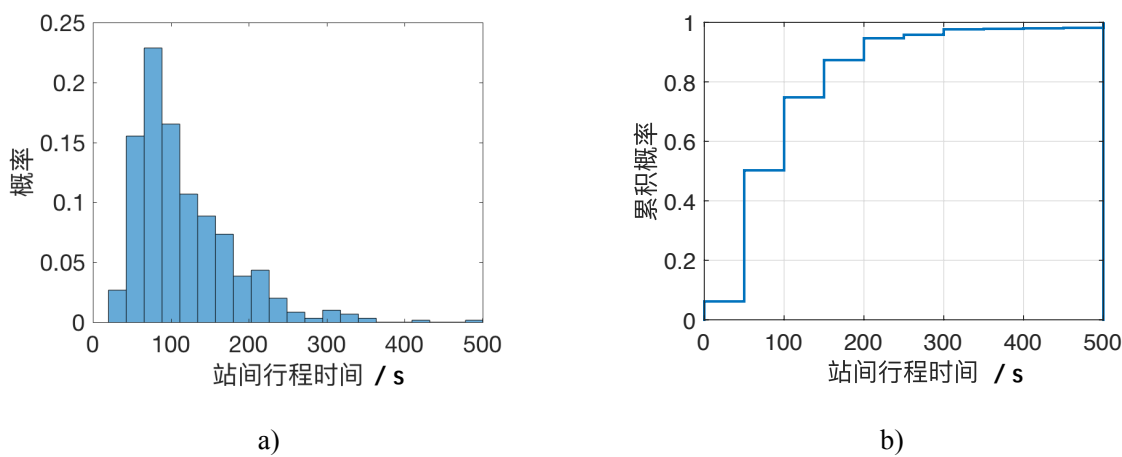

a)                                   b)

图 3-4 公交车站相邻站点间的行程时间分布

Fig. 3-4    Distribution of bus travel time within consecutive stops

由图（3-4a），站间平均行程时间分布于 60s 至 250s 的区间内；从图（3-4b）累积分布上看，超过 95%的站间行程时间超过 60s。因此只需要将相邻刷卡记录的分离阈值设置为$\varepsilon$=40s，即可有效将同一辆车同方向各站的乘客刷卡记录分离。

#### 3.2.2.2 刷卡事件信号域变换与时间误差搜索

取得车辆该行程各站点刷卡记录集合$C_1, C_2, ..., C_N$后即得到每个站点的第一个刷卡记录的所形成的有序代表集$C_R = c_1^1, c_2^1, ..., c_n^1$，由$C_R$中各元素时间戳形成的二元时间序

列$Y_C$，定义如下：

$$T_{CR} = t_1^1, t_2^1, \ldots, t_n^1$$
$$Y_C(t) = \begin{cases} 1, & t \in T_C R \\ 0, & t \notin T_C R \end{cases} \tag{3-1}$$

其中，$T_R$为$C_R$各刷卡记录对应刷卡时刻集合，该序列时间分辨率为1s。

同理，对于$C_R$对应的报站事件$V_R = v_1, v_2, \ldots, v_n$及其进站时刻集合$T_{VR}$，可得到二元制时间序列$Y_V$，定义为：

$$T_{VR} = t_1, t_2, \ldots, t_n$$
$$Y_V(t) = \begin{cases} 1, & t_k^1 \leq t < t_k^1 + t_w; \\ 0, & otherwise. \end{cases} \tag{3-2}$$

若将$Y_V$与$Y_C$中每个$y=1$的时刻延长为一宽度为$t_w$，$t_w < \frac{\varepsilon}{4}$，幅度为1的正脉冲，则$Y_V$、$Y_C$被进一步转换为离散时间信号序列$S_C$，$S_V$。定义为：

$$T_{CR} = t_1^1, t_2^1, \ldots, t_k^1, \ldots, t_n^1$$
$$S_C(t) = \begin{cases} 1, & t_k^1 \leq t < t_k^1 + t_w, \\ 0, & otherwise. \end{cases} \tag{3-3}$$

$$T_{VR} = t_1, t_2, \ldots, t_k, \ldots, t_n S_C(t)$$
$$S_V(t) = \begin{cases} 1, & t_k \leq t < t_k + t_w; \\ 0, & otherwise. \end{cases} \tag{3-4}$$

同理可知，$S_C$、$S_V$两信号时间分辨率也为1s。

将乘客刷卡事件与车辆报站事件转换为离散时间信号后，本研究假设车辆报站数据中的时间无偏移，则对应$S_C$、$S_V$两离散时间信号的相互关函数的定义为：

$$cor(\tau) = \sum_T \sum_{-\tau}^{\tau} S_C(t-\tau) \cdot S_V(t) \tag{3-5}$$

其中，$\tau$为时间推移量，即，两信号间的时间误差值，取值范围是[-3600, 3600]；T为$S_C(t-\tau)$与$S_V$在时间轴上的重叠部分，由于两序列均为二进制序列，因此乘法运算在实现时可以用相与运算取代。若$S_C$、$S_V$不存在周期性，则搜索时间误差$T_d$的过程可定义为寻找$\tau'$使公式（3-5）取得极大值的过程。得到极大值后即有$T_d = \tau'$。

得到时间推移量后，可将时间推移量加到乘客上车刷卡记录的时间戳中，用更新时间戳后的刷卡记录与 AVL 报站记录进行时间关联，进而得到不存在时间误差的乘客上车站点数据。

### 3.2.3 基于信息压缩的算法加速方案

假设$S_C$、$S_V$两个离散时间信号的长度均为$L$（若采用一天的所有数据进行分析，则$L$的长度为86400s），则对前后各一个小时的时间推移量进行搜索，则所需要进行的运算量为：

$$N_C = \sum_{\tau=-3600}^{3600} T_\tau \tag{3-6}$$

其中，$T_\tau$为每次经过时间推移后$S_C(t-\tau)$与$S_V$间的重叠部分长度。假设步长值设定为$\lambda$，$\tau_{max}$定义为时间推移量极大值，本研究中，$\tau_{max}=3600$，则$T_\tau$变为两个从1到$L-\tau_{max}$间隔为$\lambda$的等差数列。式（3-6）可进一步展开为：

$$\begin{aligned} N_C &= \frac{2 \cdot \tau_{max}(1 + L - \tau_{max})}{\lambda} \\ &\approx \frac{2 \cdot \tau_{max}(L - \tau_{max})}{\lambda} \end{aligned} \tag{3-7}$$

由（3-7）式，若$L$的长度为86400s，步长为1，则完成时间误差搜索所需要的运算量将超过5.9亿次，运算量庞大。假设$\tau_{max}$不变，则减少$L$并增加$\lambda$都可以有效降低运算量。由式（3-3、3-4），$S_C$、$S_V$两离散时间信号中，各脉冲宽度均为$t_w$，且脉冲间隔$gap > 4\varepsilon$，设原始信号可保证信息完整的最低采样间隔为$T_s$，根据奈氏采样定理，则：

$$T_s \leq \frac{t_w}{2} \tag{3-8}$$

本研究中，$t_w$取值为20s，则$T_s$取值为10s，则$S_C$、$S_V$两离散时间信号压缩后的信号$S_C{'}$、$S_V{'}$，分辨率变为10s，长度变为$\frac{L}{T_s}$。运算量$N_C$降低到：

$$\begin{aligned} N_C{'} &= \frac{2 \cdot \tau_{max}/T_s \cdot (L - \tau_{max})/T_s}{\lambda} \\ &\approx \frac{2\tau_{max} \cdot (L - \tau_{max})}{\lambda \cdot T_s^2} \\ &\text{即，} \quad N_C{'} = \frac{N_C}{T_s^2} \end{aligned} \tag{3-9}$$

由此可见，运算量已被降低$T_s^2$倍，观察式（3-9）的分母，通过无损采样提升运算效率的数学本质是增大了步长。虽然分辨率有所降低，但仍满足实际需求。

### 3.2.4 测试I：算法可靠性测试

通过公交跟车调查，记录了目标城市2016年3月25日15辆公交运营车辆真实的车载报站系统与刷卡收费系统时间误差作为测试集，验证方法可行性。图3-5给出了某车8时至15时的乘客刷卡记录与车辆报站记录转换为离散信号波形，分别如图3-5a、

3-5b 所示。

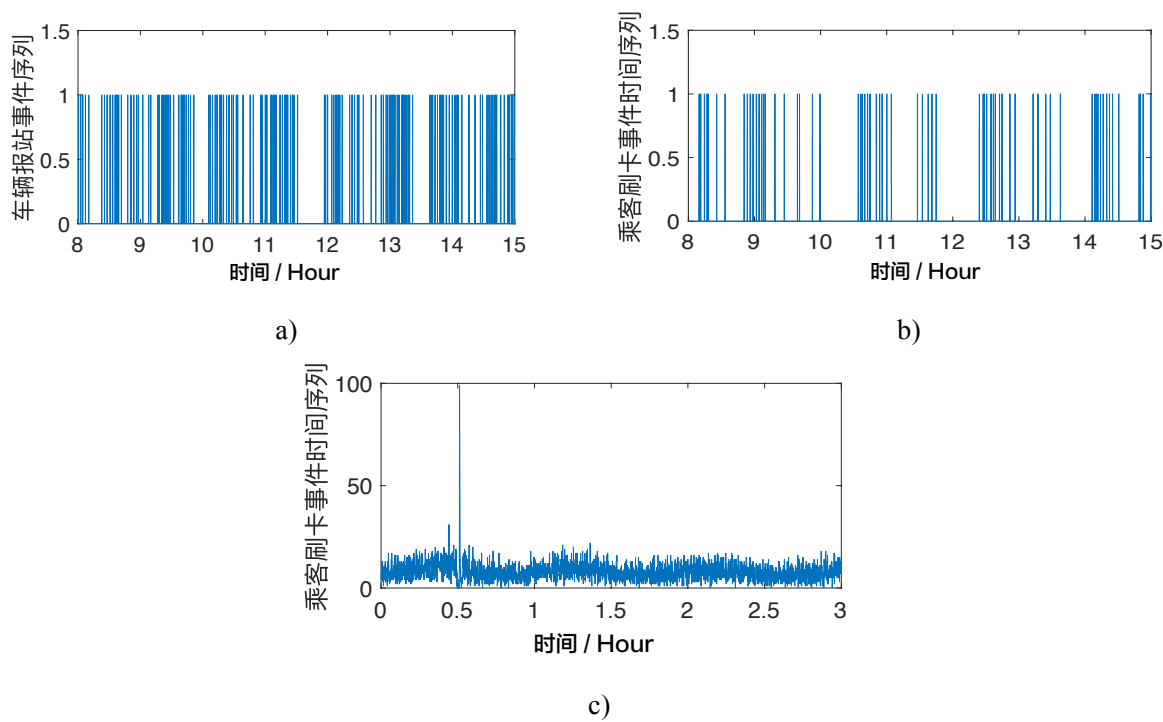

a)　　　　　　　　　　　　　　　b)

c)

图 3-5 时间误差检测试验

Fig. 3-5 Verification test of time discrepancy detection

由图 3-5，可见两路信号脉冲群间均存在明显的时间间隔，但是相比之下，由乘客刷卡数据的到的信号波形脉冲群更加稀疏，这是由于车辆在每一站都会产生报站记录，而并不能保证每一站都有乘客刷卡乘车。两序列在 0~3 小时内时域相关结果如图 3-7c 所示，由图可知时域相关的波形在 0.5h 时取得极大值，即该车 IC 卡数据源与车载报站系统数据源间的时间误差为 0.5h，该结果与人工跟车调查的结果完全相符，所调查的 15 辆公交巴士系统时间误差均被正确识别。

为进一步验证本文算法的有效性，还设计了可靠性验证试验。从信号分析上看，本研究所提出基于相关运算寻找极大值点的方法要求乘客刷卡的离散时间事件序列必须包含足够数量的脉冲簇（大部分车辆经停的站点至少有一个乘客刷卡上车），才能获得可靠的结果。由于多数情况下忽略车辆进站与该站台第一位乘客上车的时间间隔，则可得$T_{CR}$为$T_{VR}$的子集。乘客刷卡事件代表集稀疏度定义为：

$$\gamma = 1 - \frac{N_{CR}}{N_{VR}} \tag{3-10}$$

其中，$N_{CR}$为乘客刷卡事件代表集的元素个数，$N_{VR}$为 AVL 车辆报站事件个数。通过随机删除$N_{VR}$中的元素即可构造具有不同稀疏度的乘客刷卡事件代表集合。定义判定成功找到可信时间误差点的判定条件为：

$$cor(T_d) \geq \eta \cdot N_{CR} \tag{3-11}$$

其中，$\eta$ 定义为推断支持度或数据匹配度，该判定条件的物理意义为：至少有 $\eta \cdot N_{CR}$ 个报站记录成功在刷卡事件代表集中找到了对应元素。

以图 3-5 所述车辆的报站数据与乘客刷卡数据为例，假设由 0.1~0.9，步长为 0.05；每次尝试寻找时间误差 100 次，求推断成功率，得到数据统计如图 3-6 所示。

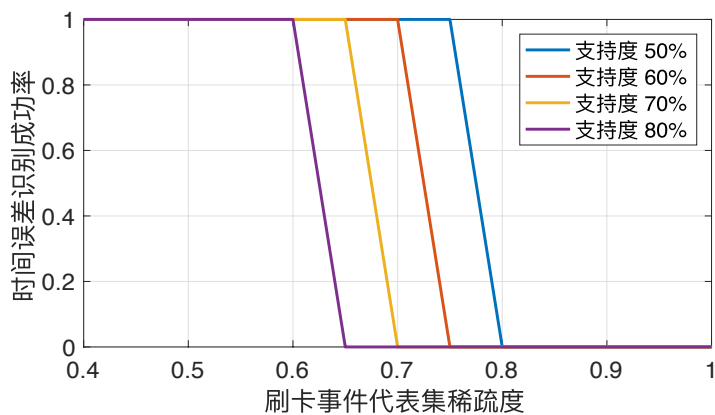

图 3-6 刷卡事件稀疏度测试对时间误差推断成功率的影响

Fig. 3-6 Test on influence of the sparsity of boarding events to time rectification success rate.

由图 3-6，随着事件代表集稀疏度增加，时间误差识别成功率出现拐点；另一方面，随着推断支持度增加，为保证成功率则要求更低的稀疏度；由图 3-6，在不低于 40%的站点有乘客刷卡上车的前提下，本研究所提供的方法能在多种情况下均达到 100%的时间误差搜索的成功率。

本方法在配置 Intel i7 2.8 GHz 4 核处理器，与 16GB 1600MHz DDR3 内存的工作站上，用 QT 编程实现，算法处理效率为平均 2.1 车每秒，满足工程应用需求。

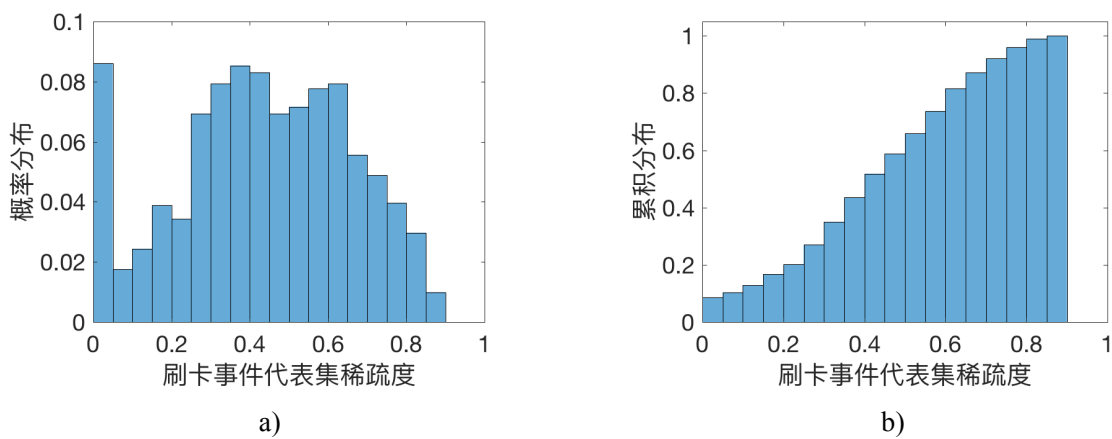

图 3-7 营运车辆乘客刷卡数据稀疏度分布

Fig. 3-7 Actual sparsity of boarding passengers' smartcard events

本研究目标城市在 2016 年 3 月 1 日至 6 月 2 日 8:00 A.M.至 11：00 A.M.时段内，所有公交运营车辆的乘客刷卡事件代表集稀疏度分布如图 3-7 所示，由图 3-7a，从概率分布图上观察，稀疏度主要分布在 0.3~0.65 区间内；由图 3-7b，由稀疏度累积分布观察，超过 45%稀疏度已覆盖 80%以上的样本。因此，本方法可以有效应用在目标城市的公交运营数据处理中。

### 3.2.5 测试 II：珠海市公交运营车辆时间误差校正测试

用本研究所提供方法计算城市 1695 辆公交车在 2016 年 6 月 3 日两 AFC、AVL 两系统间的时间误差分布，结果见图 3-8 所示。

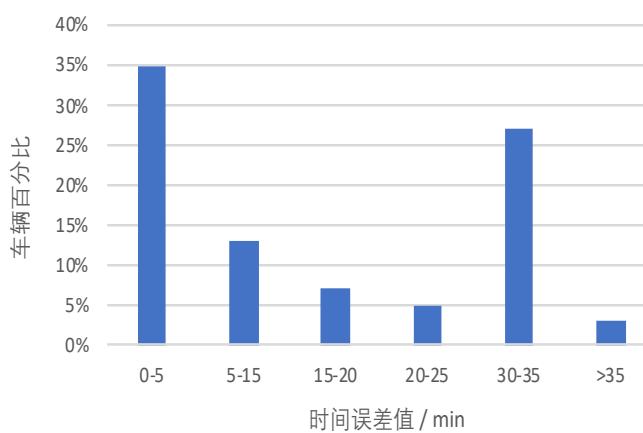

图 3-8  目标城市公交运营车辆时间误差统分布图

Fig. 3-8 Statiscal distribution of actual time discrepancies of public buses.

由图 3-8，该城市公交车辆中 AFC 与 AVL 系统的时间误差主要集中与 0~5 分钟与 30~35 分钟之间，时间误差在 0-5 分钟的车辆仅占全市公共交通营运车辆的 35%，因此 AFC 与 AVL 两系统普遍存在时间同步误差。进一步试验表明，消除时间误差前，该城市乘客刷卡数据中可找到上车站点的比率仅为 70%，采用本文所提供方法后，可找到上车站点的刷卡数据比率提升至 85%。提升了数据的利用率。

对传统的公交乘客上车站点推断方法进行改进，利用公交乘客的刷卡数据与报站数据间的关联信息，将两路数据源的原始数据转换为离散时间信号，从信号时域相关的角度求得两系统的时间误差，并在自动消除时间误差的基础上实现公交刷卡乘客的上车站点提取，不足之处在于：该方法要求每个班次沿途 45%的站点有至少一名乘客刷卡上车，未来的研究工作是设计更精细的方法，融合概率推断法与本研究提出的改进型直接推断法，降低对数据源的要求，灵活适应各种不同的情况。

## 3.3 报站数据缺失信息修补方法

车载报站系统中的位置数据对城市公交客流分析具有重大意义，其数据完整度直接决定了一系列客流分布数据挖掘过程的成功率[85]。然而，现阶段广泛使用的基于 GPS 与 GPRS 的车辆位置采集与传输系统并不能提供可靠的数据采集与传输解决方案，导致车辆报站数据缺失。报站数据缺失将对后续数据挖掘带来以下两方面的问题：

（1）导致乘客上车站点匹配失败，发生在客流热点区域的数据丢失，会导致大批乘客无法匹配上车站点。

（2）导致无法推断下车站点，在一天内，若第 k 次上车站点匹配失败往往意味着第 k–1 次公交出行的下车站点难以确定。

（3）导致无法获取行程时间，报站数据中记录了重要的车辆进站时间戳，若第 k 个站点的报站数据缺失，则对于任何$m < k$，第$k-m$站至第$k$站的行程时间将难以确定。

报站数据缺失，尤其是在客流热点区域的数据缺失将对后续客流分析带来严重影响，但此问题目前尚未在业内引起足够重视。对此，一种常用的解决方案是读取车载 AVL 设备中离线存储的轨迹点，经过降噪、停驻点识别、车站匹配后，得到车辆的停靠站记录[15,86-88]。但该方法需要手工对每一辆公交运营车辆的设备进行数据读取，费时费力，且无法得到因 GPS 系统定位失败导致丢失的数据。

本节讨论的算法建立在第 3.2 节基础之上，即已完成报站数据与乘客 刷卡数据时间误差校正并初步匹配的基础上。此时，还存在将近 20%的刷卡记录因无法在报站记录中找到与刷卡时间戳匹配的停站记录而匹配失败。将各班次的报站数据与该市公交线路表比对后做出报站系统数据丢失率的空间分布热力图以及公交站点的经停线路数热力图分别如图 3-9a，3-9b 所示，其中，热力图颜色代表数据丢失率。可以观察到，数据丢失多发生在客流密集的主城区（图中 A 区域）、城郊镇区人口聚集区（图中 B、C 区域）以及交通枢纽区域（图中 D 区），经过现场调查发现原因是在以上区域城市建筑密度同时客流量大，导致 GPS 定位精度降低的且无线数据传输（目前大多数 AVL 系统采用 2G 网络）质量变差，最终导致数据丢失严重。另一方面，虽然统计表明，报站数据缺失率整体上仅有 3.7%，但直接导致超过 12 万人次（占总数 20%）的刷卡数据无法匹配出上车站点。

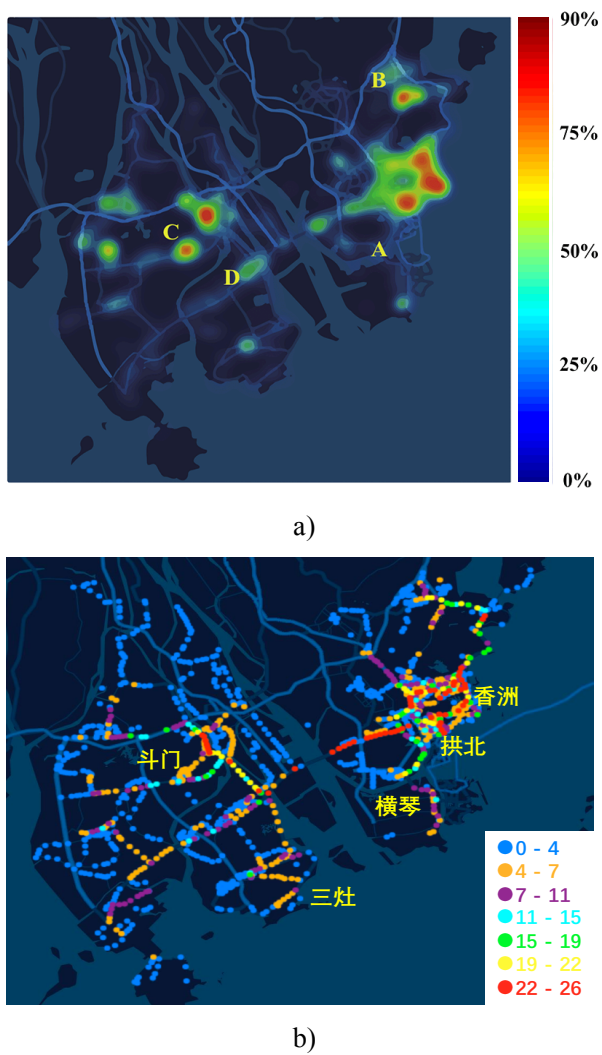

图 3-9 报站系统数据丢失率与站点线路数的空间分布热力图

Fig. 3-9 The spatial distribution of AVL data loss rate along with bus stations transit routes.

本节通过缺失数据的特征以及公交乘客的刷卡行为特点，提出了一种基于极大概率估计的报站缺失数据修补方法，并对比了不同后验概率模型下的数据还原成功率得到最优方案。可极大提高乘客公交出行轨迹还原的成功率，增加了城市公交客流分布研究的信息有效性。

### 3.3.1 算法总体流程

本节提出的缺失数据修补方法的特点为：一方面，利用刷卡乘客产生的时间戳辅助推断丢失的报站数据信息，无报站数据的情况下，完成乘客上车站点推断；另一方面，借助历史数据构建概率分布模型，在不借助乘客刷卡数据的情况下，推断出车辆从参考点到丢失失数据站点的行程时间。该算法总体流程如图 3-12 所示，该算法主要步骤如下：

**Step 1** 依照车辆调度记录，对每一辆运营车辆按每一班次，对车辆报站数据进行归

类整理，将该车辆停靠的站点按时间顺序排列；

**Step 2** 将每一班次已排序的停站数据与该车次调度数据中的停靠站列表进行对比，若站点不存在缺失，无需修复；

**Step 3** 若站点信息不完整，则说明报站数据有缺失，然后，锁定存在数据丢失的区间，以出现数据丢失前最后一个站作为参考点，从历史数据中建立参考点到区间内各站的行程时间分布模型；

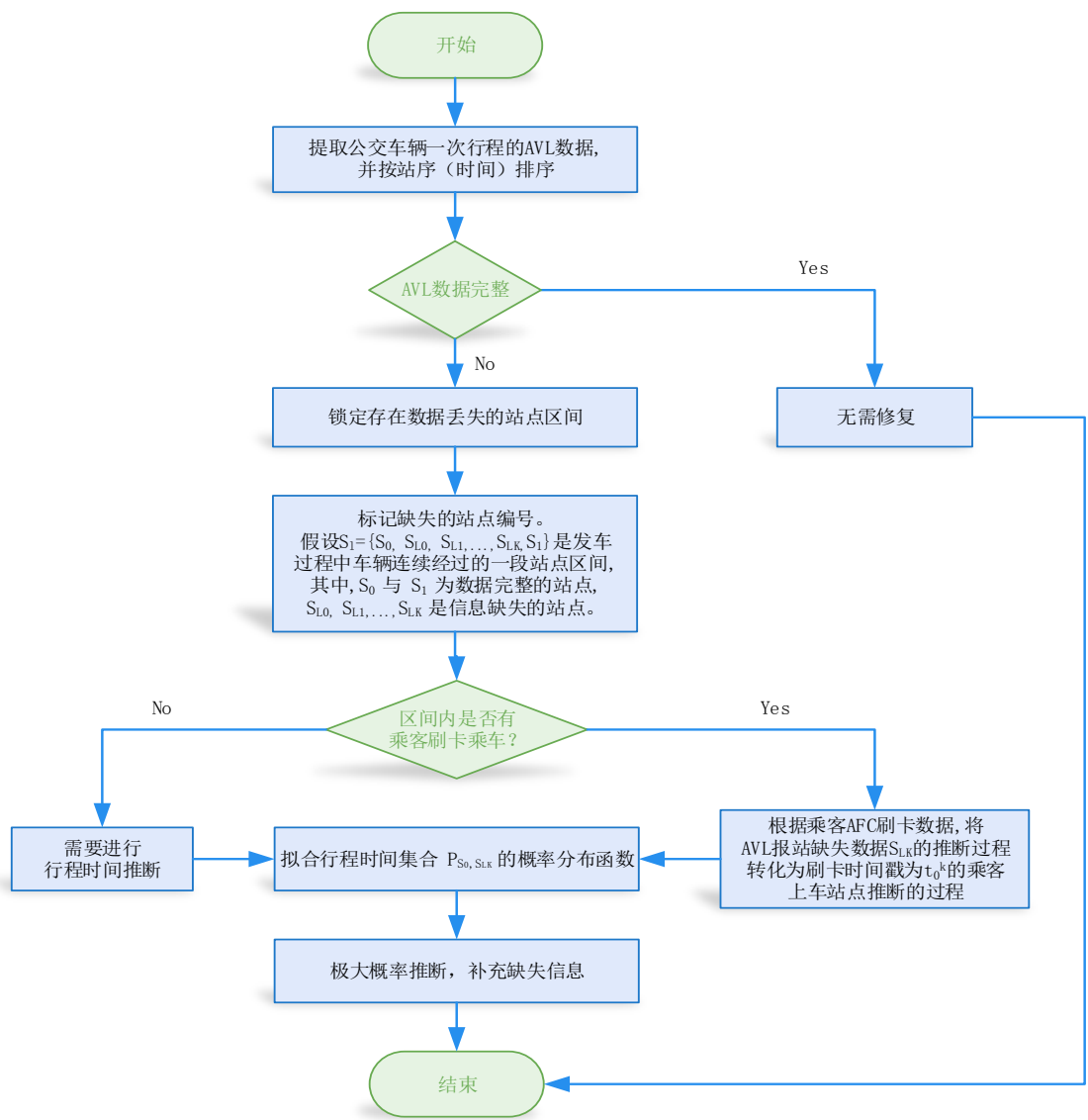

图 3-10 缺失数据推断方法总体流程

Fig. 3-10　he brief workflow for recovering missing AVL data.

**Step 4** 若该区间有乘客刷卡，则先用第 3.2.2.1 节的方法找到每站第一个刷卡的乘客，再利用第 3 步的方法判断车辆从该区间参考点出站时刻到该时间戳的行程时间内，概率最大的到达站点；

**Step 5** 若区间无乘客刷卡，则对每一个缺失的站点$k$构建以该区间行程时间为条件的行程时间概率分布模型，然后，求解模型概率极大值的时间值，作为车辆从参考点至站点$k$的行程时间；

### 3.3.2 关键参数与数学模型

对缺失报站数据统计分析表明，数据丢失率的空间分布与客流热点区域重合，但时间分布随机。假设集合$S_l = S_0, S_{L0}, S_{L1}, \ldots, S_{Lk}, \ldots, S_1$，为某一运营车辆一次行程中连续经过的站点，其中，$S_0$与$S_1$为数据完整的站点，$S_{L0}, S_{L1}, \ldots, S_{Lk}$为$S_0$与$S_1$之间缺失报站数据站点。缺失数据区间的车站间隔在分析时段内的分布如图 3-11 所示。

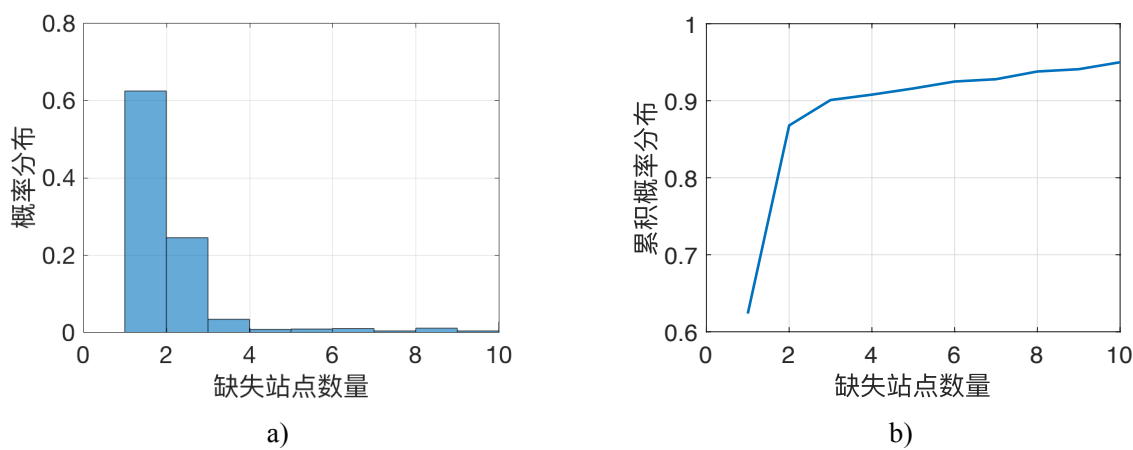

图 3-11 缺失数据区间的车站间隔分布

Fig. 3-11 Distribution of station intervals in data lost districts.

由图 3-11，超过 90%的缺失数据区间中，车站间隔在 4 站之内。假设在$S_{L0}$至$S_{Lk}$期间，乘客刷卡产生时间戳为$T_p = t_0, t_1, \ldots, t_n$，假定每站乘客刷卡只能在车辆进站之后发生，将每站第一个刷卡的乘客刷卡时刻定义为该站乘客开始刷卡时刻，通过已匹配出上车站点的乘客刷卡数据，得到乘客开始刷卡时刻与车辆进站的时差分布如图 3-12 所示。

由图 3-12，概率超过 80%的情况下，车辆进站 30s 内乘客即开始刷卡。故，乘客刷卡事件与车辆到站时刻间有很强的时间关联性。设任意乘客上车匹配的站点$S_{Lk}$，刷卡乘客在该站点产生的交易时间戳为$T_c^k = t_0^k, t_1^k, t_2^k, \ldots t_{m_k}^k$，其中$m_k$为站点$S_{Lk}$的上车人数。则在乘客刷卡上车站点匹配的前提下，$S_{Lk}$的推断的过程可表示为，车辆从$S_{L0}$出站到$t_0^k$的行程时间内，概率最大的到达站点。

设任意两站点$S_i$、$S_j$间行程时间分布可以用概率分布$P_{i,j}(t|\Theta)$表示。已知$S_0$，$t_0^k$，并借助历史数据寻找合适$S_{Lk}$的过程可以表示为以下极大值寻优过程：

$$\widehat{S_{Lk}} = \text{argmax}_{S_{Lk}} P_{S_0, S_{Lk}} P(t|\Theta, t = t_0^k) \tag{3-12}$$

其中，$\widehat{S_{Lk}}$为从$S_0$开始行程时间为$t_0^k$时，使$P_{S_0,S_{Lk}}(t|\Theta)$取得极大值的停靠站；$\Theta$为后验条件参数集合。

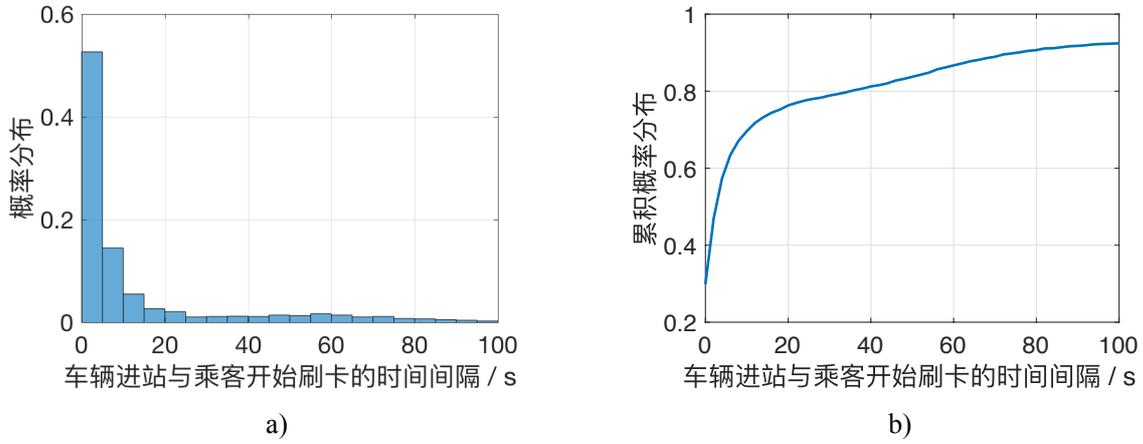

图 3-12 车辆进站与乘客开始刷卡的时间间隔分布

Fig. 3-12 Distribution of time intervals between buses' stop events and the first boarding events.

另一方面，若缺失站点的区间内无乘客刷卡数据，则必须对车辆从站点$S_0$到$t_0^k$的行程时间$T_{S_0,S_{Lk}}$进行推断。其数学过表示为：

$$\widehat{T_{S_0,S_{Lk}}} = \mathrm{argmax}_{T_{S_0,S_{Lk}}} P_{S_0,S_{Lk}} P(t|\Theta) \tag{3-13}$$

对于式（3-12，3-13）本节探讨应用如下后验条件：

（1）$\Theta_0 = \emptyset$，此时$P_{S_0,S_{Lk}}(t)$为从$S_0$至$S_{Lk}$的行程时间的概率分布。

（2）$\Theta_1 = t_{S0} \pm \Delta T$，其中，$t_{S0}$为站点$S_0$的进站时刻，$\Delta T$为时间阈值，$P_{S_0,S_{Lk}}(t|t_{S0} \pm \Delta T)$表示开始时刻为$t_{S0} \pm \Delta T$时，从$S_0$至$S_{Lk}$的行程时间的条件概率分布，本研究$\Delta T$=1200s。

（3）$\Theta_2 = t_{TRIP} \pm 0.1 \cdot t_{TRIP}$，其中$t_{TRIP}$为从该车辆本次从$S_0$至$S_1$的行程时间，$P_{S_0,S_{Lk}}(t|t_{TRIP} \pm \Delta T)$表示行程时间为$t_{TRIP} \pm 0.1 \cdot t_{TRIP}$时，从$S_0$至$S_{Lk}$的行程时间的条件概率分布。

（4）$\Theta_3 = \Theta_1 \cup \Theta_2$，$P_{S_0,S_{Lk}}(t|\Theta_3)$表示从$S_0$至$S_1$的行程时间为$t_{TRIP} \pm \Delta T$且开始时刻为$t_{S0} \pm \Delta T$时，从$S_0$至$S_{Lk}$的行程时间的条件概率分布。$\Theta_3 = \Theta_1 \cup \Theta_2$，$P_{S_0,S_{Lk}}(t|\Theta_3)$表示从$S_0$至$S_1$的行程时间为$t_{TRIP} \pm \Delta T$且开始时刻为$t_{S0} \pm \Delta T$时，从$S_0$至$S_{Lk}$的行程时间的条件概率分布。

从$S_0$至$S_{Lk}$，符合定义（1）至定义（4）的行程时间集合（定义为$T_{0,k}$）可通过扫描车辆历史报站数据得到。为简化数值寻优过程并降低计算量，本节采用高斯分布函数（式3-14）拟合行程时间集合。

$$P(t) = \frac{1}{(\sqrt{2\pi}\sigma)} \exp(-\frac{t-\mu}{2\sigma^2}) \tag{3-14}$$

### 3.3.3 测试 I：基于乘客刷卡时间戳的站点推断测试

为验证本研究所提供方案的有效性，本节将所提供方法应用于还原被人为破坏的报站数据，为尽可能接近实际情况，依据图 3-11b 的累积分布，所有人为破坏的报站数据中时间戳均加入的随机干扰，观察不同设定条件下数据还原方法的可靠性。生成测试数据集（站点推断测试数据集）的算法见图 3-13。

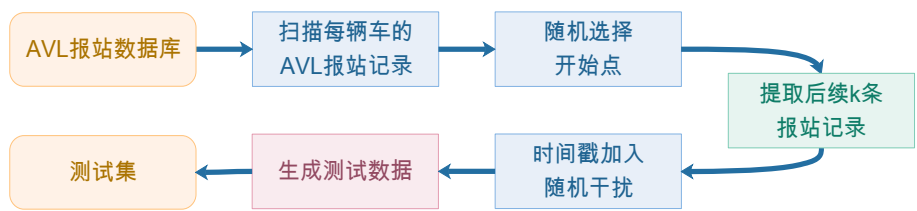

图 3-13 站点推断测试数据集生成算法

Fig. 3-13 Algorithm flowchart for generating test dat.

由图 3-13，提取 2016 年 3 月 5 日的 AVL 车辆报站数据为基础数据源，扫描 AVL 报站数据中每辆车的报站数据，以 15%的概率随机选择开始站点，取出之后 8 个站点（见图 3-11b，保证测试的情形能覆盖 90%以上的实际情况）的报站数据加入备选测试集，共计取得 94750 条测试数据。然后，对备选测试集中每一笔数据，除第一个与最后一个报站记录（$S_0$、$S_1$）外的所有报站记录进站时刻加入均匀分布的随机干扰（范围：-40s~40s），得到测试集。测试集中每笔测试数据包含的关键信息字段如表 3-2 所示：

因此，缺失数据推断测试试验任务为扫描测试集中所有记录，并根据其第一条与最后一条报站记录（$S_0$、$S_1$）及包含扰动的时间戳$T_{L0}', T_{L1}', ..., T_{Lk}'$推断$S_{L0}, S_{L1}, ..., S_{Lk}$。

表 3-2 数据集关键字段

Table 3-2 Principle information in the related data sources

| 数据源 | 关键信息字段 |
| --- | --- |
| StartRecord（$S_0$） | 模拟数据丢失发生前最后一条完整的报站记录（不加扰） |
| EndRecord（$S_1$） | 模拟数据丢失发生后第一条完整的报站记录（不加扰） |
| TimeStampList（$T_{L0}, T_{L1}, ..., T_{Lk}$） | 模拟丢失数据的进站时间戳列表（加扰） |
| StationsToPredict（$S_{L0}, S_{L1}, ..., S_{Lk}$） | 待修复的站点列表 |

按图 3-13 所生成的测试数据集测试不同后验条件下的数据还原成功率如图 3-14 所示，由图 3-14，所提供的 4 种方案数据还原成功率均高于 65%，但随着缺失数据区间的站点数增加，4 种方法的数据还原成功率均降低。其中，以行程时间为条件概率的方案还原成功率最高，在所述各种情况下均超过 85%，满足数据分析需求。

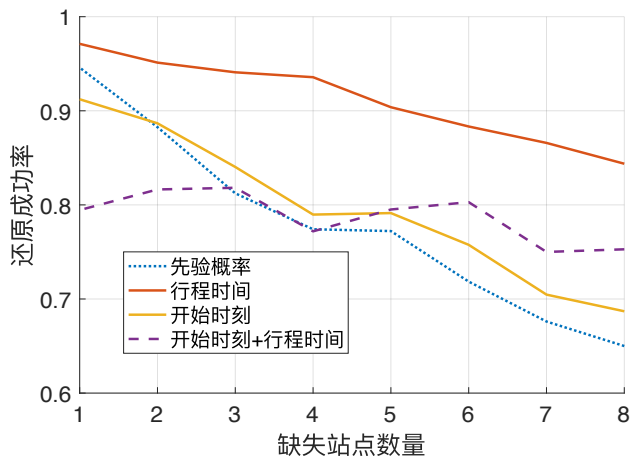

图 3-14 不同后验证条件下数据还原成功率

Fig. 3-14 Data recovery success rate with different posterior conditions.

由图 3-14，综合考虑多因素的模型（见开始时刻+行程时间曲线）并不能提高的数据还原成功率。且该曲线仅仅在缺失站点数量多于 6 个时性能变好；基于开始时刻的模型成功率仅略高于基于先验概率的模型，因此数据丢失过程的开始时刻并不能提供足够的指示信息。

为进一步分析三种限制条件（仅考虑行程时间、仅考虑开始时刻、综合考虑行程时间与开始时刻）下模型性能差异的根源，作出三种限制条件下计算高斯分布模型参数的有效数据量（单位：条）分布如图 3-15 所示。

由图 3-15，可观察到，只考虑行程时间时，大部分情况下计算得到模型的数据量均高于 100 条；若只考虑开始时刻，计算概率模型参数可用的数据量集中在 10~20 条之间；若综合考虑行程时间与出发时刻，计算正态分布模型所能用的有效数据量在绝大多数情况下均少于 10 条；因此，只考虑行程时间的条件下，得到的概率分布模型具有最高的数据支持度和最高的数据还原成功率，其余两种情况下得到的概率模型则有较高的欠拟合风险，尤其是综合考虑行程时间与开始时刻的模型，数据支持度最低。综上所述，以区间行程时间为限制条件的方法为最优方案。

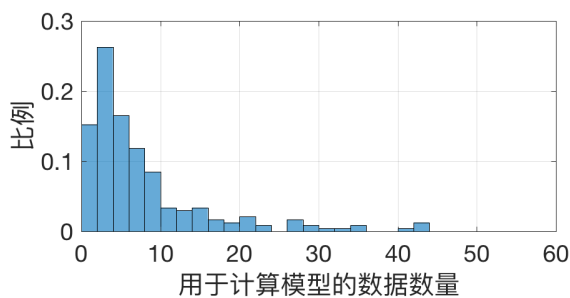

a) 考虑全程行程时间与出发时刻

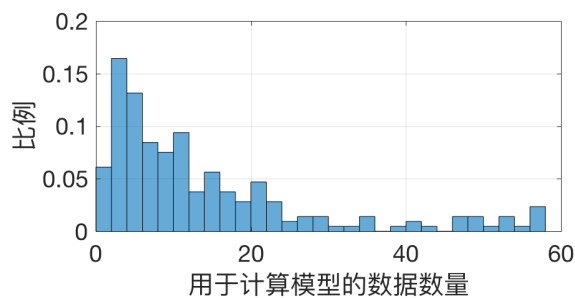

b) 只考虑出发时刻

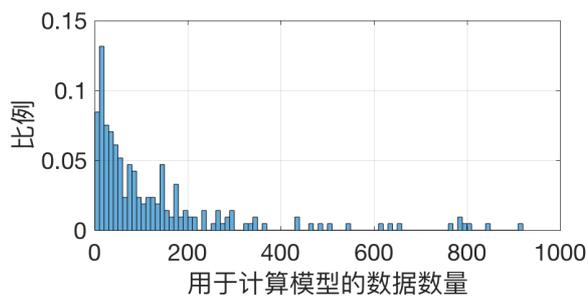

c) 只考虑行程时间

图 3-15 不同后验条件下用于计算模型参数的数据量分布

Fig. 3-15 Distribution of data available for deriving the parameters of inference mode.

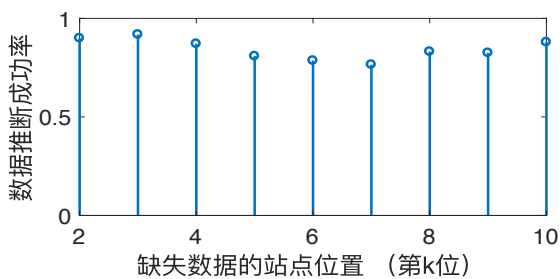

图 3-16 不同位置缺失站点的推断成功率

Fig. 3-16 Induction success rate for missing records in different positions.

设置丢失数据的站点区间长度为 12 站，应用优选方法进行，将不同位置的数据还原成功率统计于图 3-16，由图可以观察到第 6、7 位推断成功率最低，而首末两端（第 2 位、第 11 位）推断成功率最高，说明整个缺失数据的行程时间对缺失的首末站信息推断具有更高的指示价值。随着中间站点与首末站距离的增加，信息指示价值逐渐变低。

### 3.3.4 测试 II：行程时间推断测试

为验证本研究所提供方案的有效性，本节将所提供方法应用于还原被人为破坏的报站数据，观察不同设定条件下数据还原方法的可靠性。生成测试数据集（行程时间推断测试集）的算法见图 3-17。产生该数据集的方法与图 3-13 大致相同，区别是，不再需要对时间戳加入人为干扰。

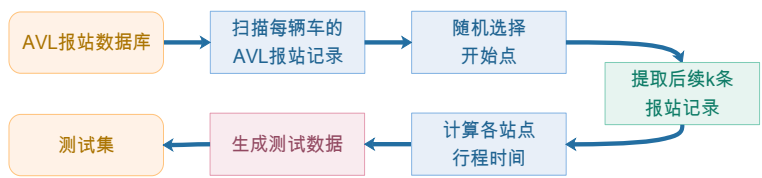

图 3-17 不同位置缺失站点的推断成功率

Fig. 3-17 Induction success rate for missing records in different positions.

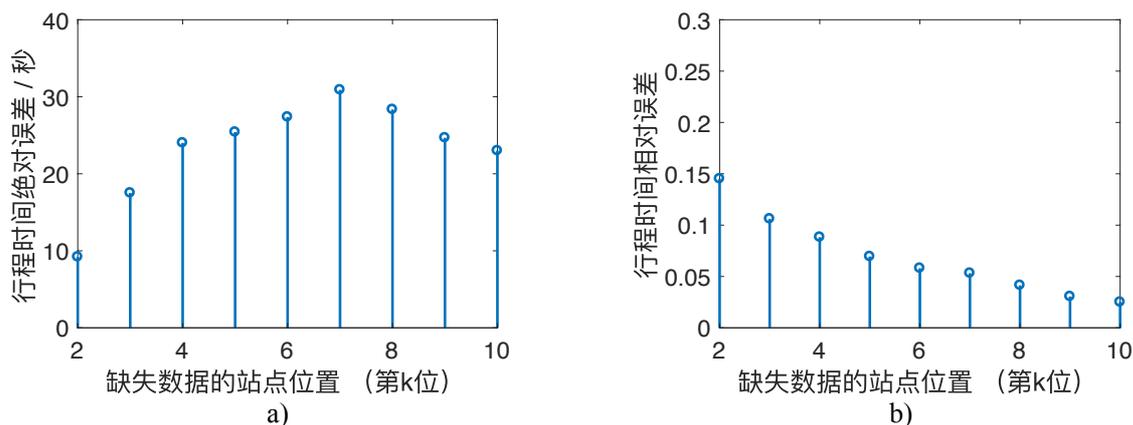

图 3-18 行程时间推断误差

Fig. 3-18  The error distribution of travel time induction.

借助第 3.3.3 节的结论，本节只讨论以行程时间为条件的高斯分布模型。设定缺失数据的区间长度为 12 站，从参考点开始，预测车辆到达 $S_{L0}, S_{L1}, ..., S_{Lk}$ 的行程时间，该模型在测试数据集上的绝对误差与相对误差与所预测站点位置的关系如图 3-18 所示。

由图 3-18a，行程时间绝对误差最大的站点为第 6 站到第 8 站，与图 3-16 相似，该处站点离有确定信息的首末端最远，因此误差最大。从相对误差分布（图 3-18b）上观察，仅有第 2 个站相对误差超过 10%，其余站点相对误差都较小，原因是由于绝对误差存在上界，总行程时间变长后，相对误差减小。同理，第 1 个站点相对误差偏大而绝对误差较小的原因是，车辆由区间第 1 个站点出发后，经过很短的时间就到达第 2 个站点，即，第 2 个站点即使行程时间推断在绝对误差不变时相对误差也表现出偏大的趋势。

本节对传统的城市公交系统缺失数据推断方法进行研究，利用公交乘客的刷卡数据与报站数据间的关联信息，建立了以行程时间为条件的概率模型。一方面，在报站数据缺失的条件下匹配乘客上车站点；另一方面，推断了丢失的车辆行程时间。不足之处在于对于连续丢失超过 10 个站点报站数据的情况，准确率会降至低于 80%，本方法的缺陷表现为：无法修补因车载报站设备损坏而导致的整车或整班次数据丢失。

## 3.4 本章小结

本章针对我国目前各大城市公交数据系统存在的缺陷，提出一套能用于在有缺陷数据环境下对一票制公交信息系统具有普适性的通用化数据预处理方案，旨在得到准确、可靠的乘客刷卡时空轨迹数据集。主要内容如下：

（1）针对公交 IC 卡收费系统与 GPS 报站系统间的时间同步误差提出了一种基于事件时空关联与信号关联匹配的时间误差校正方法，并对目标城市的公交系统乘客 IC 卡数据与 GPS 报站数据进行了逐天逐车的时间误差校准。

（2）针对 GPS 报站数据缺失，提出了一种融合乘客刷卡时间戳与区间行程时间的报站系统缺失数据推断方法。在初步匹配出乘客上车站点的基础上，完成剩余乘客的上车站点匹配，同时修复缺失的报站数据。提升数据有效性且得到了全部乘客上车刷卡时空轨迹数据集。

# 第四章 公交乘客完整出行信息提取方法

公交系统客流分布规律研究对城市交通系统具有重要意义，其中，乘客的出行规律以及出行信息具有重要价值，对于每个乘客个体而言，完整的出行信息包含每次出行的时空轨迹（上、下车站点、线路、车次、刷卡时间）和出行阶段标记（起点、换乘点、终点），这些信息也是乘客个体行为特性分析的基础。第三章和本章共同围绕着公交运营数据缺陷处理与乘客完整出行信息提取该主题。具体的，本章我们在第三章所提供乘客刷卡数据的基础上复现乘客的出行时空轨迹以及出行阶段划分信息，将孤立的刷卡事件还原成客流的基本单元——个体乘客的时空轨迹，为后续分析提供基础数据源。

在目标城市中，公交出行者占全市常住人口的比重为 1/3（观察到 75 万公交乘客，常住人口为 215 万人），因此本文的数据挖掘方法所提供的居民公交出行分布结论能够较准确的反映目标城市的客流动态。

目前广泛使用的一票制公交系统，其原始运营数据只包含一小部分的信息，运用本文第 3 章的数据预处理方法，可以得到每个乘客公交出行的上车站点、车次及刷卡时间戳，但仍然缺乏下车站点以及出行阶段标记信息。对此，传统的方法通过闭合出行链假设、最少步行距离假设以及换乘时空阈值假设来推断并乘客每次乘车的下车站点，然后识别每个出行阶段的起点、换乘点、终点等信息。

传统方法的共同特点以及应用前提，均建立在获得完整的乘客上车刷卡数据的基础上。但公交系统原始数据经过严格时间校准以及缺失数据修补后，仍无法保证提供全部乘客完整的上车刷卡数据。本研究中，数据缺失表现为每天由车载报站设备间歇性失效导致超过 12%的刷卡数据只有车次、线路号以及乘客卡号，无法进一步推断上车站点。综上所述，传统的下车站点推断方法难以直接应用，有必要对传统方法进行改进，以期适应多样化的基础数据条件。

## 4.1 整体技术流程

本章内容旨在利用多源数据融合方法，提供一套具有容错性的乘客完整出行信息还原策略。总体技术流程如图 4-1 所示。对每一名乘客的按时间顺序排列的乘车记录，该策略主要步骤如下：

**Step 1** 对于连续出行的乘车阶段，在数据无缺陷的情形下借助最小步行距离假设推断出下车站点，并加入暂不完整时空轨迹集合；

**Step 2** 若连续出行条件下的第二次出行无法提供上车站点信息，则构建该次出行的

候选下车站点集合；

**Step 3** 根据每天和次日第一次出行的乘车记录数据，判断每天最后一次出行的下车站点或构建最后一次出行的可选下车站点集合；

**Step 4** 借助相似出行日合理构造统计区间，按极大概率原则推断第 2、3 步不能确定下车站点的乘车记录；

**Step 5** 标记该乘客的出行阶段起点、换乘、终点（简称 O、T、D）。

本方法最重要的步骤为第 2 步与第 3 步，即推断每个乘客每次出行的下车站点，与在有数据丢失的情况下以极大概率推断下车站点的方案。

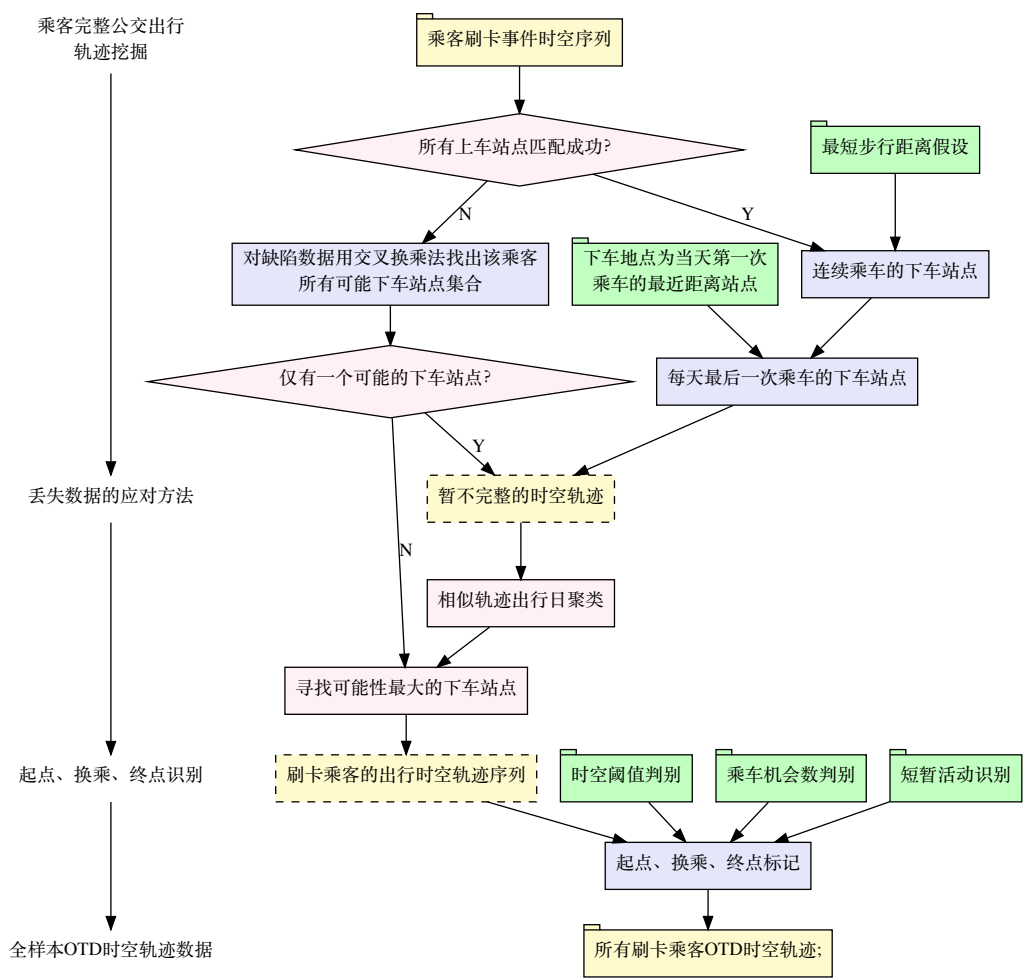

图 4-1 时空轨迹还原整体流程

Fig. 4-1 General workflow for recovering passengers' trajectories.

## 4.2 乘客公交出行下车站点推断方法

本文提供的乘客下车站点推断方法可以分为四个阶段：

第一阶段：推断连续出行的下车站点，处理数据完整的连续刷卡记录。

第二阶段：构建缺失数据的候选下车站点集合，处理时间上相邻的两条刷卡数据其

中第二条数据无上车站点的情形。

第三阶段：推断乘客每天最后一个下车站点。

第四阶段：极大概率推断，利用前三阶段得到的乘车记录，推断前三个阶段无法处理的刷卡数据下车站点。

以下分别对四个阶段展开讨论：

### 4.2.1 基本假设及关键步骤

#### 4.2.1.1 第一阶段：推断连续出行的下车站点

乘客下车站点判断有多种方法，其中，应对个体乘客出行轨迹还原常用直接推断法，该方法能够提取较完整的乘客出行信息。本方法的第一阶段即为直接推断法，该方法在推断公交乘客下车站点时，主要遵循以下假设：

**连续出行假设**：某乘客一天内连续第$k$次与第$k+1$次出行，若第$k$次出行的上车站点$B_k$与第$k+1$次出行的上车站点$B_{k+1}$均存在，且$B_k \neq B_{k+1}$，刷卡时刻分别为$t_k$与$t_{k+1}$，则查询所乘车辆在$t_k$至$t_{k+1}$时段内车辆运行记录，提取$B_k$下游站点构成集合为$AS_k$，此时，推断第$k$次出行的下车站点$\widehat{A_k}$的搜搜过程可表示为：

$$\widehat{A_k} = \underset{s \in AS_k}{\operatorname{argmin}} \, distance(s, B_{k+1}) \tag{4-1}$$

由式（4-1），如果 $B_{k+1} \in S_k$，$\widehat{A_k}$为$B_{k+1}$，即第$k+1$次乘车的上车站点；如果$B_{k+1} \notin AS_k$，则在$S_k$中寻找与$B_{k+1}$间满足步行换乘条件（距离小于700m）且距离最短的站点作为第$k$次出行的下车站点，

表 4-1 某乘客乘车记录

Table 4-1 Sample data of a passengers' transit boarding records.

| 刷卡时刻 | 车牌号 | 线路 | 上车站点编号 |
| --- | --- | --- | --- |
| 03-02 7:30 | 5xx49 | 207 路 | 无记录 |
| 03-02 10:50 | 5xx29 | 40 路 | 南屏街口 |
| 03-02 12:45 | 5xx84 | 201 路 | 湖心路口 |
| 03-02 15:23 | 5xx09 | 609 路 | 无记录 |
| 03-02 16:45 | 7xx37 | K4 路 | 斗门行政中心 |
| 03-03 7:40 | 5xx49 | 207 路 | 吉大总站 |
| 03-03 11:40 | 3xx28 | 207 路 | 南屏街口 |

连续出行假设有明显的局限性：连续出行中若缺失第$k+1$次出行上车站点则无法推断出第$k$次出行的下车站点。因此，无法应用于存在数据缺失的场景。本研究的目标城市公交系统中某乘客的乘车记录样例如表 4-1 所示：

表 4-1 为某乘客在 3 月 2 日的出行记录中，有两条记录（3 月 2 日第 1 条与第 4 条）上车站点丢失，用连续出行假设将无法得到当天第 1、4 次出行的下车站点。

#### 4.2.1.2 第二阶段：构建缺失数据的候选下车站点集合

为应对连续缺失的数据，对第 k 次出行记录缺失，在第一阶段直接推断法的基础上，本方法第二阶段采用以下步骤进行下车站点推断：

Step 1  若 $B_{k-1}$ 存在，而 $B_k$ 缺失，则提取 $B_{k-1}$ 与车辆在 $t_{k-1}$ 至 $t_k$ 时段内停靠的站点构成集合 $AS_{k-1}$，形成待处理乘车记录数据集，留待第四步处理。所述待处理乘车记录集中每个元素包含两部分：a）未能匹配出下车站点的乘车记录；b）该乘车记录的候选下车站点列表。

Step 3  若乘客当天只有一次出行记录，则下提取下车站点候选集，并在第四阶段处理。

#### 4.2.1.3 第三阶段：推断乘客每天最后一个下车站点

传统方法推断每天最后一次乘车的下车站点，依据闭合出行链假设：

**闭合出行链假设**：设某乘客在出行日 $d$ 最后一次乘车的下车站点（记为 $\widehat{A_{last}}$）的搜索过程可以表示式（4-2）：

$$\widehat{A_{last}} = \underset{s \in AS_{last}^d}{\mathrm{argmin}}\, distance(s, B_{first}^d) \tag{4-2}$$

式中，$B_{first}^d$ 为该乘客在出行日 $d$ 第一次乘车的上车站点，$AS_{last}^d$ 为出行日 $d$ 最后一次乘车所有可选下车站点构成的集合，意义为与点 $B_{first}^d$ 距离最近的站点。

闭合出行链假设的局限性在于，若乘客当天第一条刷卡记录中上车站点 $B_{first}^d$ 丢失则无法推断当天最后一次出行的下车站点，例如，表 4-1 中 3 月 2 日该乘客第一条乘车信息无上车站点，因此，无法得到当天最后一次出行的下车站点。

由式（4-2），寻找乘客每天最后一次乘车的下车站点，本质上是寻找一个参考点 $Sta_{ref}$，并从该乘客当天最后一次出行的候选下车站点中选出与该参考点距离最近的站点，即：

$$\widehat{A_{last}} = \underset{s \in AS_{last}^d}{\mathrm{argmin}}\, distance(s, Sta_{ref}) \tag{4-3}$$

对经典的闭合出行假设做出以下修正：

**假设 5.3**：顺序上相邻的出行日 $day_n, day_{n+1}$，若 $day_{n+1}$ 第一次出行的上车站点 $B_{first}^{n+1}$ 与 $day_n$ 的第一个上车站点 $B_{first}^n$ 间直线距离大于 10 千米，且 $B_{first}^{n+1}$ 刷卡时刻为早上 8：30

第四章 公交乘客完整出行信息提取方法前，则：

$$Sta_{ref}^n = B_{first}^{n+1} \tag{4-4}$$

式中，$Sta_{ref}^n$为$day_n$该乘客最后一次乘车下车站点推断参考点；$Sta_{ref}^n$为$day_{n+1}$的第一个上车站点。此外，本研究限定，$day_n$与$day_{n+1}$间隔日期不大于 2 天；

推断乘客$p$在$day_n$最后一次乘车的下车站点$\widehat{A_{last}^n}$的主要步骤如下：

**Step 1** 提取乘客$p$在观察时段内，每个出行日$d$首次刷卡的上车站点$B_{first}^d$，统计这些站点的作为每天首次出行的上车站的概率，将概率最高的站点预判为该乘客的居住地，记为$H_p$。

**Step 2** 若$B_{first}^n$，$B_{first}^{n+1}$均缺失，则推断参考点$Sta_{ref}^n = H_p$。

**Step 3** 若经过第 2 步仍无法确定$day_n$最后一次乘车的下车站点，则保存$B_{last}^n$和对应的下车站点候选集$AS_{last}^n$，并在第四阶段处理。

#### 4.2.1.4 第四阶段：极大概率推断

经过前三个阶段已推断出乘客大部分乘车记录的下车站点，但仍存在部分记录无法推断下车站点。因此，在第四阶段，本方法运用极大概率推断法，推断第二、三阶段尚未匹配出下车站点的记录。主要步骤包括相似出行日搜索与条件概率推断。

假设乘客$p$在每个出行日$d$，经过前三个阶段数据还原后得到的访问站点轨迹序列表示为：

$$sT_d^p = \{\mathbf{O_d^p}\} \tag{4-5}$$

其中，$\mathbf{O_d^p}$为当天所有的上车站点构成的集合，即刷卡乘车轨迹，在出行日$day_d$，$day_{d+k}$的访问站点轨迹序列行$sT_d^p$、$sT_{d+k}^p$，可以定义其相似度为：

$$sim_{d,d+k} = \frac{count(sT_d^p \cap sT_{d+k}^p)}{count(sT_d^p \cup sT_{d+k}^p)} \tag{4-6}$$

为计算$sT_d^p$与$sT_{d+k}^p$中的等价元素，补充以下等价站点定义：

**定义 4.1**：若两站点$S_i$、$S_j$空间距离小于 300m 或存在一条公交线路，使$S_i$、$S_j$成为相邻站点，则认为两站点等价。

对于出行日本身，给出相似出行日及相似出行日集$S_\varepsilon^d$的定义：

**定义 4.2**：$\exists\, day_d, day_{d+k}$，若$sim_{d,d+k} \geq \varepsilon$，则$day_d, day_{d+k}$为轨迹相似出行日。

本研究设置$\varepsilon$=0.7，对任一出行日$d$，其$\varepsilon$相似出行日数据集定义为

**定义 4.3**：若 $S_\varepsilon^d = \{sT_0^p, sT_1^p, \ldots sT_d^p, \ldots, sT_n^p\}$为$d$的$\varepsilon$相似出行日数据集，则：$\forall sT_i^p, sT_j^p \in S_\varepsilon^d$，$sim_{i,j} \geq \varepsilon.$

本研究所述概率推断模型的限定条件为相似的出行日及其出行数据集$S_\varepsilon^d$上。具体地，对对出行日$d$第$k$次尚未推断出下车站点的乘车记录$B_k^d$，以及候选站点集合$AS_k^x = \{A_0, A_1, \dots, A_k\}$，推断过程可以表示为以下极大值寻优过程：

$$\widehat{A_k} = \underset{s \in AS_{last}^d}{\mathrm{argmax}}\, P_{cSta}(s|\Theta)$$
$$\Theta = B_k^d, S_\varepsilon^d$$
（4-7）

其中，$P_{cSta}(x|\Theta)$为在条件为$\Theta$下，各候选站点作为乘客下车站点的概率分布。若该过程发现多个非等价站点含有相同的最大概率，则选择具有最大访问频次的站点作为$sT_k$的下车站点（等价于$\Theta = S_\varepsilon^d$）。

### 4.2.2 实证案例

本节采用跟车调查采集车辆在站点的上下车人数，并提取当天该车的运营数据，将实际调查值与推算值进行校验。对目标城市 40 路车牌号为 18374 的公交车在 2015 年 6 月 2 日早高峰时段（7:00~8:00）跟车调查统计各站点的上下车客流量与本方法推断得到的结果的对比如图 4-2：

由图 4-2a、4-2b，公交乘客上下车客流量数据挖掘结果与实际值非常接近，决定系数分别为：$R_{Boarding}^2$=0.924，$R_{Alighting}^2$=0.956，但在部分站点（如站点 4，站点 14）误差偏大，原因是这些站点存在较多使用现金的乘客，而运营数据中无记录。最终统计得到，所提方法计算的上车站点匹配与实际调查数据间相对误差为 12%，下车人数匹配相对误差为 16.4%。

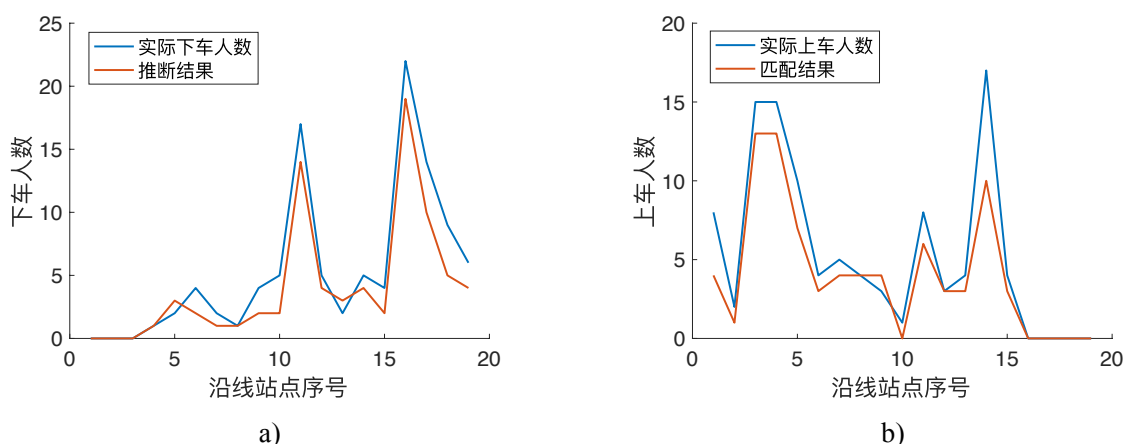

图 4-2 跟车调查与数据处理结果比较

Fig. 4-2 The results of traffic survey and data parsing.

## 4.3 换乘识别方法

在推断出乘客上、下车站点之后，还需要从乘客每天出行轨迹中，区分每次具有不

同目的出行，即出行阶段划分。最常用的方法为识别乘客出行的起点、换乘、终点站，其中换乘站点提取是该步骤的核心。

本节在前人研究的基础上，提出从时间和空间两个角度综合判断公交换乘行为。通过推断得到的 OTD 时空轨迹序列，获取乘客公交出行过程中各个节点的时间与空间信息，挖掘其停驻点的空间特征，以使换乘行为判别更加全面、合理。

乘客连续两次出行的分别记为：$T_k \rightarrow T_{k+1}$，对应线路$L_k \rightarrow L_{k+1}$，则轨迹序列即可表示为：

$$Traj_{k,k+1} = \{T_k, T_{k+1}\} = \{B_k, A_k, B_{k+1}, A_{k+1}\}, \qquad (4-8)$$

则中间过程$M_k = A_k \rightarrow B_{k+1}$，若$M_k$是换乘过程，则$T_k, T_{k+1}$为两次不同的出行，反之，$T_k,,T_{k+1}$均应归为同一出行的两个不同的乘车阶段。本研究判定过程$M_k$为换乘过程的依据如下：

**换乘距离约束**：假定乘客在$A_k \rightarrow B_{k+1}$间步行距离不应超过一段合理的步行时间可到达的最远距离，本研究定义在主城区，换乘距离空间阈值为 500m，在城郊区镇，空间阈值为 700m；

**时间约束**：换乘过程主要消耗的时间$T_{M_k}$包含乘客在$A_k \rightarrow B_{k+1}$间的步行时间$T_{M_k}^w$以在$B_{k+1}$的候车时间。因而，乘客在$B_{k+1}$实际候车时间为：

$$T_{M_k}(wait) = T_{M_k} - T_{M_k}(walk) \qquad (4-9)$$

假设乘客第$k$次出行的下车时刻（等效为所乘坐车辆进站时刻）$t_k^A$，则该乘客在车站$B_{k+1}$的候车时段为：

$$[T_{M_k}(wait)] = [t_k^A + T_{M_k}(walk), T_{M_k}(wait)] \qquad (4-10)$$

若该时段内仅有不超过 3 辆满足$B_{k+1} \rightarrow A_{k+1}$间出行需求的车辆停靠$B_{k+1}$，则判定过程$M_k$是换乘过程。时间约束隐含的该假设是乘客不会故意错过多辆满足其出行需求的车辆而耽误出行。

**重复线路约束**：若$L_k = L_{k+1}$，则过程$M_k$不为换乘。

**出行目的约束**：若在第$k$次出行站点$B_k$有从$B_k$到$A_{k+1}$的直达线路$L'_k$，且$L'_k$的候车时间不超过$L_k$候车时间的两倍，则$M_k$不是换乘过程。

本研究定义，**若只有换乘距离约束与时间约束得到满足，重复线路约束以及出行目的约束有其中一点不被满足，则认为在中间过程$M_k$发生了短时活动**。

根据以上定义的约束条件，得到出行阶段划分的步骤如下：

**Step 1** 将乘客的乘车记录按时间上车时刻进行排序；

**Step 2** 读取第$k$条与第$k+1$条出行记录$T_k$，，$T_{k+1}$；

**Step 3** 判断中间过程$M_k$是否为满足换乘距离约束、时间约束、重复线路约束和出行目的约束，若其中一点不满足则认定$T_k$，$T_{k+1}$为两次独立的出行阶段；

**Step 4** 若$M_k$是换乘阶段则继续提取下一条乘车记录$T_{k+2}$，直到提取到乘车记录$T_{k+m}$所对应的中间过程$M_{k+m-1}$不是换乘过程，此时输出整个出行阶段以站点为锚点的轨迹序列：

$$Traj_{k,k+1} = \{T_k, T_{k+1}, ..., T_{k+m}\}$$
$$= \{B_k, A_k, B_{k+1}, A_{k+1}, ..., B_{k+m-1}, A_{k+m-1}\}$$

（4-11）

经试验，本文所提供下车站点匹配与换乘识别算法在 Intel i7 2.8GHz 4 核计算平台上完成目标城市 3 个月乘客出行轨迹还原以及换乘识别用时为 7h，因此仍能够满足工程应用需求。

## 4.4 城市公交客流特性分析

### 4.4.1 客流时变特性

通过分析公交客流的时间特征，可以发现公交乘客在时间上的出行规律，公交运营部门可以根据乘客在出行时间上的需求对公交车辆进行调度以提高公交运营服务的质量，本节主要在不同粒度上对公交客流时间特征进行分析。

通过换乘上下车站点匹配以及换乘识别，得到目标城市在工作日与周末的小时客流量均值如图 4-3 所示：

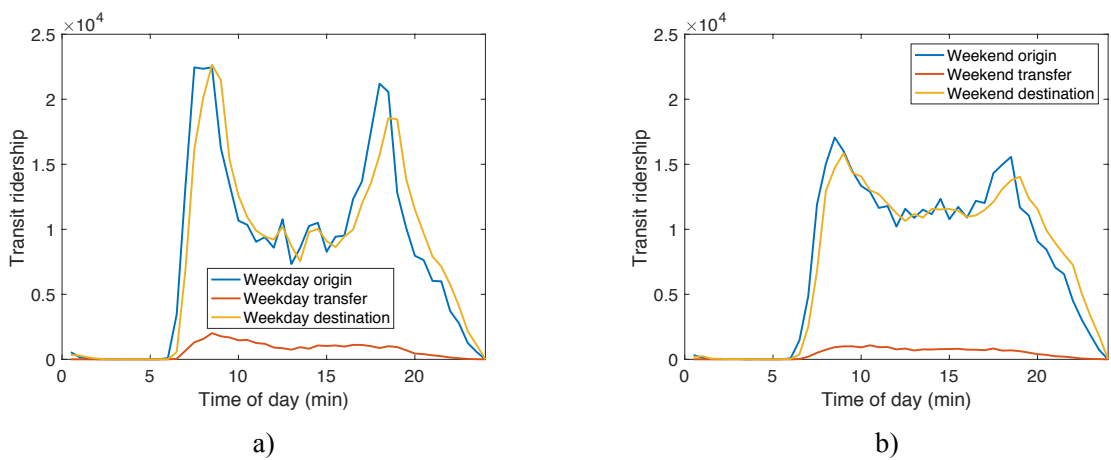

图 4-3 工作日与周末平均客流量时变图

Fig. 4-3 Compare of ridership in weekday and weekend.

由图 4-3 可知，在工作日期间，客流量均呈现明显的早晚高峰，早高峰集中在 7:30-9:00，晚高峰集中在 17:00-19:30，工作日早高峰客流持续时间短于晚高峰但流量极

大值高于晚高峰；双休日客流也呈现了较明显的早晚高峰现象，相比之下，早晚高峰客流量均值低于工作日早晚客流量。但从 9:00-17:00 的非高峰时段内，客流量明显高于工作日。从图中换乘流量上观察，可以看到换乘流量远远小于起点、终点流量，即该城市绝大多数乘客无需换乘即可达到出行目的。

进而分别统计工作日与周末不同类别持卡人在全市范围内各时段的客流量分布，结果如图 4-4 所示。

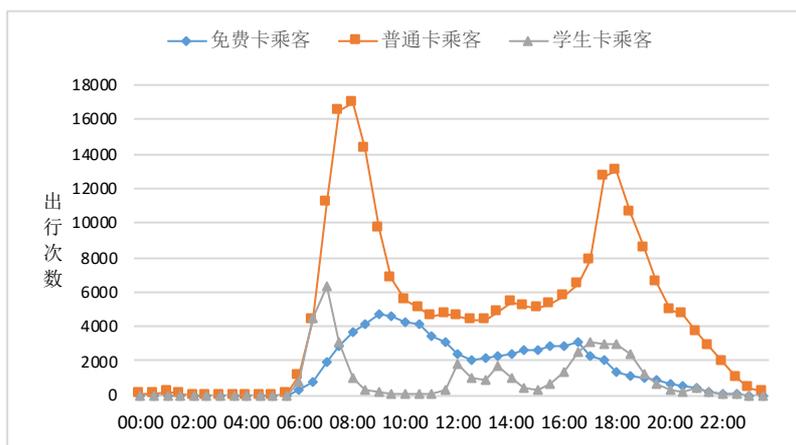

a) 工作日

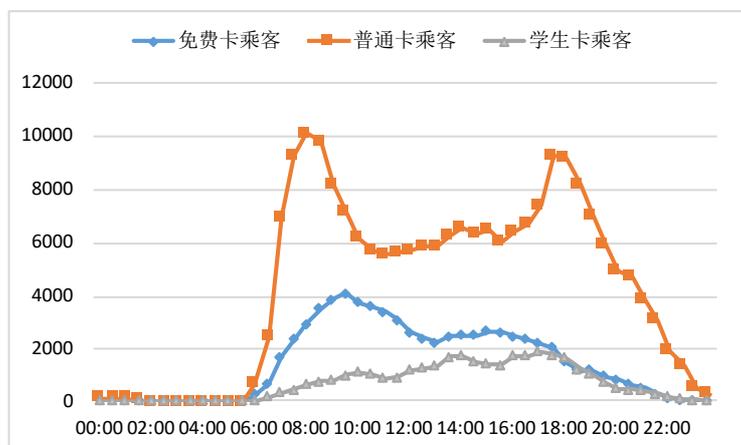

b) 周末

图 4-4 不同类型持卡人客流量时变特性

Fig. 4-4 Compare of ridership of different card holders.

由图 4-4a 观察，在工作日，普通乘客出行呈现两个明显的出行高峰期，分别为：7:30-9:00 与 17:00-19:30；学生出行具有四个高峰，其中早高峰时段为 6:30-7:30，中午有两个小高峰分布在 12:00-12:30 和 13:00-14:00，晚高峰分布在 16:30-18:00，其中早高峰跨度时间较短且出行量较大是因为乘客学生课程时间固定，需在固定时间段内出行至学校，而中午的两个小高峰说明有部分学生中午需要离校，晚高峰时间跨度较大且出行量较小；免费卡多为退休老人，出行主要集中在早 8:00-10:00，到晚上 17:00 之后，该

人群出行量锐减，符合退休老人上午外出活动，晚间较少出行的特征。

由图 4-4b，可以观察到周末普通卡出行量明显减少，免费卡乘客周末的出行时间跟工作日近似，集中在上午 8:30-10:00 之间，其他时段流量平稳，在 16:00 后逐渐下降，没明显的有晚高峰，这说明免费卡乘客的出行时间基本不受工作日和周末的影响。学生卡乘客的出行在周末大幅度减少，且不具有明显的高峰，原因是学生周末无需返校，其出行时间较为随机。

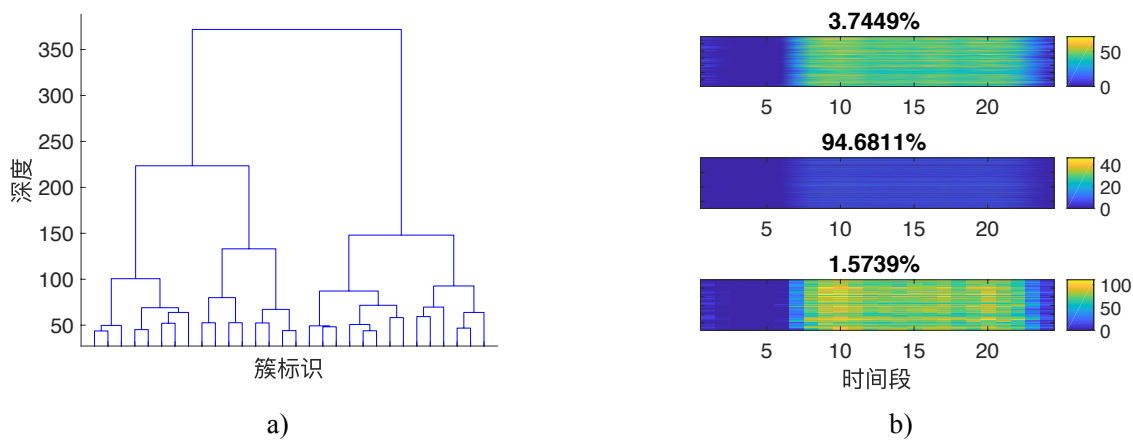

图 4-5 相邻站点区间向量时间层次聚类结果

Fig. 4-5 Clustering results of ridership vector of bus route segments.

将全市工作日所有 OD 上的客流量映射到相邻站点区间上，并以一个小时为时间窗口，统计其 24h 的客流量，即将每个相邻站点区间转化为 24 维客流时变向量，并基于层次聚类法对该 24 维向量进行层次聚类分析，选择平均距离为簇间距离，聚类过程生成的树形图（dendrogram）以及聚类结果如图 4-5 所示：

由图 4-5a，可以观察到从最底层开始聚类，当深度达到 150 层，即聚为 3 类时，簇间平均距离已高于之前迭代步骤中所有的簇间距，因此可以将最佳聚类数设定为 3 类，得到聚类结果见图 4-5b，图中，第 1 类与第 3 类 OD 客流量在时段上具有相似的特征，即存在 7:30-10:00、17:00-18:00 以及 19:30-2:30 三个高峰时段。仅是客流量上存在差异；第二类在全天范围内，每小时的客流量均不超过 30 人次。

## 4.4.2 空间分布特性

乘客出行的空间分布特性是除时间分布特性以外最重要的性质，从客流量的角度上考察，作出各类站点区间对应的轨迹，结果见图 4-5a（该数据可视化图中，轨迹颜色代表类别，透明度代表当前区间的叠加密度）。由图 4-5a，聚类结果第一类与第三类簇（黄色与红色轨迹）所对应站点区间可以认为是该城市的骨干线网，由图所示，第三类站点

区间主要分布于主城区繁华路段（B、F 区）以及该市西部城区与主城区相连的交通枢纽点上（E 区）；而第一类区间，除在主城区广泛分布外，还在该市的城郊区镇中较繁华的地段（A、C、E 区）分布，可以认为，第一类站点区间重要性仅次于第三类区间；但是，从客流量分配上考察，三类区间承载的日平均客流量分别为：9.4 万人次/天（约 20.36%），29.9 万人次/天（64%）与 6.9 万人次/天（约 15%），即大量乘客出行分布位于非主干线网上，主干线网仅承担约 37%的客流量。

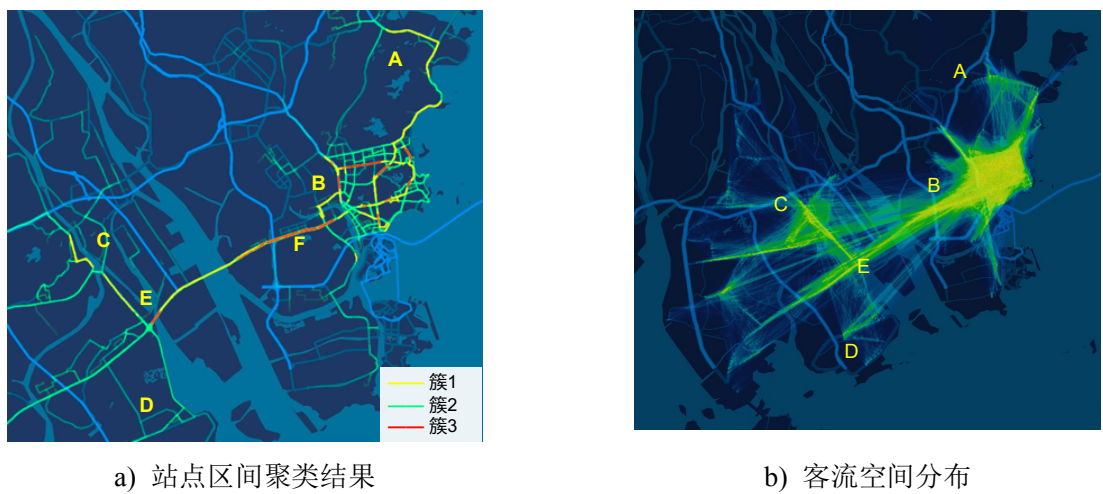

a) 站点区间聚类结果    b) 客流空间分布

图 4-6 客流空间分布特性

Fig. 4-6 Spatial distribution of ridership.

做出该城市三个月全部 OD 连线如图 4-5b 所示，由图可以观察到 B 区，即主城区，OD 间连线最密集，其次，在城郊区镇中，C 区与 E 区也产生了较密集的 OD 链接。

定义两相邻的公交车站$s_i$，$s_{i+1}$间在观察时段$T$内的满载率为：

$$o(s_i, s_{i+1}) = \frac{V(s_i, s_{i+1})}{C(s_i, s_{i+1})} \quad (4\text{-}12)$$

其中，$V(s_i, s_{i+1})$表示车站$s_i$，$s_{i+1}$间的客流量，$C(s_i, s_{i+1})$表示时段$T$内行驶于两站间的公交车运力总和，假设每辆车可乘坐人数为 30 人。

统计全市 3 个月早晚高峰以及其它时段的线网中相邻站点的区间满载率，分别如图 4-7、图 4-8 所示。

由图 4-7a，在非高峰时段，满载率超过 1 的路段主要分布于主城区（B 区），且较稀疏，从概率分布（图 4-8a）上观察，此时段内绝大多数车辆满载率低于 0.5（均值为 0.329），几乎不存在公交拥挤的区间。

在早高峰时段（图 4-7b），主城区满载率超过 1 的区间在空间分布上有所扩展，逐步连成网状，由概率分布（图 4-8b）观察，此时已有一部分车辆满载率高于 1，但全市满载率均值仅为 0.57。

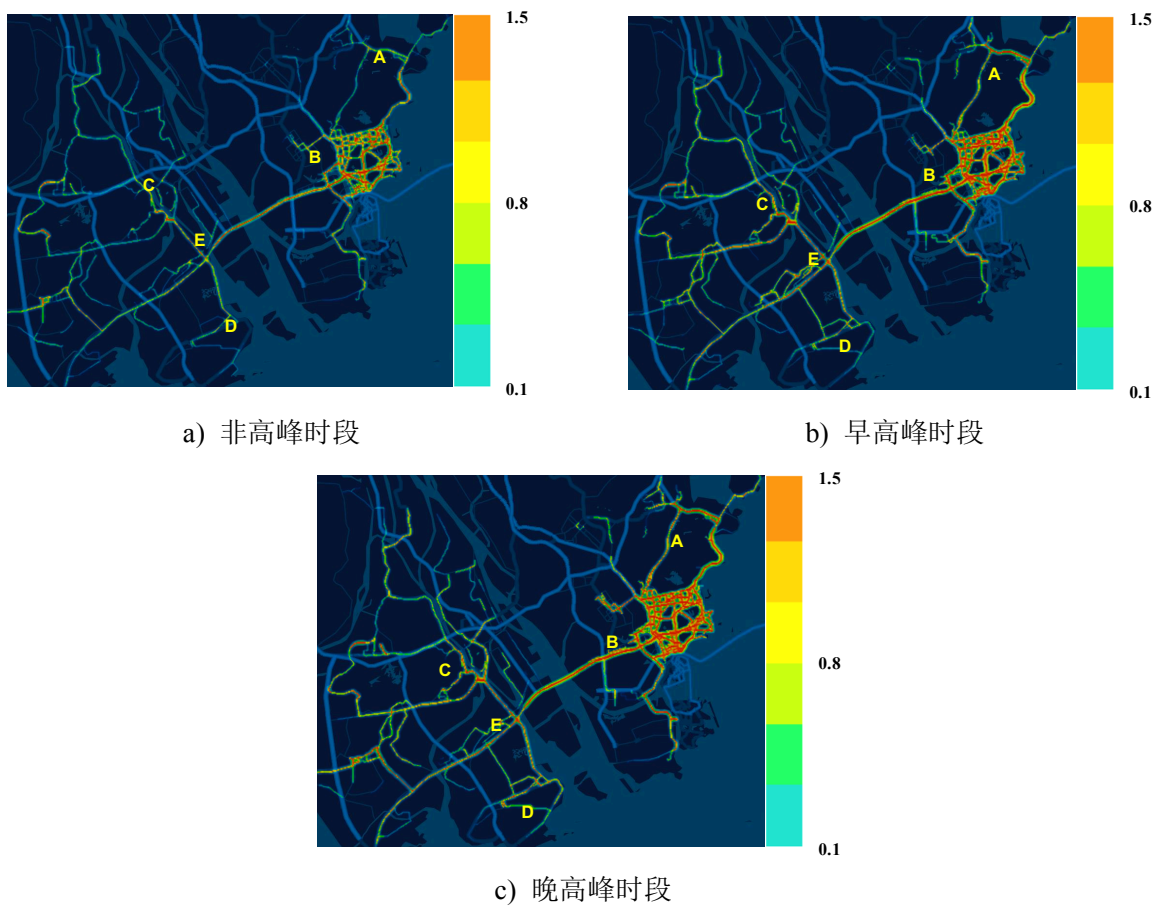

a) 非高峰时段　　　　　　　　　　b) 早高峰时段

c) 晚高峰时段

图 4-7 满载率空间分布

Fig. 4-7 Spatial distribution of occupancy ratio.

晚高峰时段（图 4-7c），拥挤区间在达到最大值，空间分布上表现为大量满载率大于 1 的区间已连成网状（图中 B 区）。此时从满载率概率分布（图 4-8c）上观察，已出现大量满载率高于 1 甚至达到 2 的区间，此时全市平均满载率为 0.73，公交线网运力面临饱和。

同时，从满载率空间分布中还可以观察到，全市唯一连通主城区与城郊区镇的路段（$B \rightarrow E$）全天满载率高于 1。结合图 4-3，晚高峰虽然客流量偏小，而满载率较高。从行程时间上分析，该城市早晚高峰以及非高峰时段平均站间行程时间分别为 116 秒，97 秒与 163s，据此，可以推断造成公交系统晚高峰拥堵的原因是道路拥堵，在观察时段内没有足够的公交车辆到达停靠站转移乘客。

## 4.5 本章小结

本章主要工作为从乘客上车刷卡中还原乘客的出行轨迹，并对目标城市公交乘客的出行分布特性进行初步研究，主要内容如下：

（1）改进了经典的下车站点推断方法，引入概率推断、居住地预估计等方法，用

第四章 公交乘客完整出行信息提取方法以在数据缺失的条件下最大程度还原乘客的出行时空轨迹，跟车调查试验表明该方法效果良好，下车站点推断相对误差不超过 15%。

（2）改进传统的换乘识别方法，在时空阈值假设的基础上引入了短暂活动识别方法。并借助 OTD 数据分析了该城市公交客流的时变特征与满载率的空间分布特性。、

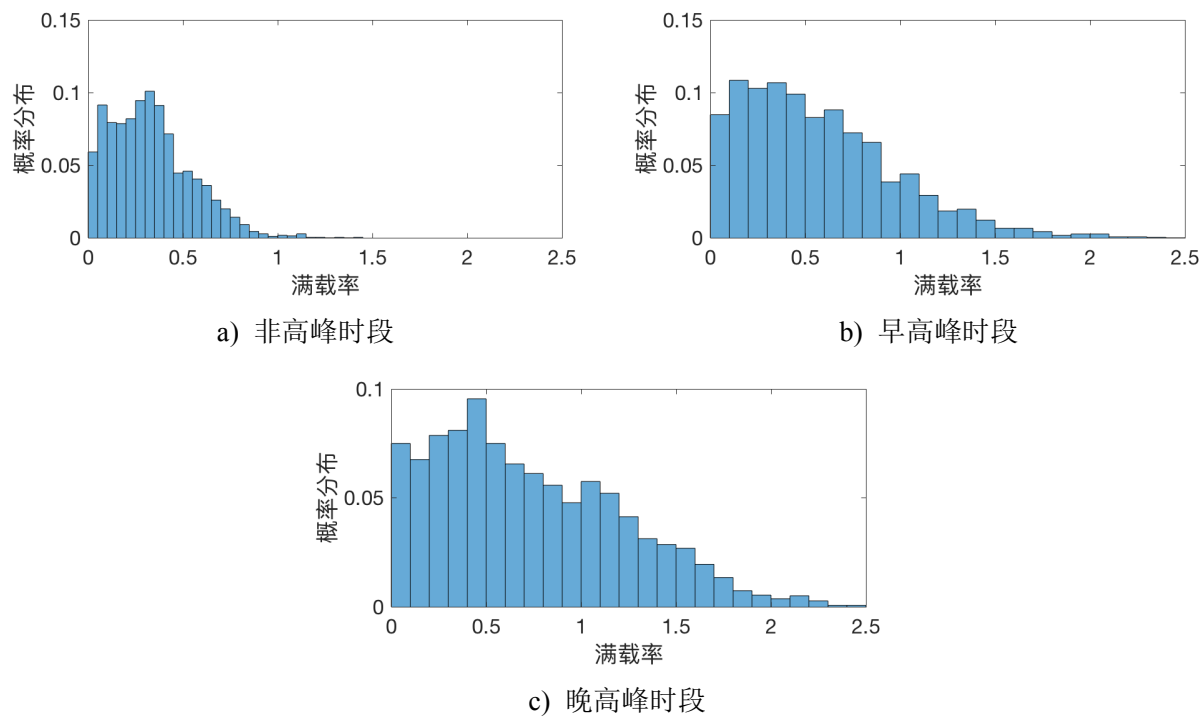

a) 非高峰时段　　　　　　　　　b) 早高峰时段

c) 晚高峰时段

图 4-8 满载率概率分布特性

Fig. 4-8  Probabilistic distribution of occupancy ratio.

# 第五章 基于时空轨迹片段的乘客闭合出行链挖掘方法

经过前面章节的数据挖掘与还原，得到了公交乘客 OTD 时空轨迹（包含时空轨迹与 O、T、D 标识），其提供了大量有价值的信息。然而公共交通不是乘客出行唯一选择，乘客在各种交通系统中切换，导致 OTD 时空轨迹仅仅反映了乘客的出行片段，并不能从根本上刻画乘客的出行规律。因此，相关文献提出了居民出行链的概念，**出行链研究是乘客交通出行行为研究的重要部分，其描述了出行者从起始点出发，经过若干个目的地，再返回起始点的出行全过程**。它摒弃了传统交通行为模型中表征每一次出行的孤立静态性，真实揭示了城市交通出行全过程的连续性特征，体现了乘客交通出行的连续动态特性[27]。即，出行链整合了 OTD 时空轨迹信息，甚至可以认为 OTD 时空轨迹是乘客出行链的外在表现，也是乘客重要的出行模式。

本章以及第六章共同围绕的主题是：公交乘客个体出行模式分析及线路选择偏好模型，具体的，本章提出了一套从碎片化的公交乘客 OTD 时空轨迹片段中，还原乘客闭合出行链的方法；且借助乘客闭合出行链分析了目标城市乘客活动的空间特性。一方面从出行链的角度揭示了乘客活动规律；另一方面，为第六章乘客个体线路选择偏好相关分析提供必要的数据基础。

## 5.1 出行链提取总体技术流程

在公共交通领域，采用数据挖掘手段对乘客的闭合出行链进行还原与挖掘分析仍是一个难点，现阶段公共交通数据挖掘及应用虽在一定程度上揭示了乘客的活动规律[89]，但大部分研究案例均未采用闭合出行链对乘客行为进行建模；另一方面，公共交通领域闭合出行链的研究，国内多处于轨迹数据还原[90]以及基于调查数据的统计建模阶段[91,92]。**究其原因，公共交通不是城市居民出行唯一的出行方式，因此，公交数据系统中只能记录碎片化的乘客出行轨迹片段，难以直接分析。**

在理想状况下，乘客的公交出行链可以通过 OTD 时空轨迹数据直接提取，但是，随着私家车、地铁交通方式的普及居民出行有更多的选择与更大的自由度。因此，乘客每一天的公交出行轨迹数据必然变得碎片化，因此需要叠加多天的出行轨迹才能有效挖掘公交乘客的闭合出行链。本章叠加乘客多天的出行轨迹数据挖掘闭合出行链及出行特征方法的整体技术流程如图 5-1 所示。主要步骤如下：

首先，对 OTD 时空轨迹进行了一次聚类归并，以合并空间属性上高度相似的出行；

其次，利用数据结构"出行拓扑关系图"对空间上不闭合的出行轨迹片段进行归并

整理，从多天数据叠加的角度构建出行阶段的拓扑关系；

再次，利用相似性关联，将数据挖掘得到的出行链与 OTD 出行轨迹进行空间关联，建立 OTD 出行轨迹片段→出行链间的映射关系；

最后，结合 OTD 出行轨迹与个体乘客出行链集合进行数据挖掘，揭示乘客深层次活动规律。

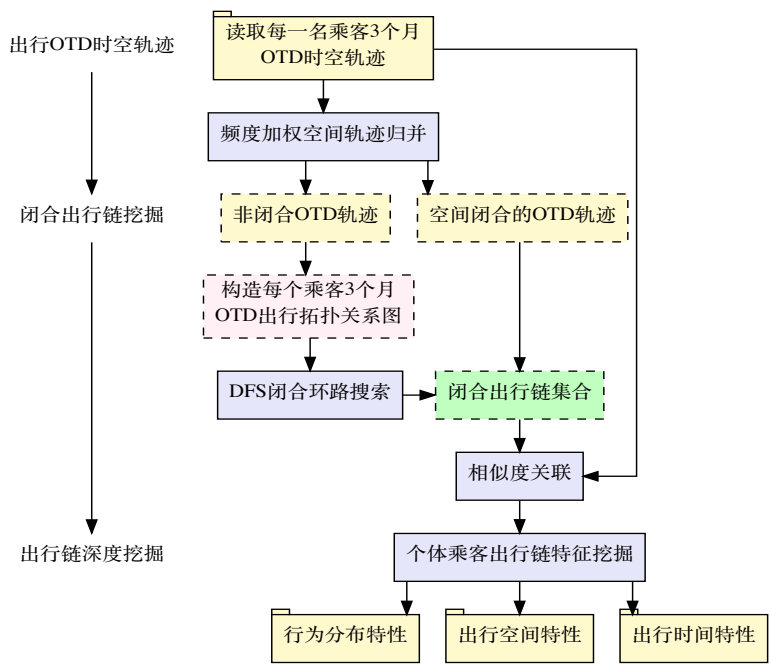

图 5-1 乘客闭合出行链挖掘方法总体流程

Fig. 5-1 General flowchart for closed transit chain mining.

## 5.2 闭合出行链定义

本节提出闭合出行链的定义，对每个公交乘客$p$，$p$在每个出行日$d$的出行记录可以用有续集的形式描述为：

$$L_p^d = [T_1, T_2, \ldots, T_n]$$
$$T_i = (s_{board}, s_{alight})$$
（5-1）

其中，$T_i$表示组成一次完整出行 OTD 的公交车站（由$s$表示）序列，例如：

$$T_i = [s_{origin}, s_{transfer.1}, s_{transfer.2}, \ldots, s_{transfer.k}, s_{destination}]$$
（5-2）

如果有序列集合 $TC = [T_i, T_{i+1}, \ldots, T_{i+n}] \in L_p^d$ 形成了一个闭合回路，则可以定义$TC$为乘客$p$的闭合出行链。实际观察中，我们发现乘客并不总返回与开始出行时候相同的站点，即，形成闭合出行链时，该出行链的第一个乘车站点（O 点）与最后一个下车站点（D 点）存在一定的空间距离差，对此，补充定义出行链的闭合距离$\delta$为：

**定义5.1**：若某出行链起点站$s_{origin}^i$与终点站$s_{destination}^{i+n}$间的空间距离$\delta$：

$$\delta = distance(s_{origin}^i, s_{destination}^{i+n}) \tag{5-3}$$

若$\delta$低于指定阈值（本文选500m）则仍然认为$TC$是空间上的有效闭合环。

对任意乘客$p$，在一段固定时间内，其出行需求在出行链空间中可以表示为：

$$D_p = [TC_1, TC_2, \ldots, TC_n] \tag{5-4}$$

如果$D_p$包含了乘客$p$所有的闭合出行链，且出行日$d$的所有出行都是公交出行，不存在数据缺失，则出行日$d$的公交出行$L_p^d$可以用于$D_p$相同维度的向量$V_p^d$描述。若该向量中每一个维度表示隶属于$TC_i$的出行发生的次数$n_i$，则已知$TC_i$所需的出行阶段后，可用$n_i$直接刻画出行链的完整度。

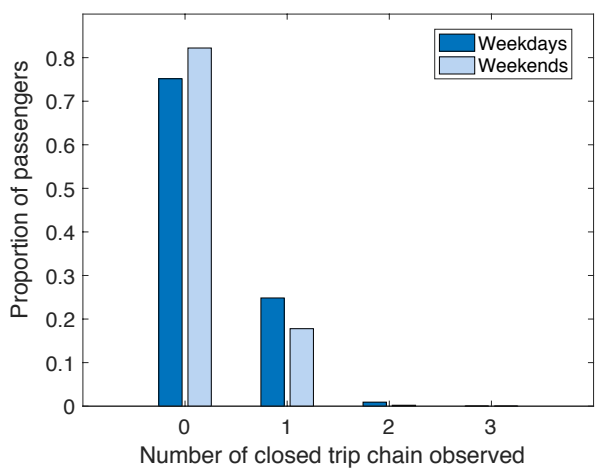

图 5-2 平均每个出行日产生的闭合出行链比率

Fig. 5-2 Initial statistic of closed transit chain.

出行链可以更加清晰的刻画乘客，尤其是个体乘客的活动规律。对个体乘客而言，频繁使用的闭合出行链是重要的出行模式，对行为分析与公共交通规划具有重要的指导意义。

本研究对目标城市从2015年3月1日至2015年6月1日的公交乘客OTD出行轨迹数据中选出最大客流量的工作日，与客流量最大的周六进行数据挖掘；考察在不进行多天数据叠加的条件下，按空间闭合原则（定义5.1），公交系统的OTD数据中乘客闭合出行链的捕获完整度。数据挖掘统计结果如图5.2所示，由图5-2，在工作日与双休日可以观察到闭合出行链的乘客比率仅仅分别占全部乘客的29%与25%。值得注意的是，在周末，能观察到闭合出行链的乘客比率低于工作日。以上统计结果说明单独一天的数据仅能得到小部分乘客的闭合出行链，因此必须采用更有效的方法提取公交乘客的闭合出行链。

本研究假定乘客$p$在一段时间内的出行（本文为 3 个月）可以表示为：

$$C_p = \{L_p^1, L_p^2, \ldots, L_p^n\} \qquad (5\text{-}5)$$

以公交站点拓扑关系上，$C_p$可以用图$G_p = \{V, E\}$表示，但在传统的交通网络分析算法中，在图$G_p = \{V, E\}$中，每个车站$s_i$可以表示为一个节点。若该乘客有发生在$s_i$与$s_j$间的公交出行，则可以定义$s_i$与$s_j$间的连边$e_{ij}$。明显，乘客$p$的出行链$D_p \subseteq G_p$，且每个闭合出行链$TC_i \in D_p$可以表示为$G_p$中满足定义 5.1 的闭合环路。但是，对多天数据构成的图应用传统的环路检测算法，完全可能提取出不存在的出行链。其中一个例子如图 5-3 所示。

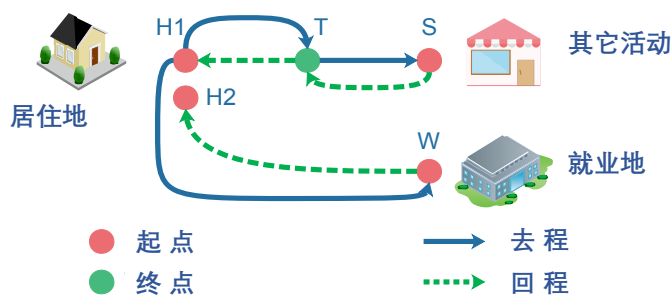

图 5-3 闭合出行链示意图

Fig. 5-3 Demonstration of closed transit chain.

在该案例中，该乘客的居住地附近有两个公交车站（H1 与 H2）并基于居住地附近的车站产生了两条出行链：$TC_{H1,S,T} = \{T_{H1-T-S}, T_{S-T-H1}\}$和$TC_{H1,W,H2} = \{T_{H1-W}, T_{W-H2}\}$。传统的环路探测算法将导致输出并不存在的出行链，例如：

$$TC_{H1,T} = \{T_{H1-T}, T_{T-H1}\}$$

$$TC_{S,T} = \{T_{S-T}, T_{T-S}\}$$

以上出行链满足定义 5.1，但无研究与讨论价值；另一方面，经典图论算法的建模方式没有利用完整的出行信息（包含起点，换乘点和终点），只利用了每个出行阶段在站点网络的空间拓扑关系，如果完整利用每次出行所包含的信息，将可以过滤大量的无效信息。例如，若将图 5-3 车站$s_T$作为换乘点隐藏，则$T_{H1-T}$与$T_{S-T}$将不存在.

针对以上讨论案例以及传统方法的不足之处，本章提出了一种基于出行 OD 拓扑图的闭合出行链挖掘方法，以期从原始数据集中准确的挖掘乘客闭合出行链。

## 5.3 乘客闭合出行链挖掘与关联方法

基于出行 OTD 拓扑图的闭合出行链挖掘流程如图 5-1 所示。该方法首先使用频度加权的空间轨迹聚类算法，将高相似度的出行合并；其次，将合并后的 OTD 数据转为 OD 拓扑图；再次，基于 OD 拓扑关系图提取其中出现的闭合出行链，并利用相似性分

析将所有非闭合的 OTD 时空轨迹与出行链关联。其中，最关键的步骤为构建出行拓扑关系图融合乘客多天的 OTD 出行记录并在其中搜索闭合出行链。

### 5.3.1 出行拓扑关系图

在本研究的目标城市中，尤其是市中心主城区，公交线网覆盖率高且站点密度较大，给乘客出行提供了多样化的选择。因此，在乘客出行记录中，会出行大量空间上相似（起点、终点均具有非常近的地理距离）但首末站名不相同的站点，为保证得到客观真实的公交出行链，本研究首先对$L_p^d$中的元素进行聚类（合并相似的出行）。其数学本质是：乘客$p$的每一条出行记录$T_i$会被归并到其最相似且最频繁使用的出行记录$T_u$中，$T_u$也可以认为是出行 OTD 轨迹集合中簇$u$的代表。对于乘客$p$，首先定义其在数据集中第$i$次与第$j$次出行的相似度为：

**定义 5.2**：对乘客$p$第$i$次与第$j$次出行记录$T_i$，$T_j$对应的乘车站点序列为$s_{T_i}$，$s_{T_j}$，则$T_i$，$T_j$的相似度定义为：

$$d_{ij} = \begin{cases} distance(s_{Destination}^i, s_{Destination}^j) \\ \quad + distance(s_{Origin}^i, s_{Origin}^j) & , if\ ang(\overrightarrow{s_{T_i}}, \overrightarrow{s_{T_j}}) \leq 90° \\ \infty & , otherwise \end{cases} \quad (5-6)$$

其中，$d_{ij}$单位为 m，表示由$T_j$、$T_i$切换所需要的步行距离。

基于相似度定义，聚类算法的主要步骤如下：

**Step 1** 统计$C_p$每个出行$T_i$的出现频次$F_i$并按降序排列。

**Step 2** 设置最小相似度阈值为 700m（设步行速度为 1.2 m/s，约 20 分钟步行时间）

**Step 3** 取出$C_p$中使用频度最高的出行记录（表示为$T_{u0}$），基于式（5-6）以及$minPts$，搜索可以与$T_{u0}$合并的出行记录，形成簇$c_{u0}$，其中，$T_{u0}$为该簇的代表元素。

**Step 4** 继续提取$C_p$中尚未被合并的且出现频次最高的出行记录进行合并聚类，直到$C_p$中所有的出行记录都已完成归类。

**Step 5** 扫描所有得到的簇，若簇$c_{u0}$所覆盖的出行低于该乘客总出行量的 3%，则放弃该簇。

经过初步过滤后，本研究将乘客$p$每一次出行看作是不可分割的元素。并定义了一种新的数据结构。乘客 OTD 出行拓扑关系图（Hyper Trip Graph）：

**定义 5.3**：乘客 OTD 出行拓扑关系图为一有向图：

$$HTG_p = \{V, E, E_c\} \quad (5-7)$$

其中，$V$为顶点集，$E$为连边集，$E_c$为每个节点的使用次数表，乘客$p$每一条完整的

OD 记录$T_i$定义为$V$中的顶点，$E$中每条边$e_{ij}$代表了$T_i$与$T_j$间的有向链接。某乘客在观察时段内公交出行记录构成 OTD 出行拓扑关系图$HTG$的示例如图 5-4：

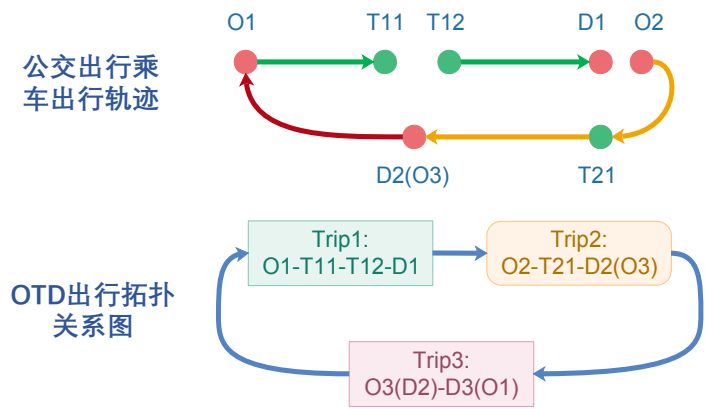

图 5-4 OTD 出行拓扑关系图示例

Fig. 5-4 An example of Hyper Trip Graph.

由图 5-4，该乘客共有 5 种不同的乘车记录，这些乘车记录组成了 3 段不同的出行(Trip1、Trip2、Trip3)，这些出行被转化成$HTG$中三个不同的节点。

为刻画出行 OD 间的关系，以下给出 OTD 拓扑关系图中节点空间可连接的一般性定义：

**定义 5.4**：对$\forall T_i, T_j \in HTG_p$，如果有：

$$distance(s_{T_i.Destination}, s_{T_j.Origin}) < \varepsilon \tag{5-8}$$

或者：

$$distance(s_{T_i.Origin}, s_{T_j.destination}) < \varepsilon \tag{5-9}$$

则定义$T_i$与$T_j$在空间上可链接，对式（5-8）建立的连边为$s_{T_i.Destination} \rightarrow s_{T_j.Origin}$，同理；对式（5-9）建立的边为$s_{T_i.Origin} \rightarrow s_{T_j.destination}$。

其次，给出乘客每个出行日的公交出行零时刻与归一化时刻的定义：

**定义 5.5**：乘客$p$在出行日$d$的公交出行零时刻为该乘客每天第一条公交刷卡记录的产生时刻。

**定义 5.6**：出行日$d$产生的所有的乘车刷卡记录产生时间戳减去该乘客在该出行日的零时刻作为归一化时刻。

推广到多天的情形，则有：

**定义 5.7**：$m$个连续出行日叠加产生的乘车刷卡零时刻为第一天的第一条刷卡记录产生时间戳，其余时间戳以零时刻为基准换算成秒数作为归一化时刻。

再次，定义出行$T_i, T_j$时间上可链接的准则为：

**准则 5.1**：至少存在 1 个独立出行日可以观察到$T_i$在$T_j$后发生。

**准则 5.2**：至少观察到 1 次存在间隔不多于$m$天的出行日$d_i, d_j$进行数据叠加，在归一化时刻背景下可观察到$T_i$在$T_j$后发生。

准则 5.2 用以应对乘客隔天返回出行链起点的情形（如周末度假）。最后，定义$T_i, T_j$可连接的条件为空间与时间上均可连接。考虑到本研究所收集数据集的限制，仅考虑$m \leq 2$的情形。

根据以上定义，OTD 出行拓扑关系图$HTG$中所有结点共同具备如下性质：

**性质 1**：如果式（5-8）成立，则$e_{ji} \neq 0$，$T_i$与$T_j$可以合并为一虚拟出行$T_v^{ij} = [T_i, T_j]$。同理，$e_{ji} \neq 0$若式（5-9）成立，且有$T_v^{ji} = [T_j, T_i]$。

**性质 2**：若$T_i$与$T_j$不能建立链接，则$HTG$中$e_{ij} = 0$且 $e_{ji} = 0$;

**性质 3**：若$e_{ij} \neq 0$且$e_{ji} \neq 0$，则$T_i$与$T_j$间产生空间闭合的环路。

**性质 4**：对于任意节点$T_k \in V$与虚拟出行$T_v^{ij}$，则：$e_{vk} = e_{jk}, e_{kv} = e_{ki}$

借助以上性质，借助进行深度优先搜索对图$HTG_p$中每个节点进行访问，即可得到每个乘客的闭合公交出行链，其主要步骤如下：

**Step 1** 从$E_c$中拷贝各节点被乘客$p$使用的频数，生成节点的可访问次数表；

**Step 2** 从$HTG_p$中任一节点$v_k$开始，先将$v_0$加入 OD 链表$ODL$，利用深度优先法则寻找下一个可连接的节点$v_{k+1}$；

**Step 3** 若$v_{k+1}$与$ODL$中所有元素均无法构成闭合链，则将$v_{k+1}$加入$ODL$；

**Step 4** 若$v_{k+n}$与$ODL$中任一元素可连接且构成闭合出行链$TC_k$，则提取该闭合链，同时，$TC_k$中所有涉及的节点可访问次数减 1；

**Step 5** 若某节点$v_m$可访问次数降为零，则从$HTG_p$与$ODL$中移除该节点。

**Step 6** 若$ODL$无法再获取到新节点且仍不闭合，则将最后一个进入链表的节点移出再次利用深度优先准则获取新节点。

**Step 7** 若$ODL$中所有元素均被移出，则重新选择开始点，直到$HTG_p$中所有可访问的节点都被当作过开始点用过为止，保证搜索到$HTG_p$中所有的连通分量。

### 5.3.2 非闭合出行轨迹与出行链的关联方法

该方法虽然得到了乘客闭合出行链的空间结构信息，但该方法得到的空间闭合环路仍缺少起点信息，因此还需要将出行链与 OTD 时空轨迹记录关联，以补充完整全部信息。

对于乘客$p$的闭合出行链集合$D_p$，可以逐天将乘客产生的非闭合出行$T_{nc} = [T_1, T_2, \ldots, T_k]$与最相似的$TC_x$进行关联，本研究采用极大相似度关联法。

**定义 5.8**：对$D_p$中每一个闭合出行链$TC_i$，其在出行日$d$非闭合出行记录集$T_{nc}$中的支持度与覆盖度分别定义为：

$$support(TC_i|T_{nc}) = \frac{TripCount(TC_i \cap T_{nc})}{TripCount(TC_i)}$$

$$coverage(TC_i|T_{nc}) = \frac{TripCount(TC_i \cap T_{nc})}{TripCount(T_{nc})}$$

（5-10）

即，支持度表示$TC_i$与$T_{nc}$共有的等价出行次数比上$TC_i$中所有的出行；覆盖度表示$TC_i$与$T_{nc}$共有的等价出行次数比上$T_{nc}$中所有的出行次数。

根据该定义，将关联问题转化为：寻找尽肯能少的闭合链$TC_x \in D_p$，将$T_{nc}$所包含的非闭合出行全部覆盖，同时要求每个$TC_x$拥有尽可能高的支持度。显然，该问题是经典的集合覆盖问题，可以借助贪婪算法求解，具体步骤如下：

**Step 1** 计算$D_p$中每个闭合链在$T_{nc}$中的支持度与覆盖度，取出支持度与覆盖度乘积最大的闭合链$TC_m$，与从$T_{nc}$中移出相应被覆盖的记录$T_{nc}^r$，同时将$TC_m$与$T_{nc}^r$建立关联。

**Step 2** 重复第1步，直到$T_{nc}$中所有的出行均已被关联。

### 5.3.3 算法性能分析

首先定义乘客$p$的一条闭合出行链为$TC_i$，该链包含了$N_t^i$次出行；对于其中每次出行$T_j$，定义其包含$N_b^j$次乘车记录，明显，$N_b \geq 2N_t$。在该乘客出行记录中$L_p$中，包含有$N_{T\_total}$次出行。

假设在研究时段内，$TC_i$发生了$F_i$次，若该乘客有$k$条闭合出行链，则：

$$N_{T\_total} \geq \sum_1^k F_i \cdot N_t^i \tag{5-11}$$

式（5-11）取等号的情形为：该乘客所有的公交出行均在该城市公交数据系统留下记录，据此，可以定义一个全局参数，用以描述该乘客出行记录完整度：

$$Cpl_p = \frac{\sum_1^k F_i \cdot N_t^i}{N_{T\_total}}. \tag{5-12}$$

为测试本文方法在缺失信息条件下的数据还原性能，按表 5-1 构造一组特殊的测试出行链：

表 5-1 虚拟乘客出行轨迹特点

Table 5-1 Configuration of trip chain of a virtual passenger.

| 类目 | 出行阶段数 | 换乘次数 | 出现频次 | 数据缺失率 |
| --- | --- | --- | --- | --- |
| 取值 | 2 ~4 | 0 ~2 | 1 ~19 | 5% ~90% |

由表 5-1，共构造 117 种不同的闭合出行链以及 2130 条出行乘车记录。对此数据集随机移除乘车记录以模拟$Cpl_p$的变化。对每个不同取值的数据完整度$Cpl_p$，重复 50 次不同的试验。出行链还原成功率与数据完整度间的关系见图 5-5

由图 5-5a 至 5-5c，随着出行阶段数增加，图中深蓝色区域面积逐渐扩大。该结果说明包含出行阶段数越多的出行链，对数据完整度要求越高；同时，被使用次数多的出行链，对数据丢失率有更高的容忍度。例如，对使用频数为 10 次的出行链，若希望其有 80%的几率被还原，则本方法可以容忍的数据丢失率为 20%。

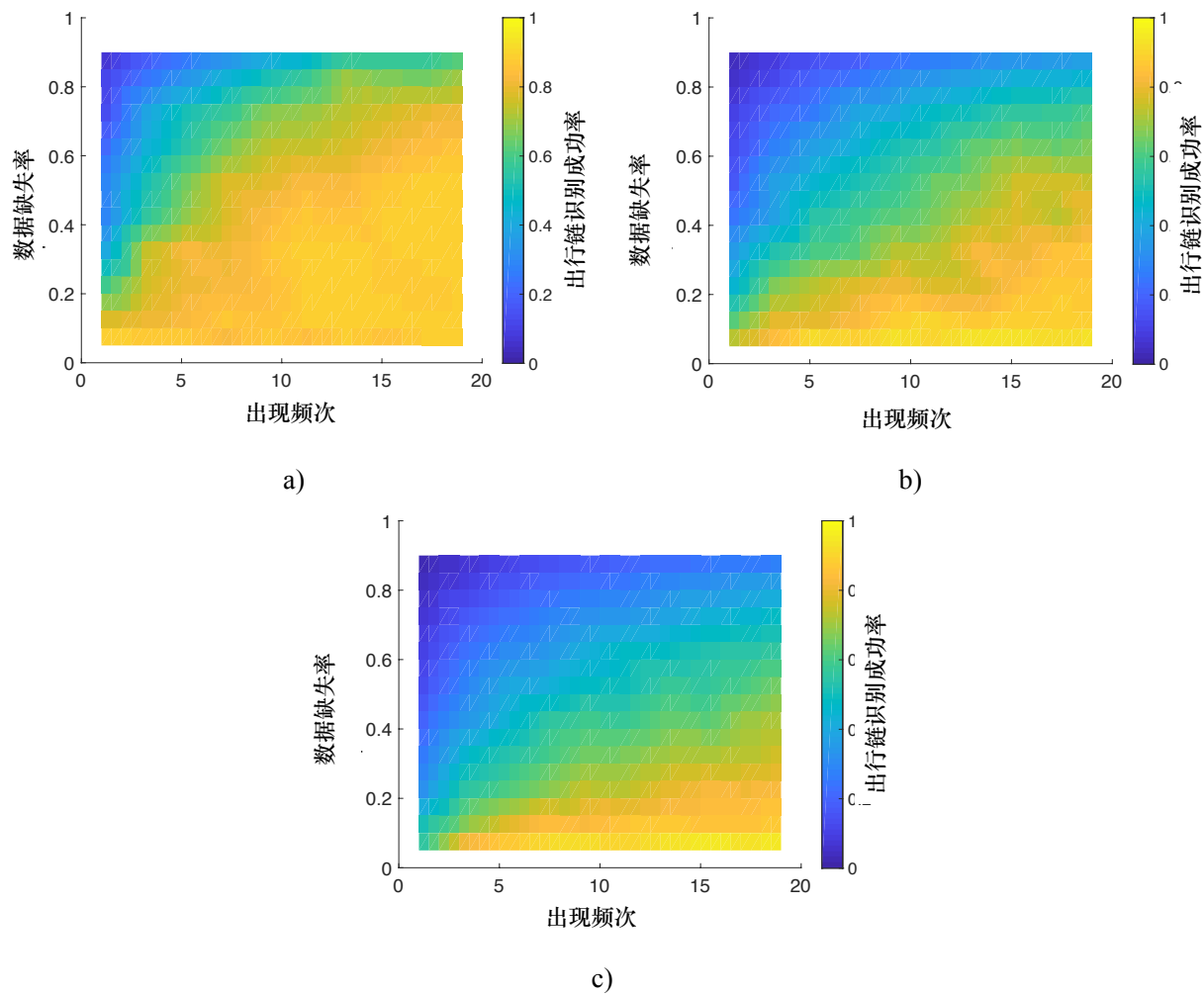

图 5-5 缺失数据对闭合出行链还原的影响分析

Fig. 5-5 The test on impact of data loss rate.

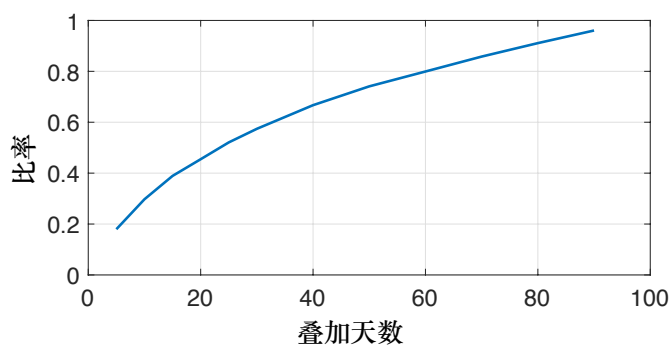

图 5-6 多天数据叠加测试

Fig. 5-6 The test on impact of data argumentation.

为测试数据增强对出行链还原效果的影响，随机选择 30%（约 18 万人）的乘客采用不同长度的采样时间窗口并记录至少找到一条闭合出行链的乘客比率，测试结果如图 5-6 所示，试验结果表明，如果用以采样的时间窗口长度少于 10 天，仅有不到 30% 乘客能找到至少一条出行链；另一方面，随着采样时间窗口长度的增加，能还原出闭合出行链的乘客比率随之提高。同时，当达到 90 天时，90%以上的乘客都能观察到至少一条闭合出行链。

## 5.4 基于闭合出行链的城市居民出行特征研究

### 5.4.1 乘客出行的宏观统计特征

本节以每一个乘客的闭合出行链为起点，从不同尺度研究目标城市居民的活动特征。该城市居民在 3 个月中的出行链数量分布如图 5-7，由图 5-7，可以观察到大多数（60%）的乘客只有少于 2 条闭合出行链。从图中累积分布上观察可以发现 90%以上的乘客所产生的闭合出行链均低于 10 条，另一方面，使用老人卡的乘客产生的出行链远远少于使用普通卡的乘客。

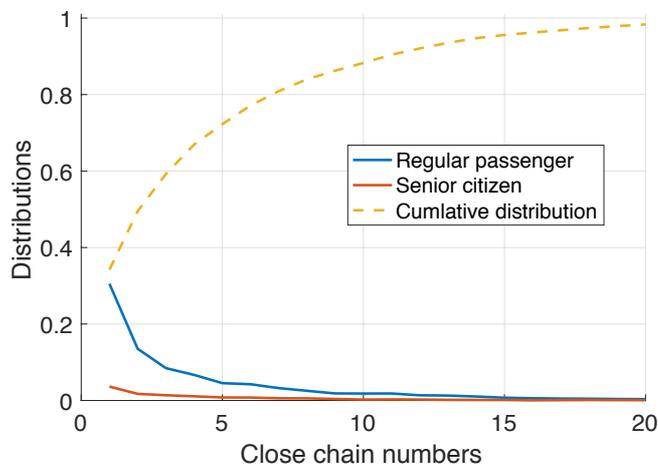

图 5-7 乘客闭合出行链数量分布

Fig. 5-7 Distribution on the number of closed transit chains.

本研究定义乘客出行链直径为距离最远的两站点间直线距离，该城市所有乘客闭合出行链直径在工作日与双休日的分布如图 5.8 所示。从全市的宏观层面上，出行链直径分布在工作日与双休日并无明显区别。值得注意的是，在周末产生长距离（10~20 千米）出行的乘客比率低于工作日。从两分布曲线的拐点（位于 9 千米处）上看，该城市乘客的公交出行链直径均低于 10 千米，即，约 1 个小时的公交车程。尽管该城市公交线网提供了长距离（超过 20 千米）交通小区间的可达性，但从出行链直径分布以及乘客职-住距离分布上观察，绝大多数公交乘客出行位于中等半径的活动空间内。

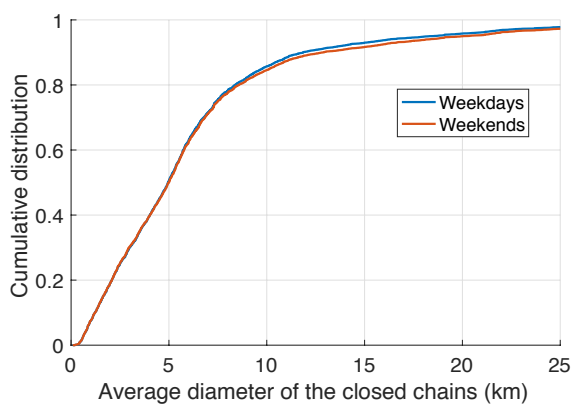

图 5-8 乘客出行链半径分布

Fig. 5-8 Distribution of closed chain diameter.

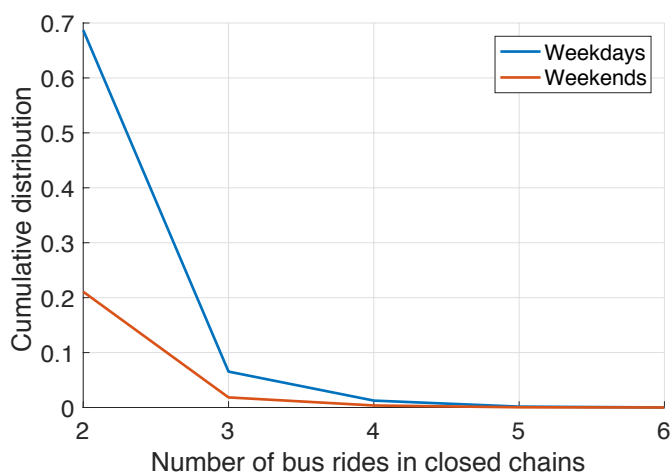

图 5-9 闭合出行链中的乘车次数分布

Fig. 5-9 Distribution of busrides in passengers' closed chains.

工作日与周末乘客形成的闭合出行链中乘车次数分布如图 5-9 所示。由图观察，大部分乘客的闭合出行链仅包含 2~3 次出行。在普通工作日，超过 70%的闭合出行链仅由 2 段乘车构成，而需要 3 次乘车才能完成的出行链仅占 10%；在双休日，与工作日类似，绝大多数乘客的闭合出行链仅由 2 个乘车阶段构成。

### 5.4.2 基于乘客出行链的职住地识别方法

为更加深入研究目标城市乘客的出行行为，除挖掘乘客的闭合出行链外，还需要从站点层面挖掘每个乘客的居住地与职业地点。首先提出以下车站可能在乘客居住地或职业地附近的车站的基本假设：

**假设 1**：居住地附近的公交站点有更大概率捕捉到乘客工作日最早的出行（最早乘车假设）。

**假设 2**： 普通工作日夜间乘客在居住地附近停留时长远大于在职业地的逗留时长（工作日白天假设）。

**假设 3**： 普通工作日白天乘客在工作地附近停留时长远高于在职业地的停留时长（工作日晚间假设）。

基于以上假设，本研究尝试将乘客 $p$ 所有的访问过站点进行向量化处理，本节后续将讨论不同站点筛选方案（起点、终点、被访问过的站点）对乘客职住地判别的指示作用。

$$s_i = \{N_e, S_d, S_n\} \tag{5-13}$$

其中，$N_e$ 代表了 $s_i'$ 成为该乘客每天最早访问的站点的概率；$S_d$ 代表了在工作日白天（8 a.m. to 7 p.m.）的累计逗留时长；$S_n$ 代表工作日晚间（5 p.m. to 7 a.m.）的累计逗留时长。 具体时间划分由全市客流在时间上的分布决定。

完成对乘客 $p$ 访问所访问的站点向量化后，将寻找 $p$ 的居住地转化为以下寻优过程：

$$\widehat{Home} = \text{argmax}_{Home} \ N_e \times exp(S_n - S_d) \tag{5-14}$$

而对 $p$ 的工作地点搜索，采用以下寻优过程：

$$\widehat{Work} = \text{argmax}_{Work} \ S_d - S_n \tag{5-15}$$

式（5-14）的物理意义是：提取 $p$ 的出行链站点列表中，晚间停留时间最长且白天有最大概率被首次使用的站点。同时，指数函数防止 $S_d$ 与 $S_n$ 相等；式（5-15）的物理意义是：提取 $p$ 出行链停驻点中白天与夜晚逗留时间差最大的站点。

为测试本文方法有效性，用该方法从 27 名志愿者的出行活动数据中提取其职住地，做出志愿者不同类别的站点及其时间上的活动热力如图 5-10 所示，图中，x 轴代表一天的不同时刻，y 轴代表志愿者所访问的车站编号。热力图的颜色表示该车站的被使用频次。

首先，由图 5-10a 可观察到，居住地与职业地的活跃时间位于完全不同的区间，具有较好的区分度，如，从居住地开始的乘车出行多发生在早上 8 点到中午 11 点的时间

段内，而从职业地开始的出行多发生在下午 15 点至晚上 20 点的时段内，

其次，从图 5-10b 中观察，虽然行程终点（D 点）也揭示了不同的信息，但从时间分布上看，职业地的到达时间段与居住地的到达时间段间区分度低于行程起点（图 5-10a），可以观察到居住地往往是志愿者最晚到达的站点。

再次，从图 5-10c 中观察，若不管站点在出行阶段中的类别，职住以及其它活动类型的站点在时间热力图上无明显特征，不适合直接用以进行职住地识别。

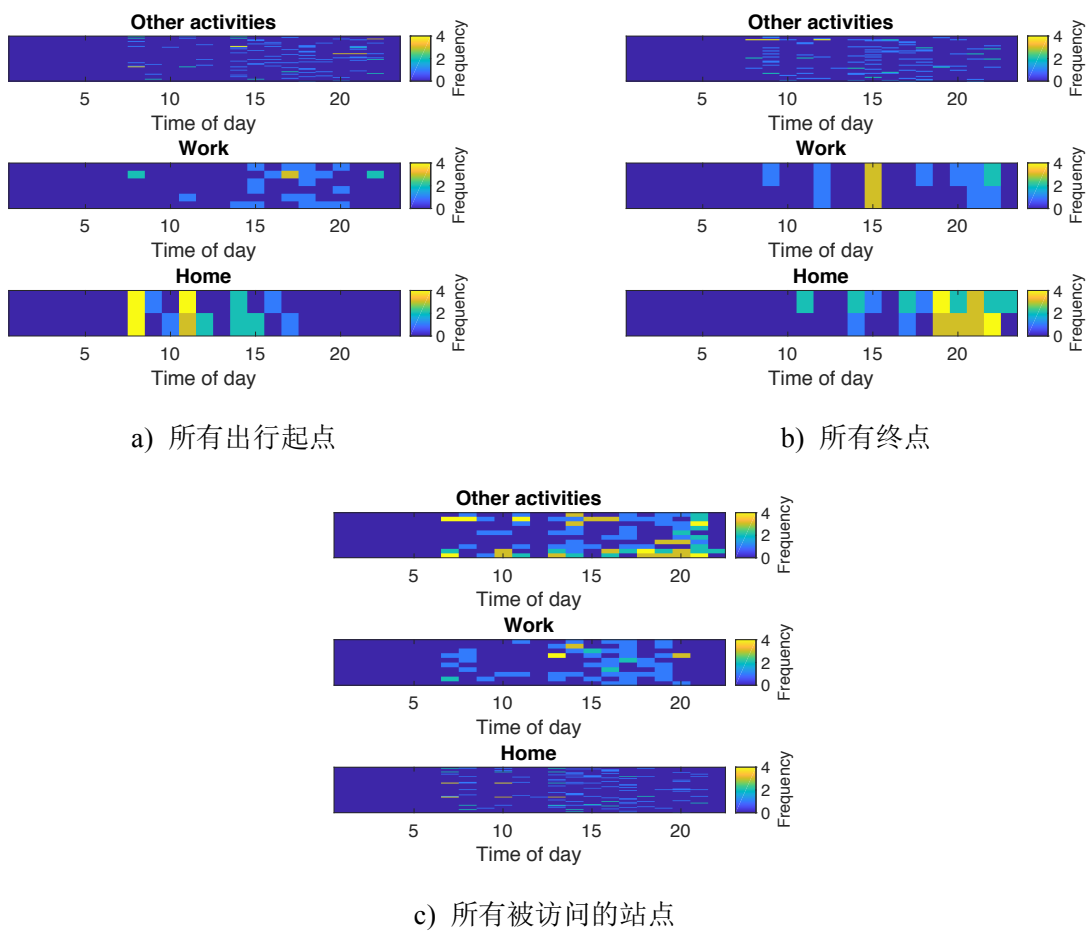

a) 所有出行起点　　　　　　　　　　　b) 所有终点

c) 所有被访问的站点

5-10 不同类别站点对乘客职住地识别的影响

Fig. 5-10 Activity heat-map of stations in different categories..

综上所述，通过志愿者提供的数据验证了职住地识别的方法，且发现用乘客的出行起点能为乘客职住地识别提供更有效的信息，这一结论也从侧面支持了 4.2 节方法中乘客居住预估计结果的可信性。

将借助职住点统计该市居民的职住距离，如图 5-11，图 5-8 对比，可以发现两曲线非常相似且拐点均位于 9 千米处，即该城市居民的职住距离分布在宏观上与出行链直径分布相似。

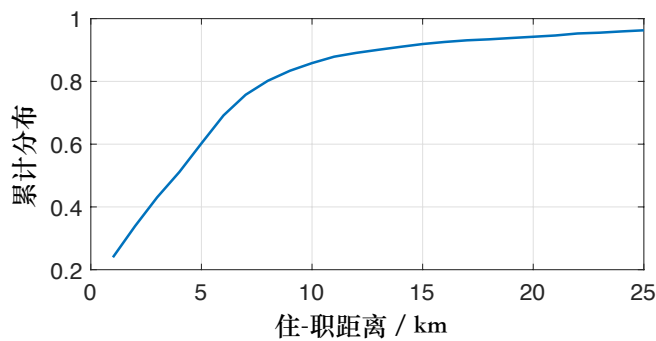

5-11 乘客职-住距离分布

Fig. 5-11　Distribution of passengers' home and work distances.

### 5.4.3 乘客出行空间分布特征

乘客的闭合出行链以及其职住点为不同区域乘客的活动特征分析提供了重要参考。珠海市 3 个的公交刷卡客流，计算月平均日客流量的空间分布如图 5-12 所示。该图说明珠海市公交客流集中分布在主城区内（图中 Downtown）。城郊区镇 A（图中 Satellite town A）为所有郊区城郊镇区中客流量最大的区域。本节将从出行链的角度对比城郊区镇 A 与主城区乘客的出行空间分布特性。

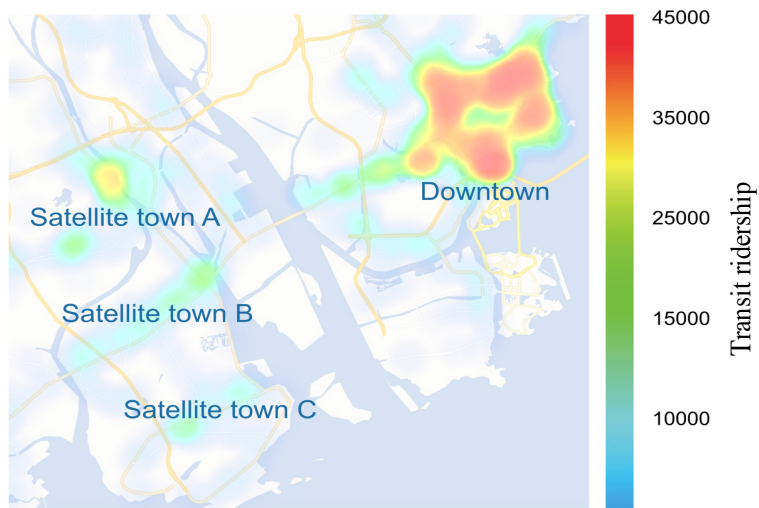

图 5-12 月平均日客流量分布

Fig. 5-12 Spatial distribution of monthly average daily ridership.

对每个站点，计算其住职比：$R_{HW} = N_{Passengers'\ homes}/N_{Passengers'\ work\ locations}$，得到该城市公交站点住职比 $HW$ 在空间上的分布如图 5-123，该图可定性说明居民的居住地具有组团特性，但 $R_{HW}$ 极大（大于 4）的站点仅占极小比率。参考图 5-11，对于该城市各城郊镇区而言，住职比高于 2 的区域就是客流产生的热点区域。

为分析居住地为城郊镇区的公交乘客与主城区间的出行需求，本节首先将每个乘客表示为三维向量：

$$\{P_{Inter}, P_{Intra}, P_{Out}\} \tag{5-16}$$

其中，$P_{Inter}$，$P_{Out}$ 和 $P_{Intra}$ 分别表示以下三种出行链在三个月中被完整使用的概率：

（1）只连接乘客居住地所在镇区内部的站点；

（2）连接乘客居住地与主城区的出行链；

（3）连接乘客居住地所在分区之外其它区域。

其次，应用 k-means++ 算法将向量化的乘客群体聚为 3 类，结果如图 5-14a、图 5-14b 所示，由于三大城郊区镇均表现出了高度相似的聚类结果，因此仅列出区镇 A 的聚类结果。

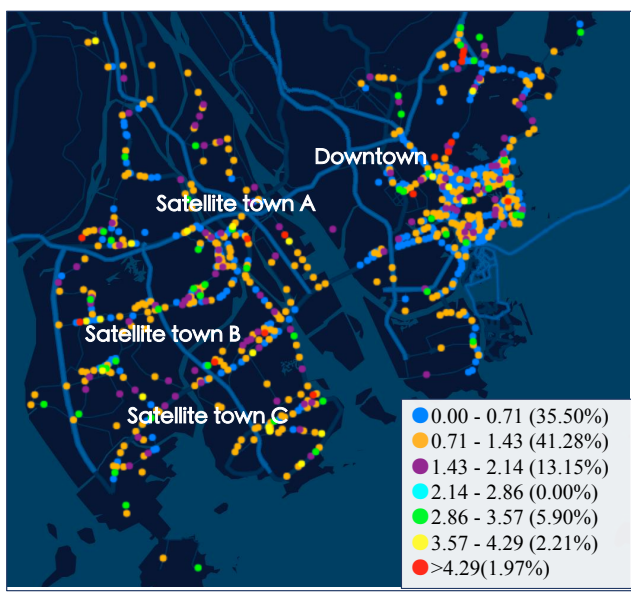

图 5-13 全市公交站点住职比分布统计

Fig. 5-13  Home and work ratio of bus stops.

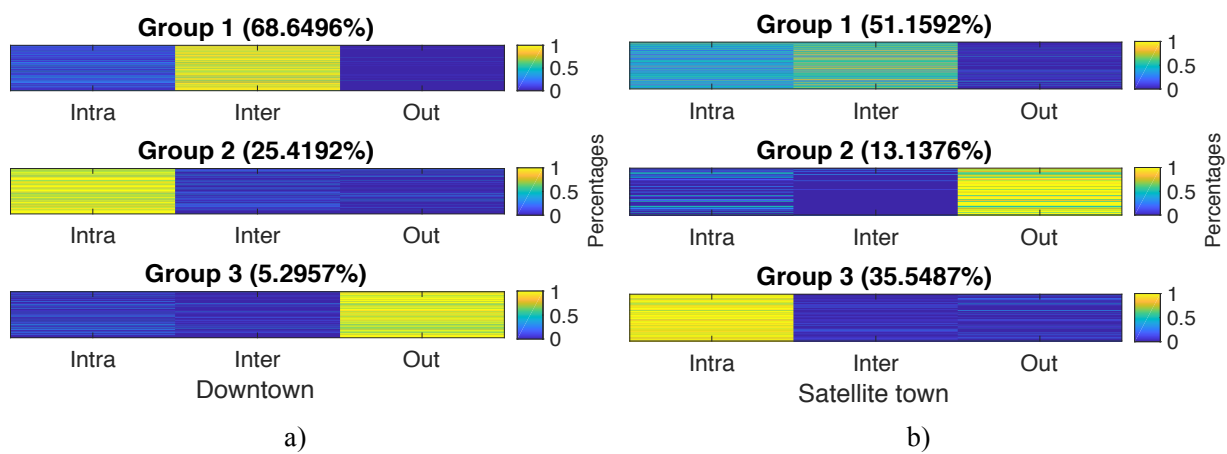

a)  b)

图 5-14 城郊镇区与主城区内具有不同活动特征的乘客分类

Fig. 5-14 Different categories of passengers in downtown and satellite town.

图 5-14 中，Intra 代表起终点均位于乘客居住地所在区镇内的出行；Inter 代表起点或终点有且仅有一个不在该乘客居住地的区镇的出行；Out 代表起点与终点都不在乘客

居住地所在区域的出行。图的颜色轴代表三种出行的占比。由图 5-14，由城郊镇区与主城区间连通性的角度挖掘，在主城区（图 5-14a）可以发现三组特征明显的乘客群体：

**群体 1：** 出行链主要用于连通居住所在镇区的乘客。

**群体 2：** 出行链主要用于连通在居住地所在区镇以及其它区镇的乘客。

**群体 3：** 出行链主要用于居住地以外区镇间连通的乘客，通过进一步统计，该群体多产生多目的地的出行。

与主城区不同的是，在城郊镇区（图 5-14b），出行链主要用于连通居住所在镇区的乘客比主城区同时也产生了更多用于连通在居住地所在区镇以及其它区镇的出行链。

## 5.5 基于出行链与行为分布的城市客流集散通道提取

### 5.5.1 公交乘客行为分布特征分析

为从更抽象的层面上分析乘客的行为分布特征，首先将乘客所有出行链按照使用频次从高到低排序，提取其中前 5 个最频繁使用的出行链的使用百分比构成向量 $\{P_1, P_2, \ldots, P_5\}$，该向量反映了乘客行为在出行链空间上的分布特性，定义为行为分布向量；其次，使用主成分分析法（Principle Component Analysis）将所有乘客的行为分布向量映射到二维空间内（见 5-15a），在该空间内，可以发现出现了 3 个分界非常清晰的乘客簇；在此空间内，直接利用 DBSCAN 算法提取出位于不同簇的乘客及其行为分布向量，聚类结果分别如图 5-15a、图 5-15b 所示。

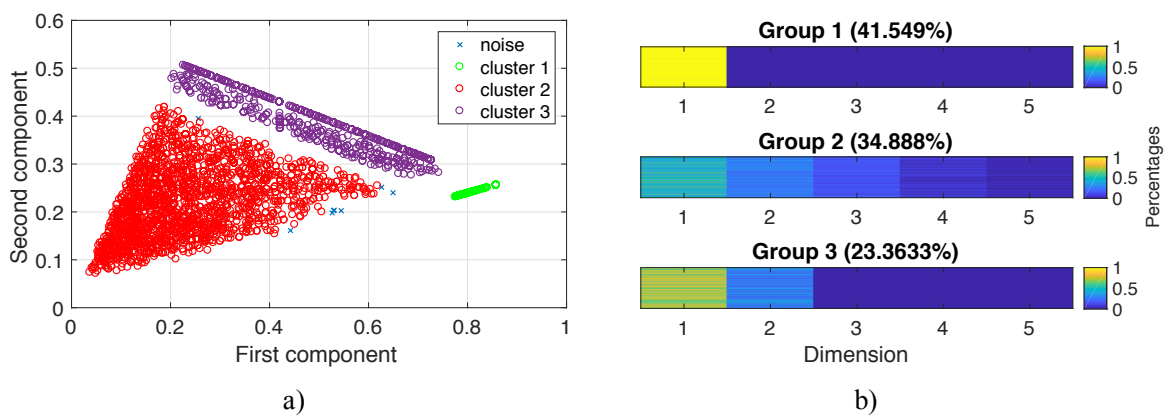

a)          b)

图 5-15 具有不同行为分布特征的乘客群体

Fig. 5-15 Passenger groups with different behavioral patterns.

根据聚类结果，乘客在行为分布特性上可以被分为 3 个群体：

**群体 1：** 行为只集中在一条闭合出行链的乘客（图 5-15a 中绿色聚集最密集且占面积最小的簇），经进一步分析，该群体主要的出行链用来其居住地与职业地；

**群体 2：** 行为分散在多条不同出行链上的乘客（图 5-15a 中红色占面积最大的簇）；

**群体 3：** 出行集中分布在前两条出行链上的乘客（图 5-15a 中紫色簇），且首要出行链使用率远远大于次要出行链。

本研究将组 1、3 总乘客定义为规律出行者，从叠加了 3 个月的数据聚类结果上看，该城市 64%的出行者具有较好的规律性，而聚类结果中第 2 组乘客则可以认为是具有弹性需求的出行者。

### 5.5.2 城市公交客流集散通道提取

首先，给出本研究中城市公交客流集散通道的定义：

**定义 5.9**：城市公交客流集散通道为覆盖全部乘客 85%的出行需求的道路网络。

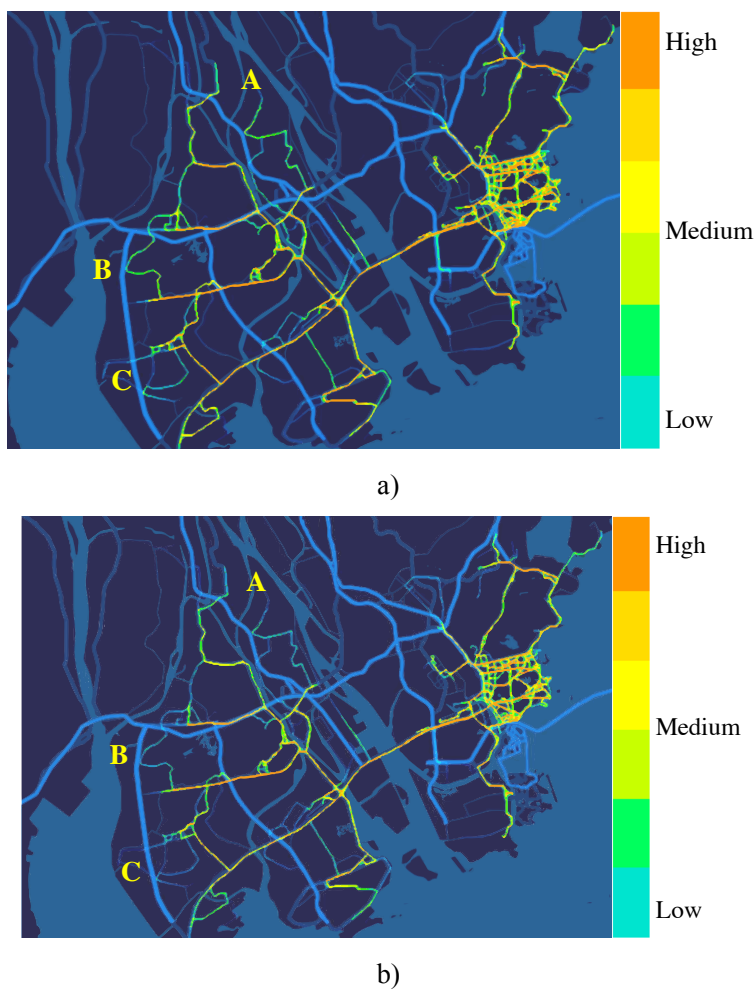

a)

b)

图 5-16 乘客出行链在路网上的映射

Fig. 5-16 Visualization of passengers' trip chain on road network.

其次，从乘客闭合出行链的角度上看，该定义可以等价转化为：覆盖每个乘客 85%出行活动的出行链在路网上的映射即为该城市的公交客流集散通道。根据第 5.5.1 节的研究成果，对每个乘客提取覆盖其 85%出行活动的出行链，并以线热力图的形式展现在图 5-16a 中，其中，热力图的颜色代表了使用该路段的乘客比率，由图 5-16a，在主城

区，被用作集散通道的道路网络分布密集，尤其是东西走向的道路比南北走向的道路具有更高的利用率；另一方面，在城郊各区镇，乘客集散通道主要沿主干道分布。

同时，做出了覆盖乘客使用率最高的出行链在路网上的分布如图 5-16b 所示，图 5-16b 与图 5-16a 并无显著区别，但是，从城郊区镇上观察（图中区域 A、B 和 C），特别是区域 A，部分图 5-16a 中的路网链接在图 5-16b 中并未出现。两图对比可以得到定性的结论：在主城区，大部分乘客最主要的出行链覆盖其 85%以上的公交出行需求；而在城郊区镇中，居民的出行结构更加复杂，居民的首要出行链并不能覆盖其 85%以上的出行需求。

## 5.6 本章小结

本章主要对基于时空轨迹片段的城市公交乘客闭合出行链进行还原以及基于闭合出行链对城市公交乘客的出行特性展开探讨，主要工作内容如下：

（1）提出了一套通过多天数据叠加在轨迹片段中还原公交乘客闭合出行链的方法，尤其是提出了一种新的数据结构：OTD 出行拓扑关系图，另一方面，给出了将乘客不闭合出行轨迹集合与闭合出行链建立关联的方法。

（2）基于全体乘客的闭合出行链集，从不同层面分析了珠海市公交乘客的特性，特别是空间分布特性。

（3）分析了城市公交乘客的行为分布特性，发现在城郊镇区的乘客出行链结构更加复杂。

（4）从公交乘客个体行为分布的角度，提出了另一种城市公交客流集散通道的方法。

# 第六章 乘客线路选择偏好挖掘方法研究

随着公交线网服务质量的提升，尤其是快速公交、大站快车以及定制公交的出现，公交运营部门能够为乘客提供多样化的出行服务。了解乘客出行需求、活动规律的基础上，进一步分析乘客的服务选择偏好，合理设置线路、安排运力配载并对乘客行为进行预测，成为了当前公交相关研究的热点，也是公交行业亟待解决的难点。

第五章的出行链挖掘提供了个体乘客深层次的空间活动模式，但是，该成果并不能直接用于指导公交线网优化，究其原因，缺失乘客线路选择偏好的关键信息，第五章与本章共同讨论的主题是：公交乘客个体出行模式分析及线路选择偏好模型，具体的，在本章一方面讨论了基于闭合出行链的乘客线路选择偏好挖掘方法，借助居民出行 OTD 时空轨迹数据以及出行链数据集，尽可能挖掘每一个可以观测到行为模式的乘客线路选择偏好。另一方面，基于目标城市 3 个月的全样本数据，分析了乘客行为偏好模型，探索乘客的行为特征分布，为下一步公交线网优化提供依据。

## 6.1 基于出行链的乘客个体线路选择影响因素提取

本研究分别针对个体乘客，提取每个出行阶段所对应的方案选择特征及其选择概率，为后续建模提供必要的基础。对于每个乘客的出行链以及相对应的公交出行信息（包含出行时空轨迹与起点、换乘点、终点），数据处理流程如图 6-1 所示：

该方法的工作流程如图 6-1 所示，本研究所采用的数据处理流程大致可分为：出行链出行阶段分解、出行阶段乘车方案提取、乘车方案特征向量构建三个步骤，最终，将该乘客在相同出行阶段的不同出行方案映射到向量空间。每种方案共提取 8 个维度的特征。其中，换乘次数、停站数、乘车里程、行程时间、乘车里程-停站数比值可以从 OTD 时空轨迹记录中获取或直接计算得到，其余关键特征定义如下：

**累计候车时间**：若每一段乘车区间 $l$，乘客所搭乘线路 $r$ 的发车间隔为 $T_l^r$，则该次乘车的候车时间可以简化为：$\frac{T_l^r}{2}$，则方案 $i$ 的累计候车时间定义为：

$$WT_i = \sum_s \frac{T_l^r}{2} \qquad （6-1）$$

其中，$s$ 为所有乘车阶段。

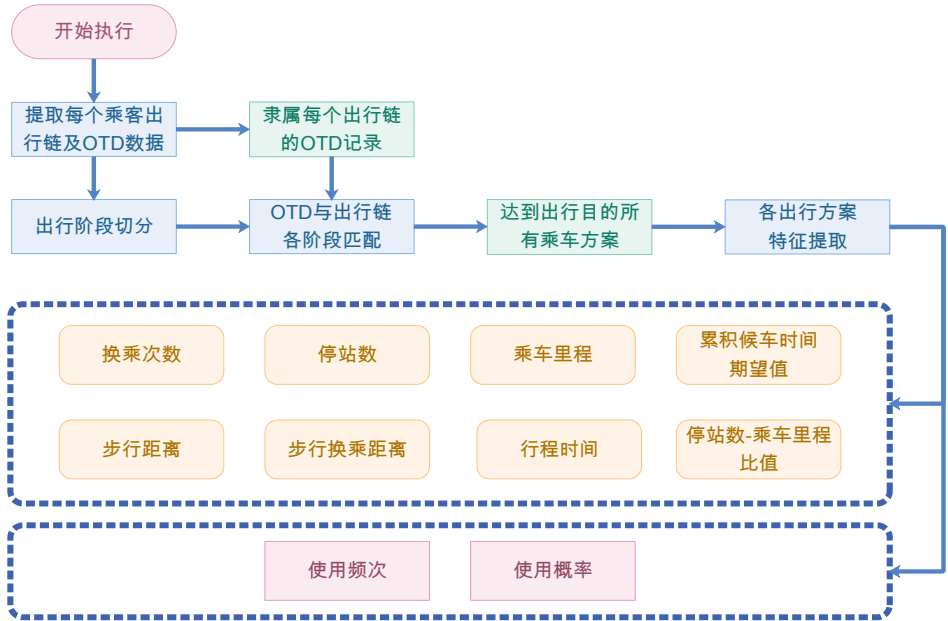

图 6-1 基于出行链的乘客乘车方案特征提取流程

Fig. 6-1 Flowchart of features deriving for individual passenger's route choice preference modeling

**步行换乘距离**：定义每一次换乘，相邻乘车区间为$l,l+1$，若第$l$段出行的下车站点$A_l$与第$l+1$段出行的上车站点$B_{l+1}$不同，则可以定义乘客在其间的步行距离为$DW_t = distance(A_l, B_{l+1})$，则累计步行换乘距离可定义为：

$$WT_i = \sum_t DW_t = \sum_t distance(A_l, B_{l+1}) \tag{6-2}$$

其中，$t$为换乘阶段，由 OTD 时空轨迹还原得到。

**步行距离**：随着线网、站点密度增加，同一出行需求$OD_x$，可以找到多种不同的站点所组成的直达方案，如图 6-2 所示：

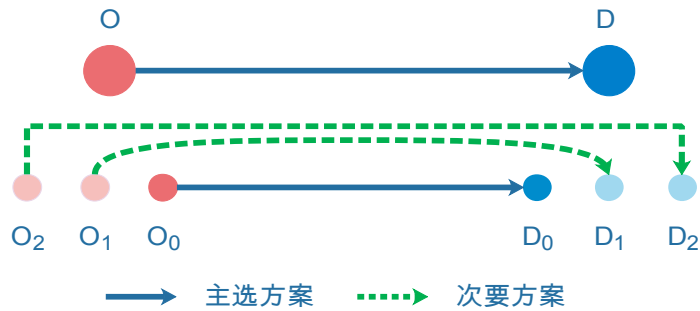

图 6-2 同一出行需求下不同的线路方案示意图

Fig. 6-2 Illustration of various plans for single transit demand.

由图，从个体乘客的角度出发，对于乘客每个闭合出行链中出行阶段$O_0 - D_0$，若存在出行方案集合：

$$PC^x = \{O_1 - D_1, O_2 - D_2, \ldots, O_k - D_k\} \tag{6-3}$$

同时，定义集合中所有元素满足：

$$\forall\, i, j,\ distance(O_i, O_j) \leq \varepsilon$$

$$\forall\, i, j,\ distance(D_i, D_j) \leq \varepsilon \tag{6-4}$$

$$\forall\, k,\ distance(O_k, D_k) \geq \varepsilon$$

其中，$\varepsilon$为乘客步行距离的阈值。对于方案$O_k - D_k$，步行距离$DW_k$可定义为：

$$DW_k = distance(O_0, O_k) + distance(D_0, D_k) \tag{6-5}$$

本研究限定$DW_k \leq 700$m，即乘客从起点到达目的地包括抵达公交车站的合理步行距离不超过 700 m，则求得$\varepsilon = 350$m。 同时规定，以下准则以更准确的判定乘客的步行距离：

**准则 1**：如果乘客的出行起点与居住地间距离小于$\varepsilon$，则：

$$distance(O_0, O_k) = distance(Home, O_k) \tag{6-6}$$

**准则 2**：如果乘客出行终点与居住地间距离小于$\varepsilon$，则：

$$distance(D_0, D_k) = (Home, D_k) \tag{6-7}$$

**准则 3**：如果乘客的出行起点与居住地间距离小于$\varepsilon$，则：

$$distance(O_0, O_k) = distance(Work, O_k) \tag{6-8}$$

**准则 4**：如果乘客出行终点与工作地间距离小于$\varepsilon$，则：

$$distance(D_0, D_k) = distance(Work, D_k) \tag{6-9}$$

该城市在本研究所涉及时间段内并未实施票价优惠政策，因此票价直接与换乘次数相关，无须单独提取。

本步骤最后输出按乘客 IC 卡识别号分组的影响因素特征向量。输出数据逻辑结构示例如图 6-3 所示。

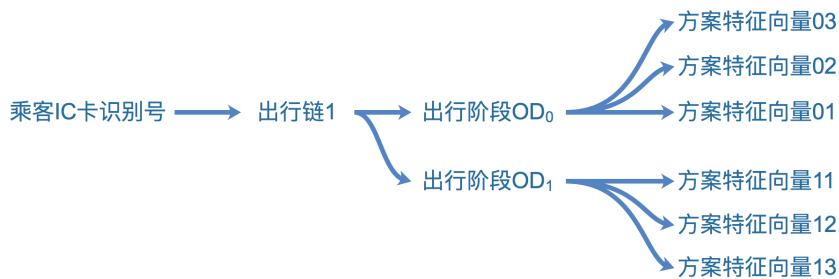

图 6-3 个体乘客乘车方案向量数据结构

Fig. 6-3 Structure diagram of route plan features.

如图 6-3，该例子中，本方法该乘客的只有 1 条出行链，并从该出行链中提取出了两个不同的出行阶段，且每个出行阶段均具有 3 种不同的乘车方案。

## 6.2 基于 NML 模型的乘客线路选偏好建模

### 6.2.1 关键参数与数学模型

假设乘客$p$面对闭合出行链中的出行阶段$OD_x$，有$n$种可选方案，每种选择方案有$k$种不同属性，即每种方案表示为向量$\mathbf{\Theta}^i = (\theta_0^i, \theta_1^i, \ldots, \theta_k^i)$，且各属性互为独立，则乘客对第$i$种方案的选择意愿可用线性效用函数表示为：

$$S_p^i(\mathbf{\Theta}^i) = b_p - \sum_1^k \omega_k^x \cdot \theta_k^i \cdot \eta_k^x \qquad (6\text{-}10)$$

其中，$b_p^x$为常数项；$\theta_k^i$表示第$i$种出行方案的特性，如：时间消耗、等候时间、步行时间等；$\theta_k^i \geq 0$，$\omega_k \geq 0$；$\eta_k^i$为哑元，当数据输入数据中第$k$维不具备拟合条件时$\eta_k^x = 0$，否则$\eta_k^x = 1$，令$\mathbf{\eta}^x = (\eta_1^x, \eta_2^x, \ldots, \eta_k^x)$，则$\mathbf{\eta}^x$为哑元向量，该向量一定程度上反映了公交系统对该出行需求的服务多样化水平。由此，给出场景差异化向量的定义：

**定义 6.1**：对出行阶段$OD_x$，有$n$种可选方案，每种选择方案有$k$种不同属性，则可将$n$种方案的属性表示为一$n$行$k$列特征矩阵$\mathbf{P_X}$，$\mathbf{P_X}$中每一行代表一种方案的属性向量。定义$OD_x$对应的$k$维服务场景差异化向量$\mathbf{\eta}^x$，若$\mathbf{P_X}$中第$m(m \leq k)$列元素取值全部相同，即所有出行方案中该选项无差异，则$\eta_k^x = 0$，否则$\eta_k^x = 1$。

若$\mathbf{P_X}$经过按列均一化，及各种属性的取值范围相同，则可给出本文乘客出行偏好向量的定义：

**定义 6.2**：$\mathbf{P_X}$经过按列均一化，所有列取值范围相同，则式（6-10）中的系数$\omega_k^x$定义为乘客$p$对$OD_x$中每种属性的权重，此时，若令$\mathbf{W_p^x} = (\omega_0^x, \omega_1^x, \ldots, \omega_k^x)$，则$\mathbf{W_p^x}$每一个维的数值表示乘客$p$对该项因素的敏感度系数（还可看作是响应系数）。同时，定义$\mathbf{W_p^x}$是乘客在出行需求$OD_x$上对公交线路选择偏好向量。

根据 NML (Multiple Nominal Logit)离散行为选择模型[97-98]，乘客对满足需求出行需求$OD_x$的每一种出行方案$i$的选择概率$P_i^x$还可以进一步表示为：

$$P_i^x = \frac{\exp S_p^i(\mathbf{\Theta}^i)}{\sum_{i=1}^n \exp S_p^i(\mathbf{\Theta}^i)} \qquad (6\text{-}11)$$

根据式（6-11），每一名乘客出行偏好建模过程转换为根据各方案的选择概率分布，寻找$\omega_k^x$的过程。本研究使用集成多种选择模型的 Biogeme 软件包对个体乘客以及目标城市交通小区内乘客的宏观线路选择偏好特性进行求解。对于每一条出行链中的出行阶段乘车$OD_x$，处理方法如下：

从公交线路表以及乘客出行链、出行 OTD 记录，读入$OD_x$及每种方案（包含乘客

没有使用过的方案）的特征向量，构成特征矩阵：

$$\begin{aligned} \boldsymbol{P_X} &= [\boldsymbol{\Theta^1} \quad \boldsymbol{\Theta^2} \quad ... \quad \boldsymbol{\Theta^n}]^T \\ &= [\boldsymbol{C^1} \quad \boldsymbol{C^2} \quad ... \quad \boldsymbol{C^k}] \end{aligned} \quad (6\text{-}12)$$

检查$\boldsymbol{P_X}$每一列，计算第$m$列所对应哑元变量$\eta_m$的取值：

$$\eta_m^x = \begin{cases} 1, & rank(\boldsymbol{C^m}) > 1; \\ 0, & otherwise. \end{cases} \quad (6\text{-}13)$$

若$\eta_m^x = 0$，即第$m$列特征不满足拟合条件，模型不考虑该特征。

用式（6-14）将所有待考虑的特征进行列内均一化处理，将数值缩放到[0,1]区间内：

$$\boldsymbol{C^{m\prime}} = \frac{\boldsymbol{C^m} - min(\boldsymbol{C^m})}{max(\boldsymbol{C^m}) - min(\boldsymbol{C^m})} \quad (6\text{-}14)$$

此时，$\boldsymbol{P_X}$转化为$\boldsymbol{P_X}' = [\boldsymbol{\Theta^{1\prime}} \quad \boldsymbol{\Theta^{2\prime}} \quad ... \quad \boldsymbol{\Theta^{n\prime}}]^T$，同理，$\boldsymbol{\Theta^i}$转化为$\boldsymbol{\Theta^{i\prime}} = (\theta_0^{i\prime}, \theta_1^{i\prime}, ..., \theta_k^{i\prime})$。

调用 Biogeme 软件包对所有待考虑特征对该方案的选择概率进行拟合，得到出行需求记录$OD_x$所对应的乘客出行线路选择特征偏好模型：

$$\begin{cases} P_i^x(\boldsymbol{\Theta^{n\prime}}|\boldsymbol{\eta}, OD_x) = \dfrac{\exp S_p^i(\boldsymbol{\Theta^{n\prime}})}{\sum_{i=1}^n \exp S_p^i(\boldsymbol{\Theta^{n\prime}})} \\ S_p^i(\boldsymbol{\Theta^{n\prime}}) = b_p^x - \sum_1^k \omega_k^x \cdot \theta_k^{'} \cdot \eta_k \end{cases} \quad (6\text{-}15)$$

由式（6-15）提取出模型系数即可得到对于$OD_x$以及服务场景差异化向量$\boldsymbol{\eta^x}$影响下的乘客线路选择敏感度向量$\boldsymbol{W_p^x}$，同时记下该偏好对应的客流量与服务多样性向量$\boldsymbol{\eta^x}$。

进一步，对个体乘客而言，该乘客完成所有出行阶段及其服务场景下的行为选择偏好拟合后，将相同场景但不同出行需求（$OD_x$）下的行为偏好向量按使用频次（客流量）加权平均，最终得到该乘客在不同服务场景$\boldsymbol{\eta}$影响下的线路选择偏好模型：

$$P_i(\boldsymbol{\Theta^i}|\boldsymbol{\eta}) = \frac{\exp S_p^i(\boldsymbol{\Theta^i})}{\sum_{i=1}^n \exp S_p^i(\boldsymbol{\Theta^i})} \quad (6\text{-}16)$$

最后，将相同场景下的乘客偏好模型中的敏感度系数加权平均，得到该场景的宏观乘客行为偏好模型。乘客线路选择偏好模型是后续线路优化时公交客流重分配的基础。

### 6.2.2 客流重分配推演算例

为验证本文提出的乘客线路选择偏好模型，本研究提取目标城市 504 路公交车沿线各站点的 OD 矩阵以及每个乘客的选择偏好模型，并将 504 路作为一条新线加入原有站点集合中，分别利用基于广义时间效用模型的客流分配方法与本文基于个体乘客偏好的分配方法进行客流重新分配计算，最后将两种方法重分配后得到上下车客流分布矩阵的

与原始上下车客流分布矩阵对比以验证本方法的有效性。测试方法的具体流程为：

**Step 1** 提取 504 路公交沿线所有站点保存至列表$S_{504}$，并按先后顺序对站点编号。

**Step 2** 根据乘客乘车 OTD 记录计算出 504 路公交的乘客上下车客流分布矩阵$M_{504}$。

**Step 3** 根据乘客乘车 OTD 记录计算沿线站点$S_{504}$的上下车客流分布矩阵$M_{S_{504}}$（含非 504 路公交车的客流），即网络客流分布矩阵。

**Step 4** 假定 504 路公交为穿过$S_{504}$的新线，给定站点及发车班次，由已知客流分布$M_{S_{504}}$，分别利用广义时间效用模型与本文所提出乘客个体偏好模型推演计算新的 504 路车上下客流分布矩阵$M'_{504}$，并将推演得到的客流分布 OD 矩阵与原始矩阵比较。

步骤 3 所述广义时间效用模型简述如下：该模型首先对满足可达性的$k$种乘车方案，分别计算出第$i$种含有$s$阶段乘车方案的特性转化为广义时间：

$$T_g^i = \sum_s T_{journey}^i + \sum_s T_{transfer}^i + \sum_s T_{wait}^i \tag{6-17}$$

$T_{transfer}^i$为各换乘阶段步行时间，计算时假定乘客步行速度为 1.2 m/s。

$T_{journey}^i$为行程时间，估算时将常规公交速度设置为 20 km/h，快速公交速度设置为 30 km/h，每个车站的停站时间设置为 30 s。

$T_{wait}^i$为乘客所有阶段的候车时间，估算时取发车班次间隔的一半。

其次，为保证数值稳定，将所有$T_g^i$按式（6-14）进行归一化到[0,1]区间，得到$T_{g0}^i$。

最后，按式（6-18）计算每种乘车方案的被选择概率：

$$P_g^i = \frac{exp(-T_{g0}^i)}{\sum_k exp(-T_{g0}^i)} \tag{6-18}$$

经过客流重分配后的客流分布矩阵进行可视化后得到结果如图 6-4 所示，图中，$x$，$y$ 坐标轴对应的数值为站点编号，矩阵中网格的颜色代表客流量。由图观察可知，三种客流重分配数据推演方法都能大致还原客流的空间分布。由广义时间效用函数推演的结果（图 6-4b，$R^2 = 0.716$）与真实值（6-4a）相比较，部分站点广义时间效用函数推算出的客流量明显偏大，表现为图中出现超过数值超过 35 人次的亮斑；而本文所提供的乘客个体偏好模型推演得到的结果（图 6-4c，$R^2 = 0.823$）则更接近真实值。

对结果进一步分析，可观察到本文所提供方法推演得到的客流矩阵在部分站点客流仍然偏大，原因是部分乘客的行为偏好未能拟合，这部分乘客选择乘车方案时并不遵循效用最大原则。本试验进一步说明了本文提供方法能够为车辆班次调整以及新线规划中提供有益指导。

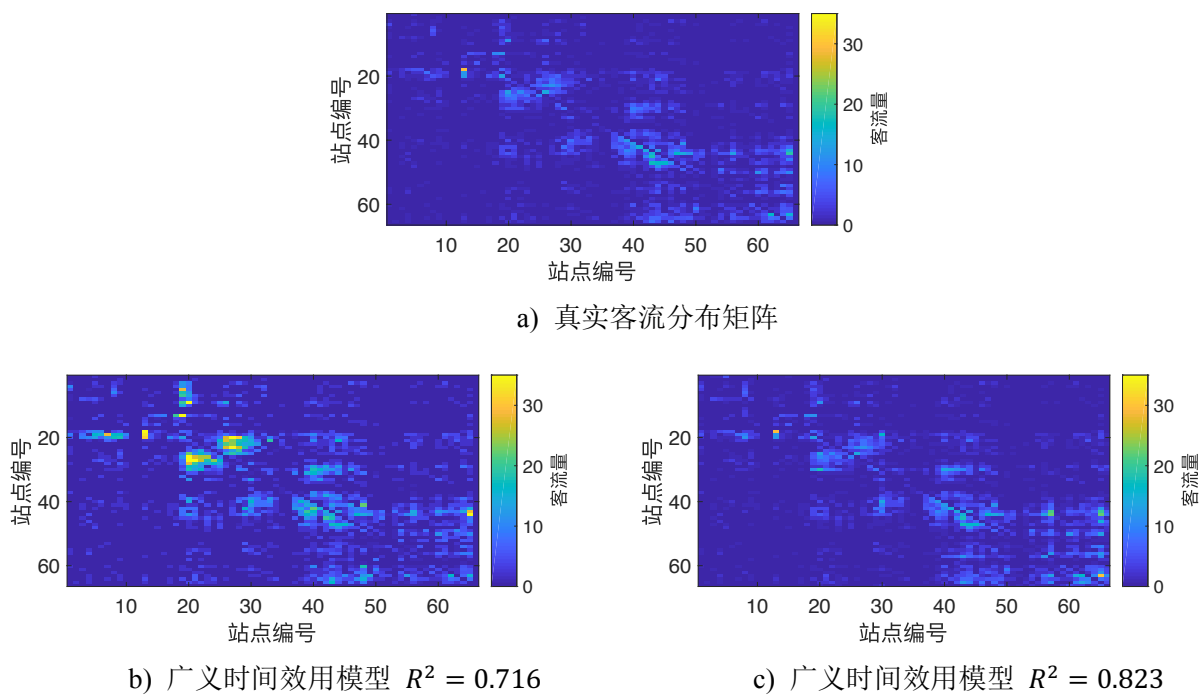

a) 真实客流分布矩阵

b) 广义时间效用模型 $R^2 = 0.716$

c) 广义时间效用模型 $R^2 = 0.823$

图 6-4 不同客流分配模型得出的上下车客流矩阵对比（珠海 504 路）

Fig. 6-4 Compares of ridership distribution matrix from different distribute models

## 6.3 乘客线路选择偏好分析

### 6.3.1 差异化服务场景下乘客的行为偏好

对该城市三个月内刷卡次数超过 30 次的乘客的出行行为进行模型拟合，提取决定系数$R^2 \geq 0.5$的模型进行分析；从服务差异化向量$\boldsymbol{\eta}$及其所对应客流量占比上，将所有情形进行数据可视化，结果如图 6-5 所示。

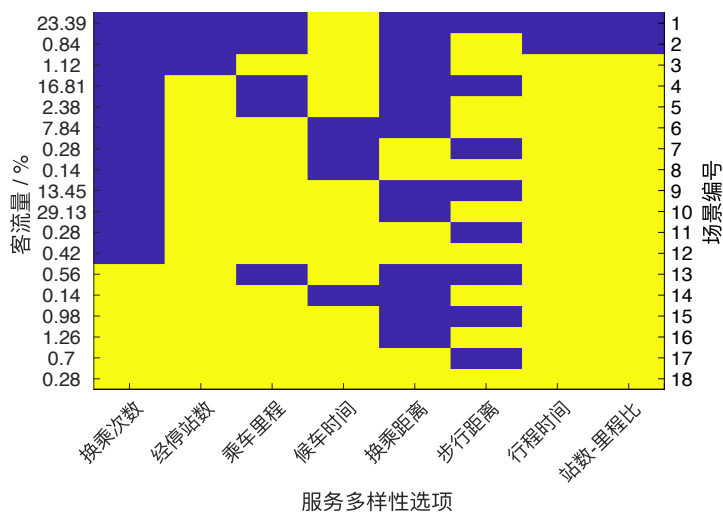

图 6-5 线网服务多样化选项及客流比重

Fig. 6-5 Distribution of ridership on transit service with various options

由图 6-5，该市公交线网提供多达 18 种不同的服务差异化场景向量（矩阵中蓝色代表该选项无差异，黄色代表该选项有差异），统计范畴内 23.39%乘客出行发生在仅有候车时间差别的区间内（场景 1）；所有选项均呈现多样化的场景编号为 18，但其仅仅占全部乘客的 0.26%。将同场景不同乘客的敏感度系数求平均值，即可得到每种场景对应的乘客的出行偏好均值向量，并结合客流量以数据可视化的形式展现如图 6-5，图中颜色轴代表敏感程度系数值。

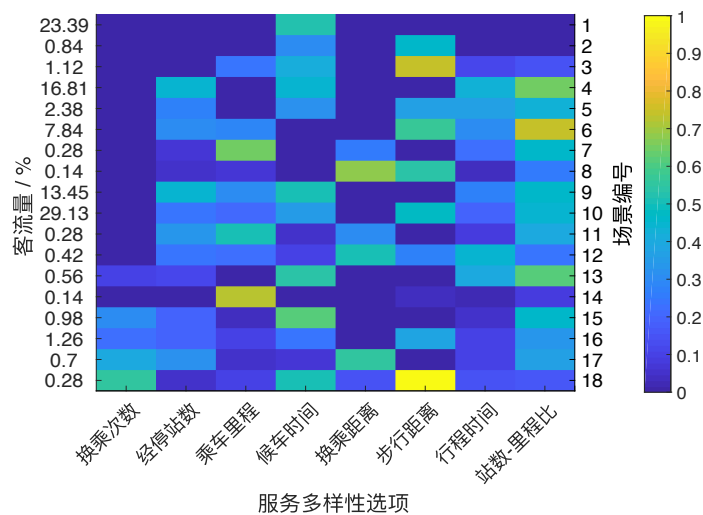

图 6-6 不同场景下乘客线路选择行为偏好权重

Fig. 6-6    Distribution of ridership on transit service with various options

由图 6-6，乘客在不同服务场景下线路选择偏好具有明显差异，即，不同的服务选项共同作用下，乘客会表现出不同的行为特性，以下按服务特性分类讨论：

（1）乘客对换乘次数最敏感的场景为 18，该场景下所有服务选项均有差异，但该场景下，乘客敏感度系数最大值位于步行距离偏好条件下，结合场景 13、15~17，表明相对于换乘次数而言，其它因素具有更高的影响权重。

（2）对停站数表现较敏感的场景为场景 4、9，但在场景 4，站数-里程比敏感度远高于停站次数；在场景 9，站数、站数-里程比、候车时间三者敏感度接近，而乘客对候车时间更敏感，以上现象说明停站数并不是乘客选择线路考虑的主要因素。

（3）对乘车里程表现敏感的场景为 14、其余场景内，乘车里程均不是主要敏感项。

（4）候车时间与乘客步行距离在多达 5 种场景下为最敏感项，说明其在宏观上对乘客选择线路具有非常明显的影响。

（5）行程时间是唯一一个在所有场景均不成为极大敏感度的因子，说明宏观上该市乘客在线路选择上并不会把行程时间作为最主要的考虑因素。

（6）最后，站数-里程比在 3 个场景（4、6、13）内成为了高敏感度指标，该城市

有较大比率乘客偏好经停站少的快速公交线路。

（7）根据以上观察，将敏感度取得唯一极大值（首要极大值超过次要极值的绝对值20%）的场景摘出，并统计其客流占比，如表 6-1 所示：

表 6-1 数据源关键字段

Table 6-1 Principle information in the related data sources

| 极大权重项 | 极大值场景 | 客流量 / % |
|---|---|---|
| 换乘次数 | 无 | 0 |
| 经停站数 | 无 | 0 |
| 乘车里程 | 7、11、14 | 0.65 |
| 候车时间 | 1、9、15 | 39.29 |
| 换乘距离 | 8、17、12 | 1.32 |
| 步行距离 | 2、3、10、16、18 | 32.27 |
| 行程时间 | 无 | 0 |
| 站数-里程比 | 4、6、13 | 24.08 |
| 合计 / % | | 97.58 |

由表 6-1，超过 90%的乘客的线路选择偏好中存在主导因素（极大敏感项），因此，可以根据极大值将乘客分类。另一方面，为对乘客出行特征做进一步分析，按图 6-6，将众多服务选项所对应的场景归并为 4 大类，如图 6-6 所示。

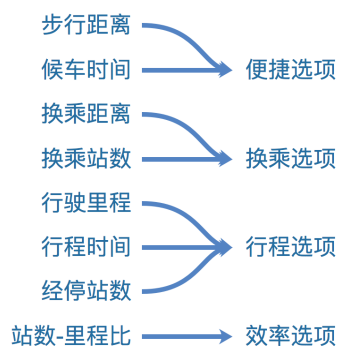

图 6-7 服务差异化选项归并分组

Fig. 6-7 Integration of service options

由图 6-7，将 8 种线网提供的差异化服务选项归纳便捷选项、换乘选项、行程选项、效率选项。其中，便捷选项用以描述乘客到达公交站台的步行距离，以及候车时间；换乘选项用于描述乘客从起点到终点的换乘站数以及累计换乘距离；行程选项描述乘客乘车过程；效率选项表示乘客对出行效率的关注程度。

## 6.3.2 乘客线路选择偏好的时变规律

分别做出工作日与周末 24h 内对 8 种服务选项偏好存在唯一极大值的客流比重，并按图 6-7 将数据进行分组，得到结果如图 6-8 与图 6-9。总结出以下规律：

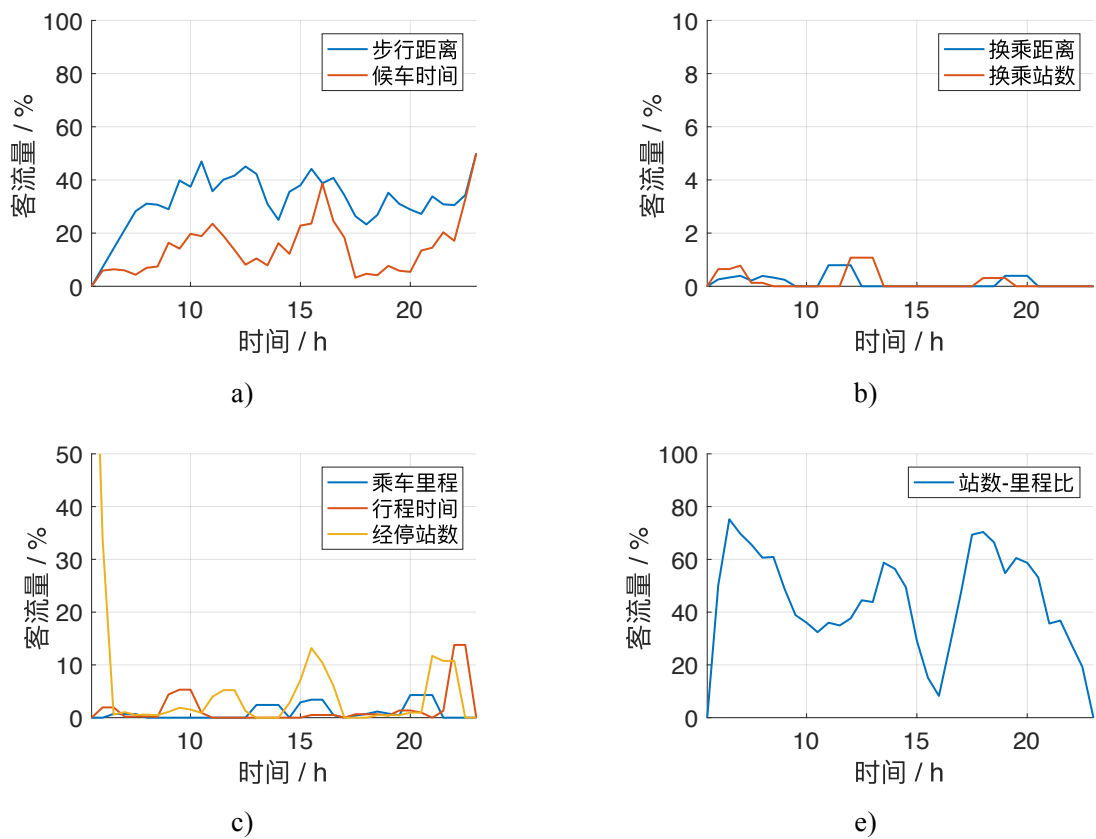

图 6-8 工作日具有不同线路选择偏好乘客的出行时变规律

Fig. 6-8 Proportional variation of ridership with different route choice preferences on weekdays.

由图 6-8a，在工作日，对步行距离具有高敏感度的乘客在 24 h 内保持较稳定的比率，该群体客流量占比保持在 20%到 50%的区间内，同时可以观察到在 22:00 之后，对步行距离敏感的乘客比率明显提高；对候车时间具有最强敏感性的乘客具有三个明显的活动高峰：9:30 至 12:00，14:30 至 17:00，22:00 至次日零点。在 22:00 之后，对候车时间敏感的乘客比率达到极大值，通过查询调度数据，此时，部分班线已处于末班车阶段，乘客对其它因素（特别是效率选项）的敏感度下降。总体而言，乘客对到达车站以的步行距离的敏感度高于对候车时间。

由图 6-8a，在周末，对步行距离与候车时间表现敏感的乘客比率高于普通工作日；相比而言，对步行距离表现最敏感的乘客比率高于对候车时间表现敏感的乘客。

由图 6-8b 与图 6-9b，在工作日与周末，对换乘次数、换乘距离表现敏感的乘客在全天客流量中占极少比率，这是由于该城市公交线网提供的乘客直达率高达 97%，换乘

客流极少，仅占全部客流的 2%。

由图 6-8c，从行程选项上看，工作日凌晨 5:00 至 6:00 出行的乘客主要考虑因素为行程选项，只选择经停站少的线路，导致该现象的原因是此时公交线网只有部分夜间线路运行，该类型的线路相比其它线路，少停站数少，具有区间车的特点。其余时段内，该类型乘客仅在 15:00 至 16:30、21:00 至 22:00 时段内出现比重高于 10%。

由图 6-9c，在周末，将行程选项作为极大敏感项的乘客比率高于普通工作日;；将经停站数作为极大敏感项的乘客仍高于对行程时间敏感的乘客。

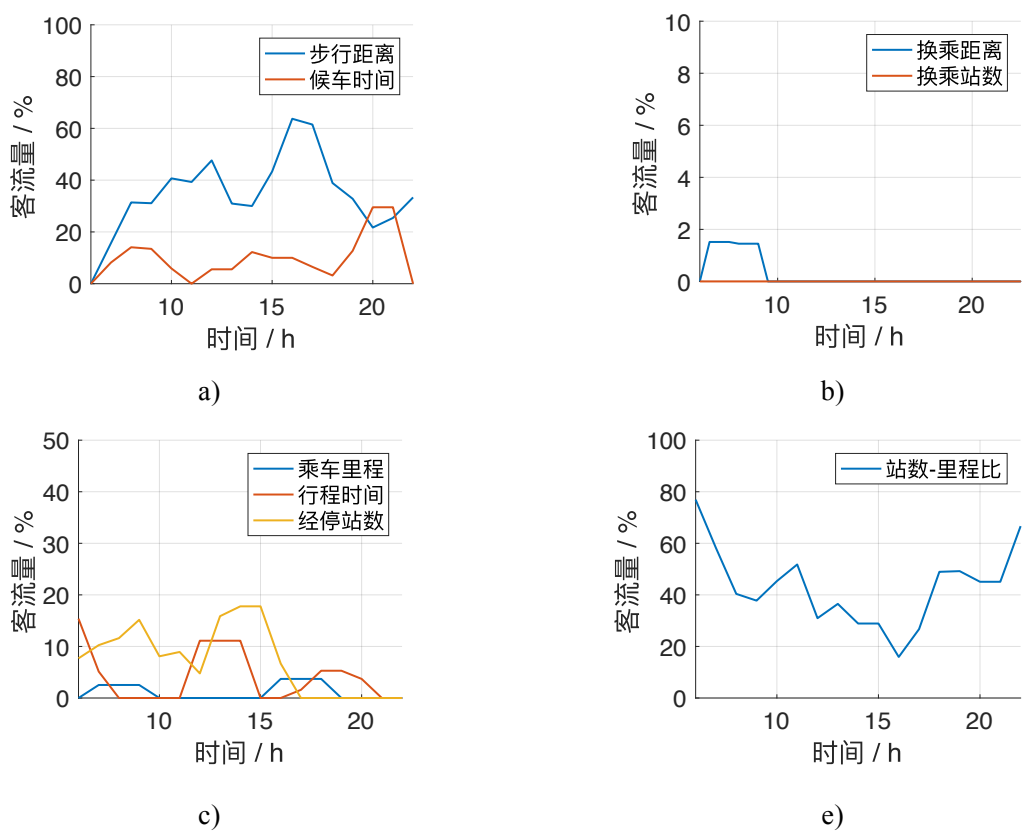

图 6-9 周末线路选择偏好具有唯一极大值的客流时变规律

Fig. 6-9 Proportional variation of ridership with different route choice preferences on weekends.

由图 6-8d，从站数-里程比上看，工作日 6:00 至 9:00 时段内，对该项表现敏感的乘客占比达到全天极大值（75.19%），同时在 12:30 至 14:30 时段内，也呈现峰值，最后，在 17:00 至 21:00 时段内，出现了接近全天极大值的峰值（73.59%）。该现象表明，出行效率为多数乘客，尤其是高峰期出行乘客线路选择的重要参考项。

由图 6-9d，同样，在周末将该选项视为极大敏感项的乘客占总体客流的较大比重与工作日的规律相似；同样在下午 16 时，对站数-里程比表现最敏感的乘客比率出现最低值。与工作日区别最大的现象在于，该部分乘客并未出现在晚上 21 时。

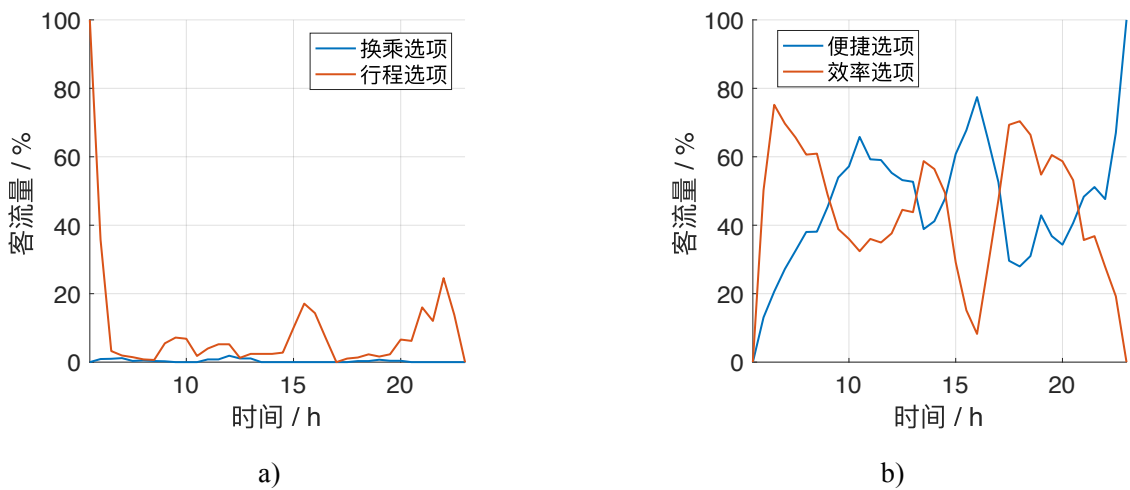

图 6-10 工作日归并线路选择偏好的客流时变规律

Fig. 6-9 Proportional variation of ridership with integrated route choice preferences on weekdays

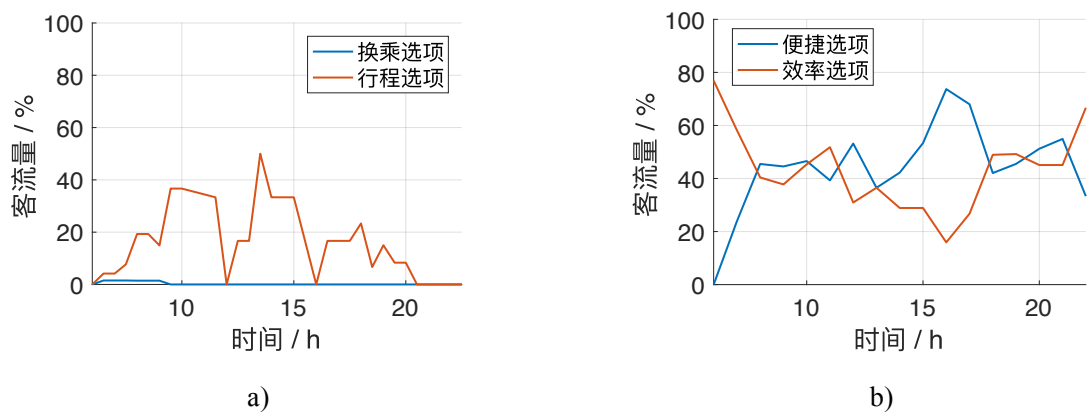

图 6-11 周末归并线路选择偏好的客流时变规律

Fig. 6-10 Proportional variation of ridership with integrated route choice preferences on weekends.

同时，分别给出将四大综合选项作为极大敏感项的乘客比率时变规律如图 6-10 和 6-11，由图可得到以下结论：

首先，由图 6-10a，对行程以及换乘选项最敏感的乘客比率在工作日全天内均处于较低的水平，相比而言，对行程选项具有最高敏感性的乘客比率高于对换乘选项敏感的乘客比率；由图 6-11a，同样的乘客群体在全天的客流占用率远高于工作日。

其次，由图 6-10b，对效率选项敏感的乘客在工作日出行具有明显的早、午、晚高峰特征，尤其是在 7:00 开始急剧上升，但在 21:00 之后，该部分客流量急剧下降；由图 6-11b，在周末，将效率选项作为极大敏感项的乘客在客流中的比率低于该群体在工作日中的比率，

再次，在工作日与周末将便捷选项作为极大权重项的乘客比率与将效率作为极大敏感项的乘客比率出现了此消彼长的现象；在工作日，此现象开始于上午 9:45，而在周末则没有明显的开始时刻。此结果表明将便捷选项作为最高权重项的乘客对行程效率关注

度并不高,同时,对便捷选项具有高关注度的乘客出行高峰位于 10:00 至 12:00 以及 14:30 至 16:30 之间。与对效率关注度高的乘客相比,该类型乘客活动高峰位于下午 14:30 开始的两小时内。

## 6.4 本章小结

本章基于乘客出行链数据以及 OTD 时空轨迹数据,挖掘乘客的公交出行线路选择偏好并建立数学模型,并分析了具有不同特征的乘客群体的出行时变规律。具体工作内容如下:

(1)借助 NML 模型以及珠海市 3 个月的全样本客流数据,提取了 8 大影响出行线路选择的特征,拟合了乘客的出行偏好模型,尤其是发现了公交系统中乘客在不同服务差异化场景下会表现出完全不一致的出行线路选择偏好。

(2)将乘客按线路选择敏感因素分类,并分析了不同种类的乘客在工作日与周末不同时段在全市客流量的占比。

(3)本章提出的乘客出行线路选择偏好模型为后续线网优化和从乘客个体的角度模拟客流充分配以及影响评估提供了必要基础。

# 第七章 基于乘客选择偏好的公交线路优化方法研究

公交线网是目前我国各大城市提供可达性服务的主要手段，很长一段时间内，政府以公共投入保证公交运营机构的合理盈利。近年来，在"政府购买公共服务"的政策引导下，公交线网的服务效益已经逐渐成为地方政府以及公交系统运营服务单位关注的焦点，根据乘客出行分布以及线路选择偏好合理的规划、设计公交线成为了现阶段的研究热点。

由第二章的文献综述，现阶段的公交线路乃至线网优化方法均只是利用客流时空分布规律，而未考虑乘客的出行线路选择偏好，存在一定的改进空间。究其原因，一方面，当前通用的公交系统优化方法往往采用满足可达性且客流量最大化的方法对公交线路站点位置以及发车班次进行优化，通常不考虑乘客的出行线路选择偏好。难以真正设计出符合实际需要的公交线路，同时该做法在本文所讨论的目标城市会出现大量停站数过多的站点，导致乘客出行时间效率偏低。另一方面，传统方法未提供从海量运营数据中挖掘乘客出行线路选择偏好的方法，而无法根据乘客的线路选择偏好对公交线路进行调整。

**本章，我们将第三至六章数据处理和乘客行为模式挖掘两大主题针对个体乘客行为模式数据挖掘的成果应用于线网优化中**，具体的，我们对目标城市公交线网服务质量进行深入分析的基础上提出了一套基于逐乘客出行线路选择偏好推演和粒子群优化算法的公交线路优化设计方法，具体内容如下：

首先，我们从乘客出行效率以及服务里程转化率、服务效益三个方面提出了城市公交系统的服务质量评价指标，同时，分析了目标城市公交系统的服务质量的变化规律与关联关系。

其次，提出了一种基于乘客个体出行偏好的客流重分配模型，以预测线路经停站点以及发车班次发生改变后的客流。

再次，基于客流数据推演与粒子群优化算法，提出了利用原有线路基础设施设计新的线的方法以及合理的目标函数和约束条件。

最后，评估了新线开通后对旧线路的影响以及新旧线路乘客出行效率变化，客流数据推演试验表明，所选方案提供的新线能有效提高沿线居民的出行效率，提升公共投入至服务里程（全体乘客的出行里程累加）的转化率。

## 7.1 公交线路服务质量评价指标与优化目标

首先，对线网中任一条公交线路$r$定义其运营成本为：

$$E_r = E_s + E_d \tag{7-1}$$

其中，$E_s$为固定成本（5 元/公里），$E_d$为线路的动态成本(1.2 元/分钟)。

其次，定义线路$r$的服务里程转化率如下：

$$R_s = \frac{m_r}{E_r} \tag{7-2}$$

其中，$m_r$表示该线路承载的服务里程，即所有乘客在该线路上的出行里程数之和，单位为：公里。$R_s$表示服务里程与运营成本之间的转化率，单位为：公里/元。

再次，对于该线路上每个乘客$p$，假设其乘车轨迹为：A→B，在此区间内，定义其的出行成本为：

$$E_p = T_w + T_{bus} + T_s \tag{7-3}$$

其中：$T_w$为乘客的候车时间；$T_{bus}$为乘客的行程时间；$T_s$为途径各站点的累计停留时间。则可以定义乘客$p$在此区间内的时间利用效率（简称时间效率）为：

$$r_T(AB) = \frac{T_{bus}}{E_p} = \frac{T_j}{T_w + T_{bus} + T_s} \tag{7-4}$$

式（7-4）中，$T_s$可以从车辆历史轨迹数据中获取，在新线规划时可以近似估算。

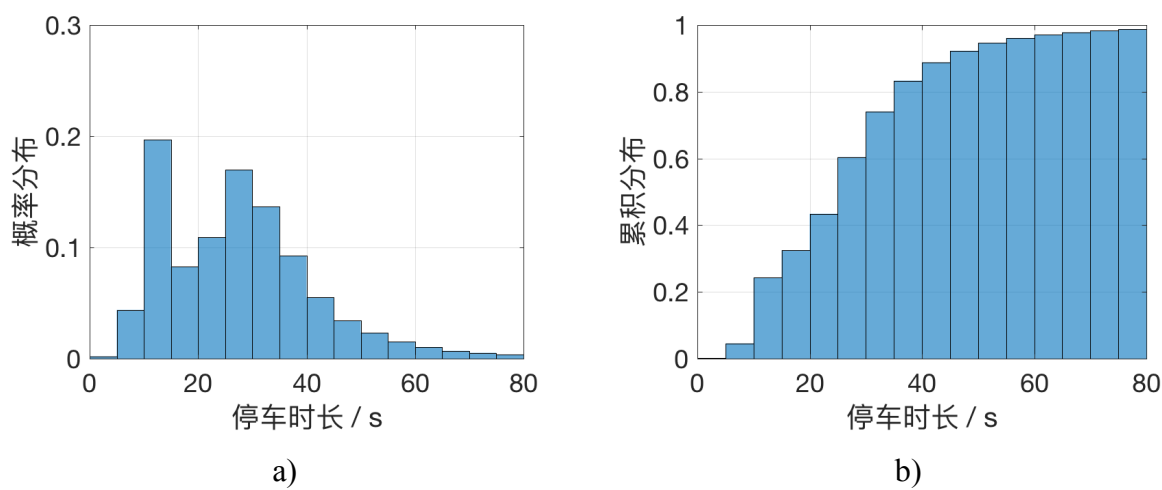

a)            b)

图 7-1 车辆在普通车站的停站时间分布

Fig. 7-1 Distribution on vehicles' length of stay in common stops

本文统计了目标城市 3 个月内全市公交车辆在除总站以外其他站点的驻站时刻得到车辆停站时长的概率分布与累计分布如图 7-1a 和图 7-1b 所示，由图可以观察到，公交车辆停站时间分布最集中的数值区间是 25 ~35 s 之间，尤其是在 15 s 附近存在峰值，

经过现场调查，引起该现象的原因是乘客先上车后刷卡付费。同时，从观察停站时间的累积分布，可以观察到取驻站时间 45 s 即可覆盖超过 85%的情形。

对于公交线路$r$而言，该线路在研究时段内产生的乘客上下车序列组成的集合为$C_{OD} = \{od_0, od_1, ..., od_N\}$，则在研究时段内，乘客在该线路上的累计出行时间效率为：

$$R_T = \frac{\sum_{k=1}^{N} T_{bus}^k}{\sum_{k=1}^{N} E_p^k} \quad (7\text{-}5)$$

还可定义线路的有效服务里程为：

$$M_r = m_r \cdot R_T \quad (7\text{-}6)$$

最后，本研究定义公交线路$r$的综合服务效益为：

$$P_s = R_T \cdot R_s = \frac{M_r}{E_r} \quad (7\text{-}7)$$

式（7-7）表达的物理意义为线路服务里程转化率与乘客平均出行效率的乘积。

对于已有的公交线路$r$及其沿线站点集合$\{S_r\}$而言，本文所讨论公交线路优化思路为：与原有线路始发站、终点站保持一致，利用部分旧线路的站点，保持既有线路不变同时新加入一条停站较少的快速公交线路$r_n$，用旧线保证沿线交通小区的可达性，用新线提供较高的出行效率以满足乘客多样化的服务需求。新加入公交快线的设计问题可以转化为受约束的整数型优化问题，且目标函数为：

$$\max F(r_n)$$
$$\text{s.t.} \; T_n, S_r, OD_{sr}, P_{sr}, MP_{sr} \quad (7\text{-}8)$$

其中，$T_n$为新加入线路可选的发车班次间隔；$S_r$为可选站点列表；$OD_{sr}$为乘客在站点集合$\{S_r\}$中形成的 OD 矩阵；$P_{sr}$为$OD_{sr}$中每个 OD 对上的乘客标识；$MP_{sr}$为$P_{sr}$中每个乘客的线路选择偏好模型。$F(r_n)$为目标函数，不同的目标函数将导致不同的线路设计以及发车班次方案，**本文分别讨论服务效益、服务里程转化率、行程时间效率、服务里程 4 种优化场景对应的线网优化结果**。

## 7.2 目标城市公交线网服务指标现状分析

本节借助目标城市三个月的公交乘客 OTD 轨迹数据以及车辆 GPS 报站数据，挖掘线网的服务指标，并分析其分布规律，旨在为公交线网优化提供前期理论依据。

本文利用该城市的公交乘客 OTD 轨迹数据以及车辆 GPS 报站数据分别对工作日与双休日全市每一条公交线路的服务效益、出行效率、服务里程转化率进行分析，得到这些指标的概率分布图如图 7-2。同时，将计算各指标分的时段平均值总结于表 7-1。可得

到以下结论：

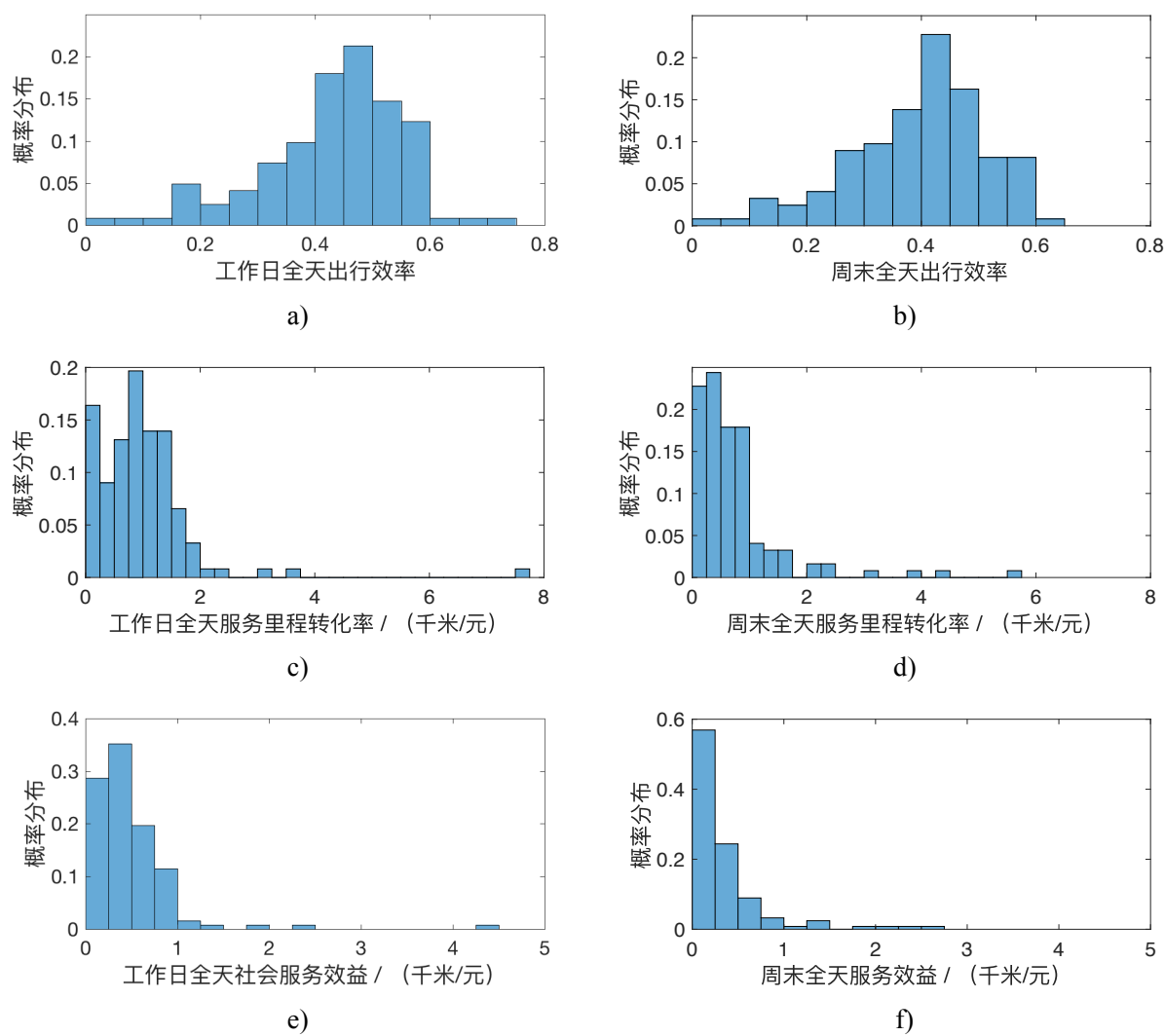

图 7-2 珠海市公交线网综合服务指标分布

Fig. 7-2　Numerical distribution of integral social service indicators of Zhuhai

由图 7-2a，图 7-2b 可观察到出行时间效率在工作日与双休日的数值有显著区别，工作日居民出行效率数值集中分布在 0.4~0.6 区间内，而双休日出行效率则集中分布在 0.35~0.5 区间内，同时由表 7-1 可观察到工作日的出行时间效率略高于双休日出行时间效率。

由图 7-2c，图 7-2d，从服务里程数与运营成本转化率概率分布上看，可以观察到该线网工作日与双休日有明显区别。在工作日，数值分布具有明显的"两极分化"现象，即转化率低于 0.5 公里/元的线路与转化率介于 0.75~1.5 公里/元的线路均在概率分布上出现峰值；但服务里程数-运营成本转化率高于 1 公里/元的线路占比极少；但在周末，峰值位于[0.1,0.5]区间内。根据表 7-1，周末各时段的服务里程数-运营成本转化率均低于普通工作日。

由图 7-2c，图 7-2d，由图 7-2e，图 7-2f，在工作日，服务效益数值分布集中在[0.5,2]区间内，同时，双休日则收缩到[0.5,1]区间内，结合表 7-1，工作日全市公交线网的服务效益仍明显高于双休日。

最后，结合图 7-2e，图 7-2c 以及图 7-2f，图 7-2d，可观察到考虑停站与候车引起的时间效率损失后，公交线网服务效益远低于服务里程-运营成本转化率。引起该现象的一个主要原因是由于珠海市在双休日时段长距离出行的客流少于工作日，这部分减少的客流量直接导致大站快车等效率较高的线路客流量降低。

由表 7-1，可进一步观察到居民的出行时间效率在早晚高峰以及其它时段没有显著差别，同时在工作日与双休日两种情形下，服务里程数转化率以及服务效益均在早高峰达到最大值，晚高峰次之，在其它时段呈现极小值。同时，工作日全时段出行效率没有显著区别，而周末早高峰时段乘客的平均出行效率最低。

表 7-1 服务指标时变规律表

Table 7-1 The summary of temporal variation of service indicators

| 类目 | 工作日 | | | 周末 | | |
| --- | --- | --- | --- | --- | --- | --- |
| | 早高峰 | 晚高峰 | 其它时段 | 早高峰 | 晚高峰 | 其它时段 |
| 平均出行时间效率 | 0.423 | 0.426 | 0.422 | 0.394 | 0.408 | 0.402 |
| 平均服务里程转化率 | 1.005 | 0.719 | 0.444 | 0.758 | 0.674 | 0.359 |
| 平均服务效益 | 0.496 | 0.354 | 0.214 | 0.361 | 0.334 | 0.170 |

在得到以上服务指标分布规律的基础上，进一步提取每一条公交线路的平均站距（单位：米）与站数里程比（单位：站/千米），尝试探讨公交线路结构特征与服务效益、出行时间效率以及服务里程转化率间的关系。结果总结如图 7-3。

由图 7-3a，图 7-3b，可以观察到站数里程比对服务里程转化率以及服务效益之间大致呈负相关关系，具体的，站数里程比决定了服务里程转化率以及服务效益的上界。图 7-3a 中，服务里程转化率的分布完全位于直线 $y = 400 - 133x, x \in [0,3]$ 以下的三角形区域内；而图 7-3b 中，服务效益指标数值均位于直线 $y = 250 - 83.33x, x \in [0,3]$ 以下的三角形区域内。

由图 7-3c，随着平均站距离增大，乘客的平均出行时间效率也随之增加，但平均站距大于 1500m 后，增幅开始减缓。尤其是超过 2000m 后甚至开始出现下降的情况，此时由于班线站点过于稀疏导致无法满足乘客出行的可达性需求，进而表现为客源不足。

由图 7-3d，居民的出行效率与线路的运营成本-服务里程转化率呈正相关（拟合模

型：$y = 0.0821ln(x) + 0.0993$，$R^2 = 0.634$），表明合理规划线路能够同时提升服务里转化率的同时提升乘客的出行时间效率。

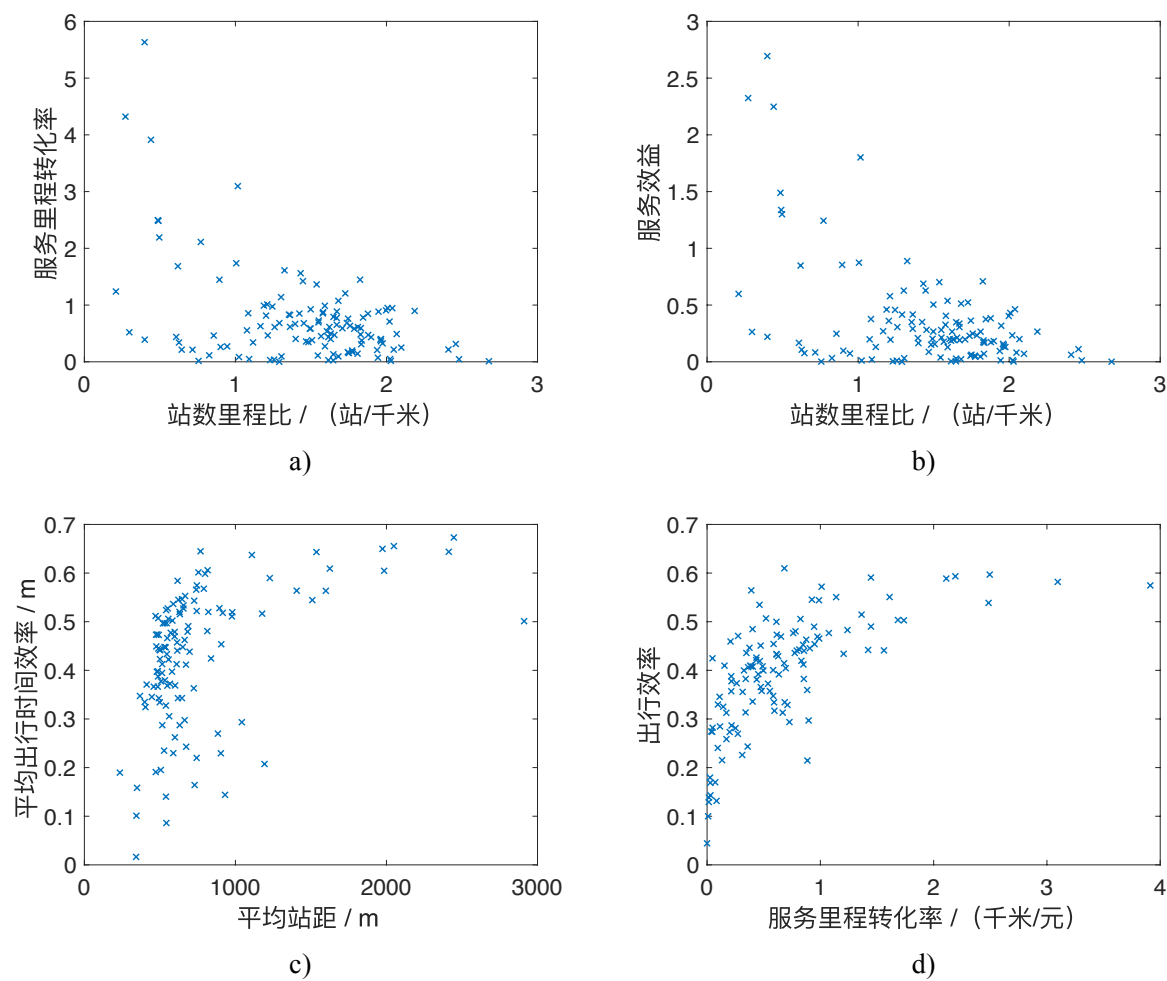

图 7-3 珠海市公交线网综合服务指标关联关系

Fig. 7-3 Numerical correlation of integral social service indicators of Zhuhai

综合本节所有数据挖掘、分析结果，得到以下结论：一方面，公交运营部门可以通过合理掌握乘客需求并结合其出行选择特性合理调整站点布置以及车次安排，可以在提升乘客出行时间效率与提升运营成本-服务里程转化率两方面实现双赢；另一方面，可以发现公交线网整体的乘客出行时间效率较低（大多数情况下不超过 0.55），尤其是站数量里程比较大的运营线路中，乘客出行时间效率、服务里程转化率以及服务里程转化率均处于低水平。结合 6.3.2 中乘客线路选择偏好，该城市应引入快速公交以及定制公交线路对线网进行升级改造，以提升服务效益与乘客出行效率。

## 7.3 线路参数求解方法

### 7.3.1 算法总体流程

线网优化问题包含较多不连续变量（如站点列表）无法用常规的数值优化方法求解，

因此采取基于粒子群优化的方法来寻找新线的优化方案。本文提出的算法总体流程如图 7-4 所示。主要步骤如下：

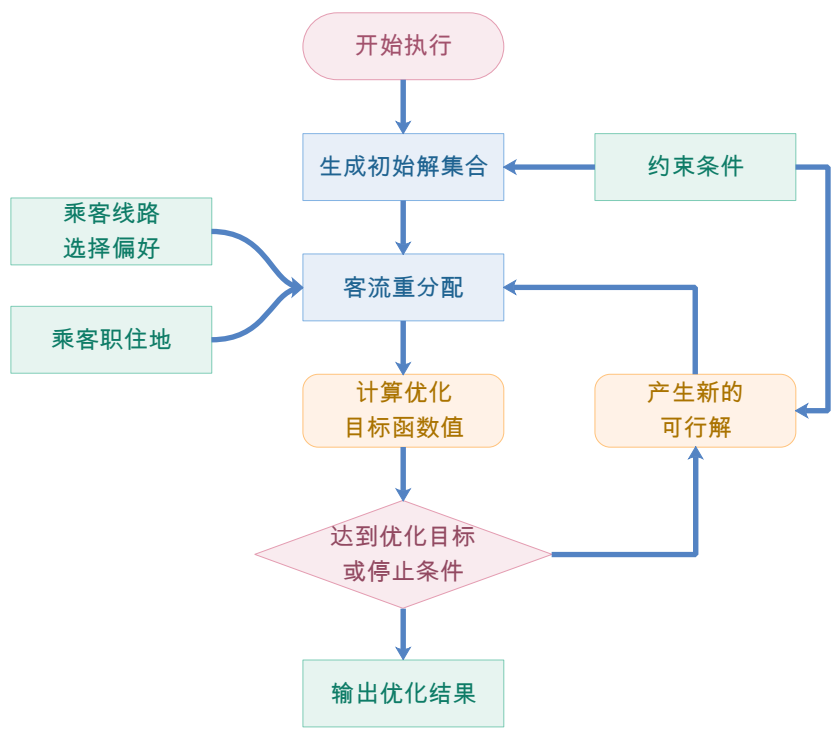

图 7-4 优化算法总流程

Fig. 7-4 Workflow of optimization

**Step 1** 利用随机初始化与约束条件生成初始解集合；

**Step 2** 对所有初始解，结合乘客线路选择偏好以及职住地、出行链计算每一名受影响的乘客在新场景下对新线路与旧线路的选择概率，并将乘客分配到相应的线路上；特别的，如果某乘客没有在此场景下的线路选择偏好模型，则套用在该场景下全体乘客的宏观选择模型进行线路分配（客流重分配模型验证见 6.2.2 节）；

**Step 3** 根据重新分配的客流，计算目标函数值，若目标函数值达到优化目标，则输出此时的解并结束优化过程；若超过最大迭代次数，则输出当前粒子群获得群体最优解；否则继续运行第 4 步；

**Step 4** 对每个粒子，根据群体最优解与自身最优解更新速度，并转移到新的位置，返回第 2 步。

### 7.3.2 粒子群优化算法运行流程

按优化算法整体流程（图 7-4），粒子群算法产生新的可行解以不断进行反馈调整。粒子群算法的根据不同的设计思想包含多种不同类型，针对本文的优化目标，采用全局粒子群算法，能够更加全面地对多变量问题进行求解寻优。粒子群优化算法的重要设计

思想简述如下：

首先，粒子群优化算法将每个初始可行向量解表达为一个粒子$pt$，设群体$G$共有 M 个粒子(通常取 10~50)；每个编号为$m$的粒子分别具有如下属性：

**粒子位置**： $\mathbf{S^m} = (\mathbf{Intv_{all}^m}, I_1^m, I_2^m, \cdots, I_k^m)$，位置为当前解向量。式中，将两种优化策略下的发车班次间隔统一定义为向量$\mathbf{Intv_{all}^m}$以方便叙述。

**粒子速度**： $\mathbf{V^m} = (\mathbf{v_{intv}^m}, v_1^m, v_2^m, \cdots, v_k^m)$，代表粒子下一次迭代中位置的变化方向与幅度。

其次，粒子在解空间运动时候，需要不断更新每个粒子自身经历的最优位置$\mathbf{P_{best}}$和群体的历史最优位置$\mathbf{G_{best}}$：

$$\mathbf{P_{best}} = (\mathbf{Intv_{all}^c}, I_1^c, I_2^c, \cdots, I_k^c)$$
$$\mathbf{G_{best}} = (\mathbf{Intv_{all}^G}, I_1^G, I_2^G, \cdots, I_k^G)$$
（7-9）

其中，$\mathbf{P_{best}}$并不跟随群体最优解$\mathbf{G_{best}}$改变，可以认为粒子自身拥有一定的记忆特性。

再次，在计算过程中需要不断更新每个粒子的速度，对编号为$m$的粒子，已知当前速度$\mathbf{V_i^m}$，与位置$\mathbf{S_i^m}$，结合粒子群的全局参量，其新速度为：

$$\mathbf{V_{i+1}^m} = \omega \cdot \mathbf{V_i^m} + c_1 \cdot r_1(\mathbf{P_{best}}^m - \mathbf{S_i^m}) + c_2 \cdot r_2(\mathbf{G_{best}}^m - \mathbf{S_i^m})$$
（7-10）

其中，$\omega$称为是 PSO 的惯性权重（inertia weight），它的取值介于[0,1]区间，用以刻画该粒子对当前速度的偏好权重，本文采用自适应权重变化，即$\omega$从 0.9 逐步递减直至达到最大迭代次数前 80%时候递减至 0.1，然后保持不变；参数$c_1$和$c_2$称为是学习因子（learn factor）其中，$c_1$表示粒子对自身最佳位置的偏好权重，$c_2$表示粒子对群体最佳位置的偏好权重，一般设置为 1.4961；而$r1$和$r2$为介于[0,1]间均匀分布的随机数。根据新速度，粒子$m$的新位置为：

$$\mathbf{S_{i+1}^m} = \mathbf{S_i^m} + \mathbf{V_{i+1}^m}$$
（7-11）

粒子群优化算法通过以上步骤，假设每个粒子在解空间中不断探索，每次改变位置时，都会参考自身最佳位置与群体发现的最佳位置。因此，该算法具有一定的知识记忆性，能避免遗传算法随着种群大规模变异后丢失所有历史最优解信息的缺陷。

最后，本文用以评价位置的适应度为该可行解经过客流重分配后的服务指标评价函数值，其表达的物理意义为线路运营成本—服务效益与乘客平均出行效率的乘积。

## 7.3.3 初始解编码方法与约束条件

对任何一条新增线路$r_n$以及旧线路$r$，假设$r$从起点到终点经过的站点（不考虑上下行停靠不同站点的情形）构成顺序集合$C_{Stops}=\{s_0,s_1,\cdots,s_n\}$，由该集合出发，首先，构建长度为$n$的站点列表向量：

$$\mathbf{V_{Stops}} = (sf_0, sf_1, \cdots, sf_n)$$

$$\text{s.t.} \quad sf_k = \begin{cases} 0 &,\text{选择第}k\text{个站作为停靠站;} \\ 1 &,\text{otherwise.} \end{cases} \quad (7\text{-}12)$$

其中，$sf_k$为布尔变量，表示在第$k$个站点是否停车。即，需要确定新线的发车班次间隔与站点列表，假设可用的解向量为：

$$\begin{aligned} \mathbf{S_{comp}} &= (Intv_{rn}, \mathbf{V_{Stops}}) \\ &= (Intv_{rn}, sf_0, sf_1, \cdots, sf_n) \\ &\text{s.t.} \, Intv_{rn} \in [2,20] \end{aligned} \quad (7\text{-}13)$$

其中，子向量$\mathbf{V_{Stops}}$维数较高，且每一位的数值类型都为布尔型，不适合直接输入到已有的智能优化算法程序集中进行求解，本研究将$\mathbf{V_{Stops}}$中各站点停靠标志位每 8 个一组进行编码处理并构造相应的约束条件。将$\mathbf{V_{Stops}}$转换为：

$$\mathbf{V_{Stops}} = (I_1, I_2, \cdots, I_k) \quad (7\text{-}14)$$

其中，$I_k$为编码变量，作用举例如下：

**例 5.1**：假设某新路有 18 个可选站点，首末站分别位于第一个和最后一个站，则可以用 3 个 8 位的无符号整形变量$I_1,I_2,I_3$进行编码处理。对于协同优化策略，新线经过其首末站，即第一个 8 位变量（$I_1$）从左至右第 1 位置 1，其余位无限制；最后一个 8 位整形变量（$I_3$）从右起第 2 个位置置 1 从右起第 3 位开始全部置 0，该变量其余位置无限制；其余中间变量（本例中为$I_2$）不需要限制其范围。将以上叙述转为如下数学描述：

$$2^7 \leq I_1 \leq 2^8 - 1$$
$$2^1 \leq I_3 \leq 2^2 - 1$$

最后，将最后一个整形变量（本例中$I_3$）从最高位向最低位反转。

根据例 5.1 可总结出不失一般性的编码方案，即对任意$n$个站点以及字长为$m$且$n>m$的编码变量，若$r$为$n$除以$m$的余数，若$r \geq 1$：

$$\forall r \geq 1, \begin{cases} 2^{m-1} \leq I_0 \leq 2^m - 1 \\ 2^{r-1} \leq I'_k \leq 2^r - 1 \\ 0 \leq I_2, I_3, \cdots I_{k-1} \leq 2^m - 1 \\ I_k = I'_k \text{按 MSB} \rightarrow \text{LSB 反转} \\ m < n \end{cases} \quad (7\text{-}15)$$

如果$r = 0$：

$$\forall r = 0, \begin{cases} 2^{m-1} \leq I_0 \leq 2^m - 1 \\ 0 \leq I_2, I_3, \cdots I_{k-1} \leq 2^m - 1 \\ I_k \geq 1 \\ m < n \end{cases} \quad (7\text{-}16)$$

### 7.3.4 基于乘客个体偏好的客流重分配模型

新线加入，直接导致公交线路走廊内的服务多样化场景发生变化，对乘客个体而言，若其出行 OD 对应的站点被新线所覆盖，则公交系统为其提供了新的到达选择。新线与已有线路将构成竞争关系，乘客将按其偏好特性综合考虑各种因素，并按各种方案的效用值排序按概率做出选择[98-99]。本研究依照 7.2 节取得的模型，对所涉范围内所有已拟合得到偏好特征的乘客个体计算其线路选择概率，将其按概率分配到其希望选择的线路上。具体的，对于任一乘客p，计算流程如图 7-15 所示，对每个乘客，该算法主要可分为读取所涉区间内 OD，分析服务选项与场景变化，选择偏好模型，重新计算线路选择概率，概率分配五大步骤。

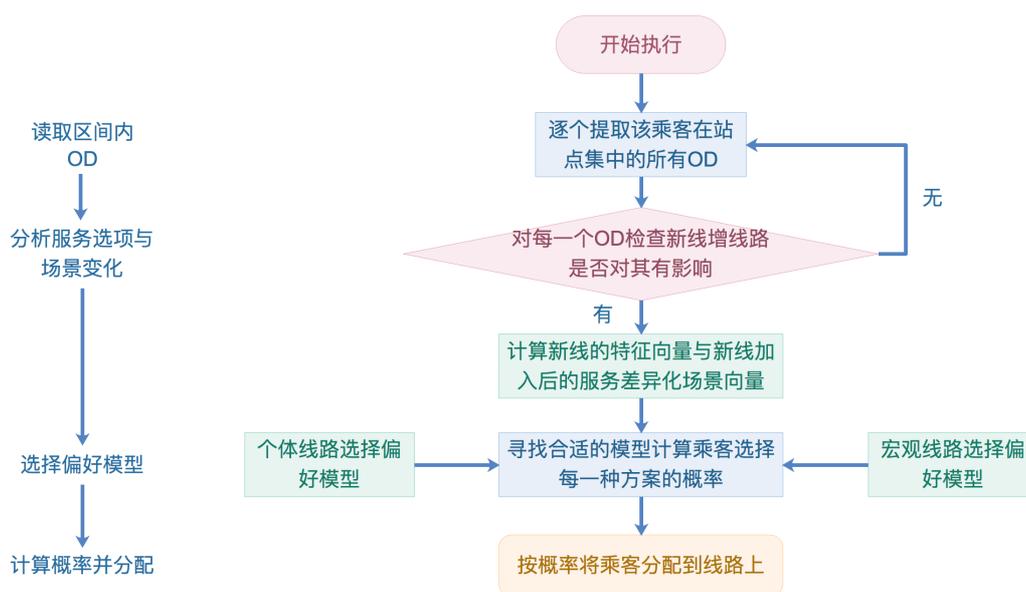

图 7-5 客流重分配流程

Fig. 7-5 Workflow of ridership re-distribute.

其中，设已有的乘车 OD 为 A→B，新加入线路$r_n$对其产生影响的判断依据为：

$$\exists\ station_m, station_n \in r_n,\\ distance(station_m, A) + distance(station_n, B) \leq 2\varepsilon \quad (7\text{-}17)$$

式中，$station_m$, $station_n$为在 A→B 相同方向上新线$r_n$经过的站点，该定义的物理意义是：乘客可以在其原来的站点（A 点）直接乘坐快线车到达目的地（B 点）或者借助快线车并步行一段合理的距离后到达其目的地。其中，累计步行距离$\varepsilon$ = 350 m（与

6.2 节一致）。

图 7-5 所述流程中，选择模型计算乘客新方案与旧线路选择概率的具体步骤如下：

**Step 1** 计算新线加入后乘客 OD（A→B）的服务差异化场景向量$\boldsymbol{\eta}'$，与新线的服务选项特征向量；

**Step 2** 查找该乘客是否有该场景下对应的乘车方案选择偏好模型；

**Step 3** 如果有对应该场景的权重偏好模型，则直接使用该模型重新计算乘客对每种乘车方案的选择概率。

**Step 4** 若不存在则调用在该场景下的全体乘客的宏观偏好模型，将其作为该乘客的选择偏好模型。计算该乘客对每种方案的选择概率。

## 7.4 案例分析

本文分别基于珠海市 504 路（图 7-6a）与 207 路（图 7-6b）公交所涉及的站点进行快速公交线路设计案例研究。504 路为珠海市直接连通西部两城郊镇区（金湾区与斗门区）的线路，同时也是斗门区唯一一条可直达珠海国际机场的公交线路；207 路为珠海机场与高铁站间唯一的直连公交线路，同时也是金湾区航空工业园与拱北主城区间唯一的直达公交线路。因此，本文所选择的公交线路具有极其重要的区域经济地位。**同时，由 4.2.2 以及 5.2.2 中的数据可视化图，207 路和 504 路均为珠海市公交骨干线网以及客流集散通道的重要组成部分。**

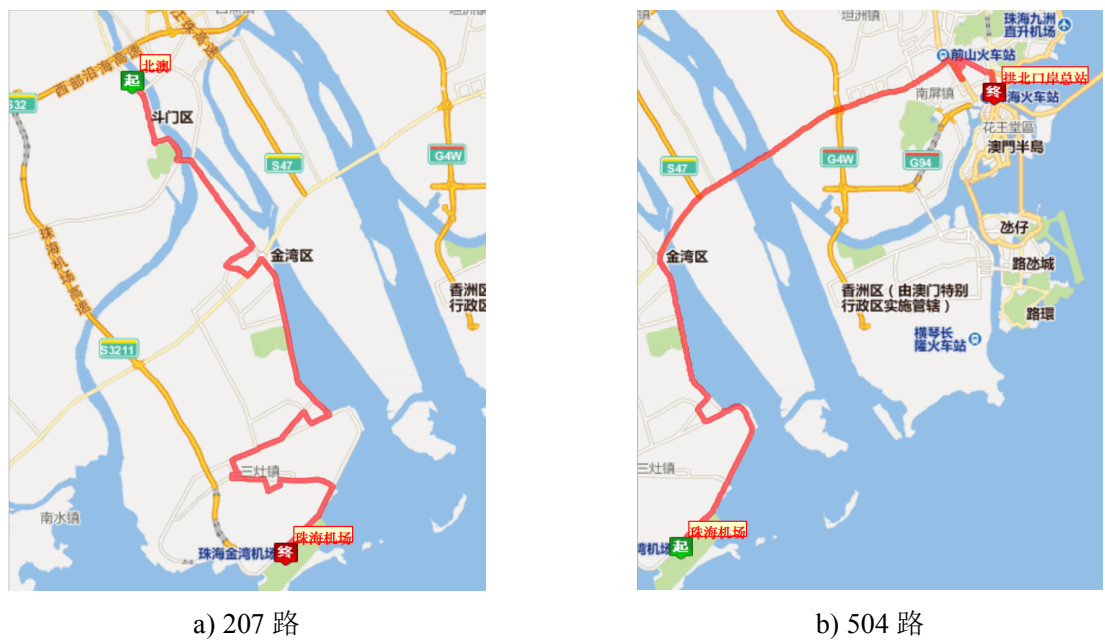

a) 207 路　　　　　　　　　　　　b) 504 路

图 7-6 珠海市 504 路与 207 路公交路线图

Fig. 7-6 Map of public bus line 504 and 207 in Zhuhai

207 路与 504 路两线路的基本信息如表 7-2 所示，由表，所选线路的站数量均超过 60 个，站数里程比均大于 1.26 站/千米，但服务里程转化率低于 0.78 千米/元。因此，组成这两条公交线路的站点网络可以尝试引入快线以提升服务里程转化率与乘客的出行效率。

提取 2015 年 3 月 2 日乘客出行信息（含乘车轨迹与出行阶段划分），做出两线路乘客的乘车里程分布与出行效率分布如图 7-7，由图可观察到 504 路公交乘客出行距离集中分布在[0.5,10]千米的区间内，而 207 路乘客的出行乘车里程除了在[0,5]千米区间内集中分布外还有约 35%分布于[5,20]区间内，两线路均有较多乘客出行分布在[0.6,0.8]的区间内；经统计，504 路平均乘客平均乘车里程为 10.19 km，207 路乘客乘车里程为 10.83 km；相应的，两公交线路的平均乘客出行时间效率均低于 65%，相比而言，207 路有更多乘客出行距离在 10km 以上。

表 7-2 所选线路的基本信息
Table 7-2  Brief information of selected bus routes

|  | 线路里程(千米) | 停站数 | 班次间隔(分钟) | 服务里程转化率（千米/元） | 总服务里程（千米） | 客流量（人次） |
| --- | --- | --- | --- | --- | --- | --- |
| 504 路 | 51.6 | 65 | 15 | 0.79 | 35002 | 3665 |
| 207 路 | 49.7 | 62 | 15 | 0.63 | 29737 | 2518 |

为防止粒子群优化算法找出极端解（如：客流量不足 50 人但距离长且无停站的线路），根据 504 路与 207 路的现状，还补充以下约束条件：**优化后的线路全天总客流量不低于 1500 人次，乘客平均出行效率不低于 75%。**

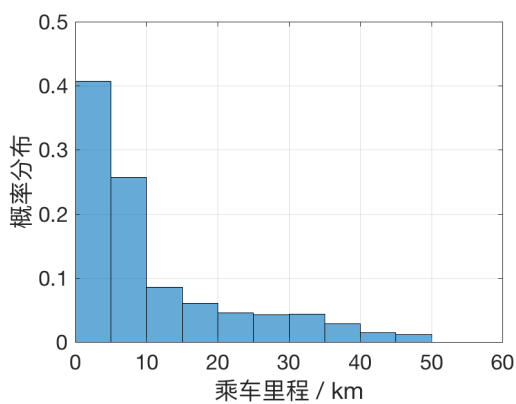
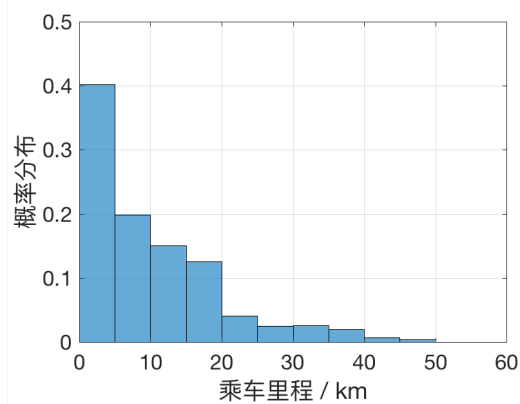

a) 207 路                                   b) 504 路

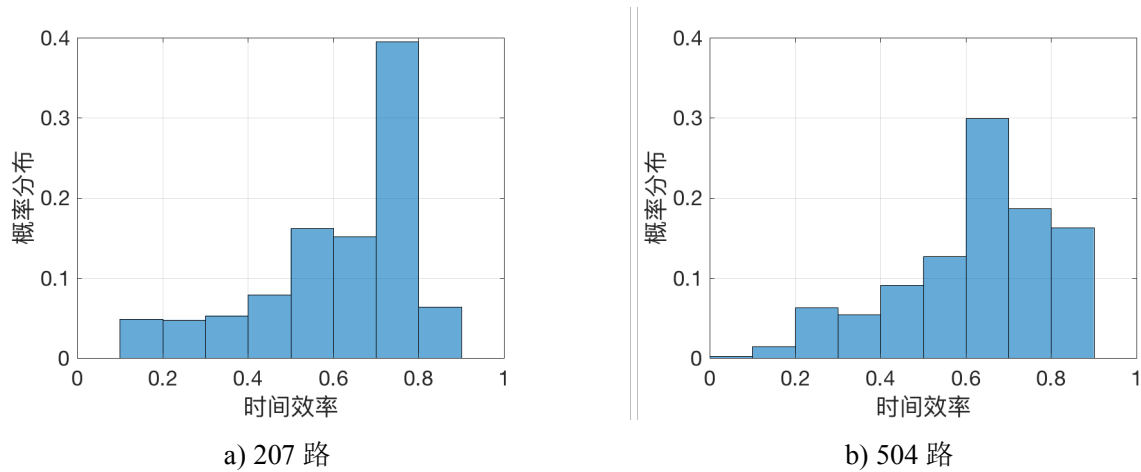

a) 207 路　　　　　　　　　　　　　　b) 504 路

图 7-7 优化前乘客乘车里程与时间效率分布对比

Fig. 7-7 Distributions passengers' travel distance and time efficiency of transit route 504 and 207

分别采用服务效益、服务里程转化率、服务里程、时间效率四项指标为目标函数，按 7.3.3 节的方法对可行解空间进行编码并启动粒子群优化算法分别寻找合适的快速公交线路。

四种优化目标下 504 路新线与 207 路新线的发车班次间隔与中途停站数对比如图 7-8a、7-8b 所示，由图 7-8a，在以服务里程最大化与时间效率最大化为优化目标的情形下，优化算法选择了限制范围内最小的发车间隔；而在以服务效益与服务里程转化率为优化目标的情形下，发车间隔均在 10 分钟以上，经过对比分析，优化算法为降低运营成本，在保证乘客出行效率的前提下尽可能提高了发车班次间隔；从站点数分布上观察，以服务里程最大化为目标函数的情形下，两条新线的站点数达到极大值（超过 50 个）；以时间效率为优化目标函数的前提下，两条线的站点数均为最小值，此情形下 207 路新线的站点数仍远小于 504 路。

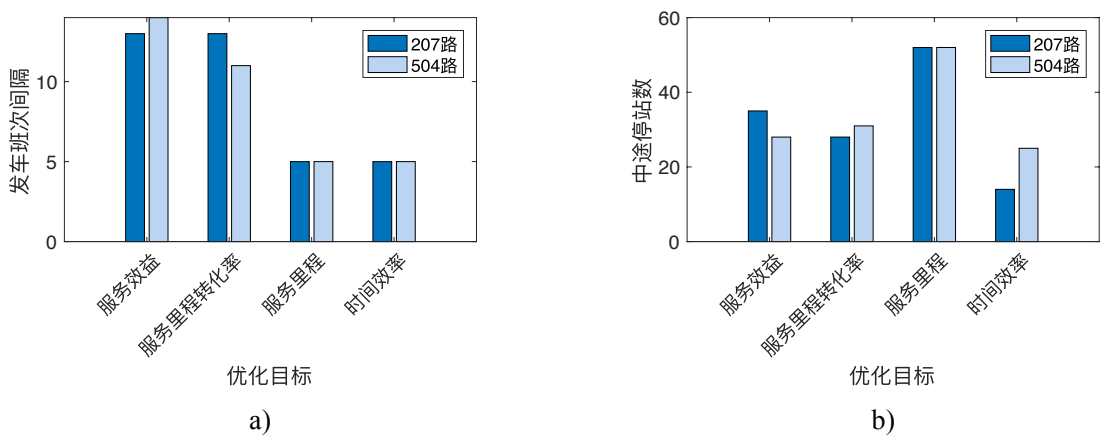

a)　　　　　　　　　　　　　　b)

图 7-8 不同优化目标下的最优解服务指标对比

Fig. 7-8 Compare of service metrics of optimal solutions with various target functions

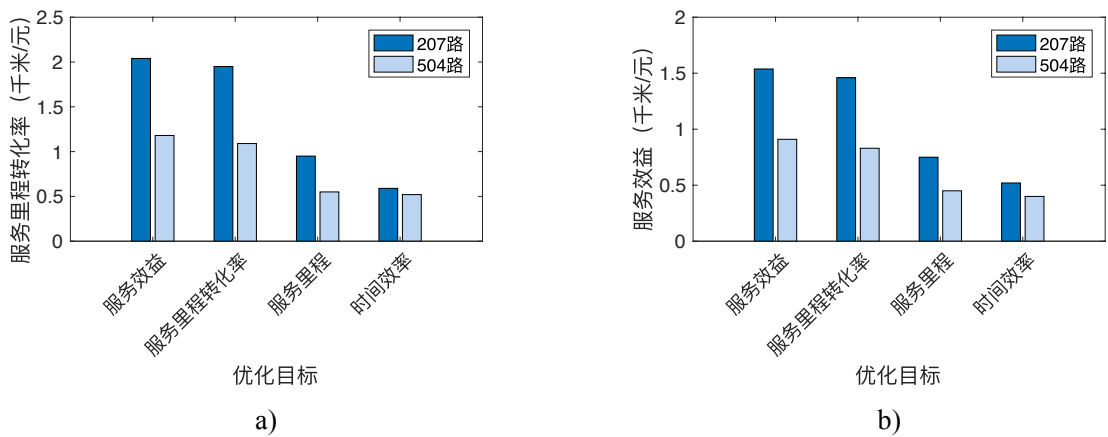

a)                              b)

图 7-9 不同优化目标下的服务效益与服务里程转化率对比

Fig. 7-9 Compare of profit ratio and revenue transformation rate of optimal solutions with different targets

四种目标限制下所得新线路的服务里程转化率与服务效益对比如图 7-9 所示，对比图 7-9a、7-9b，四种优化目标作用下的最优解中，以极大服务效益与服务里程转化率为目标的寻优方案均得到较高的服务里程转化率与服务效益值；相反，以极大时间效率为目标的优化方案均导致低于旧线路的服务里程转化率与服务效益。同时，优化后的 207 路新线服务效益与服务里程转化率均高于优化后的 504 路。

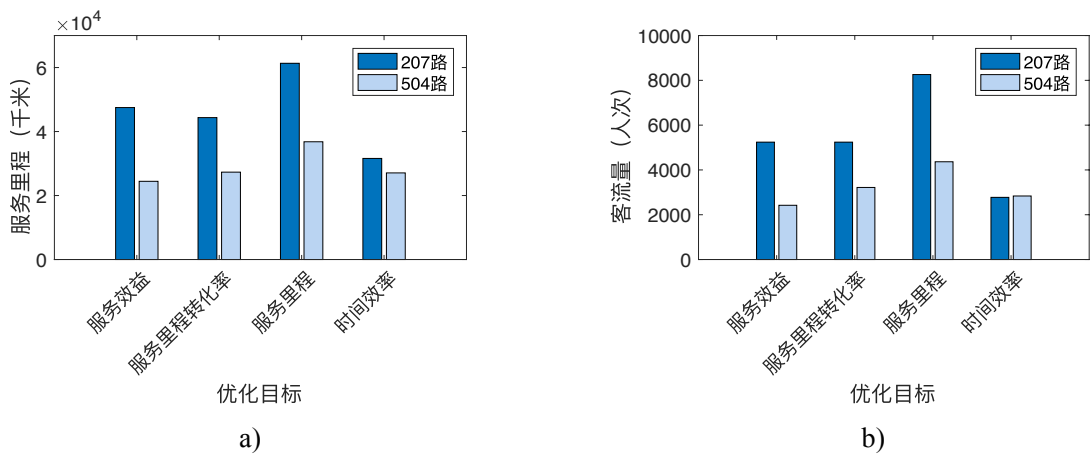

a)                              b)

图 7-10 不同优化目标下的客流量与服务里程对比

Fig. 7-10 Compare of ridership and service miles of optimal solutions with different targets

四种情形下客流量与服务里程对比如图 7-10，由图 7-10a，图 7-10b，可观察到客流量与服务里程均在以极大服务里程为目标函数的寻优场景下取得极大值，但在以极大时间效率为目标的寻优场景下取得极小值。同时，207 路新线的客流量除在以效率为目标的优化场景外，其它场景下均高于原线路的客流，而 504 路新线仅在以服务里程最大化为目标的寻优场景下客流量高于原线客流持平，其它场景下客流均低于旧线路。207 路新线客流大于原先线路客流的原因是线路经过珠海市主城区，新线提高了出行效率，因此吸引了更多乘客从其它线路转移至新线，也从另一个侧面说明，全部站点位于西部城

郊的 504 路沿线的乘客对出行效率并不敏感。

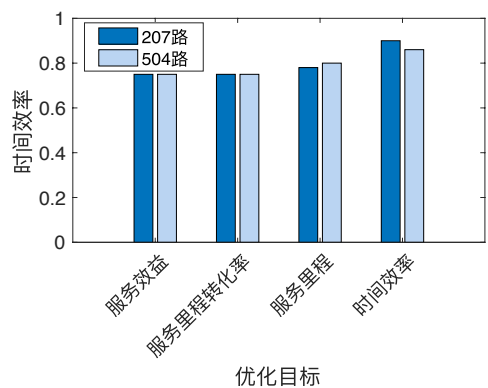

图 7-11 不同优化目标下的乘客出行时间效率对比

Fig. 7-11 Compare of time efficiency of optimal solutions with different targets

四种场景下最优解的乘客出行时间效率对比如图 7-11 所示，由图可观察到，粒子群优化算法得到的全部最优解时间效率均达到 75%以上，高于原线路的水平。其中，以极大服务效益以及服务里程转化率为目标的优化方案中，寻优算法仅将时间效率保持在 75%，即设定的下限值；而以最大化时间效率为导向的优化方案中，两条线路乘客的乘车时间效率均高于 90%。

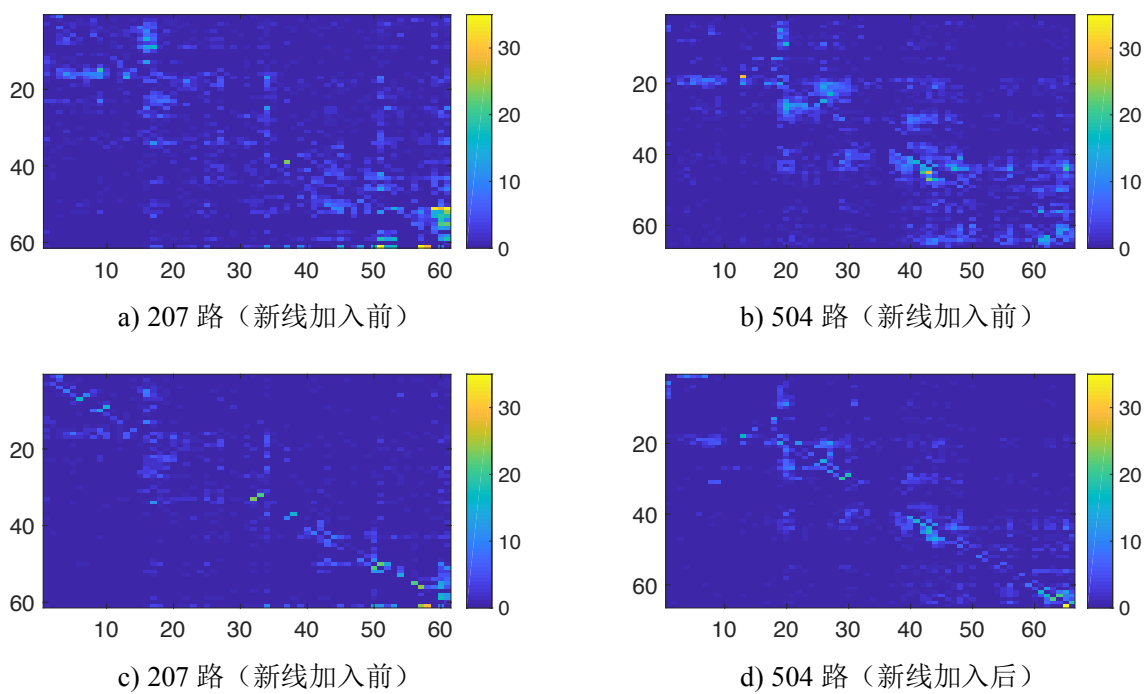

a) 207 路（新线加入前）      b) 504 路（新线加入前）

c) 207 路（新线加入前）      d) 504 路（新线加入后）

图 7-12 新线加入后客流分布变化

Fig. 7-12 Simulation of ridership distribution before and after the establishment of new routes

综合考虑以上多方面因素，本研究采用极大服务里程转化率为寻优目标所产生的新线作为 207 路与 504 的新线方案。该方案保证的乘客乘车时间效率高于 75%，同时，也

能吸引足够的客流。另一方面，以时间效率最大值为目标的寻优方案导致停站数过少，客流量不足，但该方案的站数最少，可作为未来开通定制班线的参考方案。

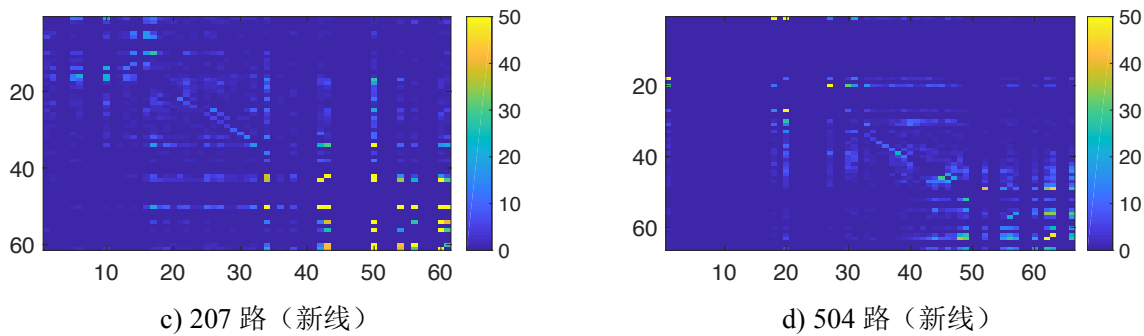

c) 207 路（新线）    d) 504 路（新线）

图 7-13 新线客流 OD 矩阵

Fig. 7-13  Simulation of ridership distribution of new routes

运用客流重分配算法推演得到的 504 路与 207 路新线加入后的客流 OD 矩阵分布改变如图 7-12 所示：由图 7-12a，图 7-12c 对比以及图 7-12b，图 7-12d 对比可以观察到，新线加入后，OD 矩阵主对角线附近的客流受影响较小，而离开主对角线越远的矩阵单元客流衰减越大，该现象的原因是，主对角线附近的矩阵单元代表短距离出行的客流，因此较少受到停站少的新线影响。同时，相比于 207 路，504 路客流受到新线加入影响较少，原因是 504 路的乘客多为短距离出行者，OD 矩阵中表现为主对角线附近客流量较大。

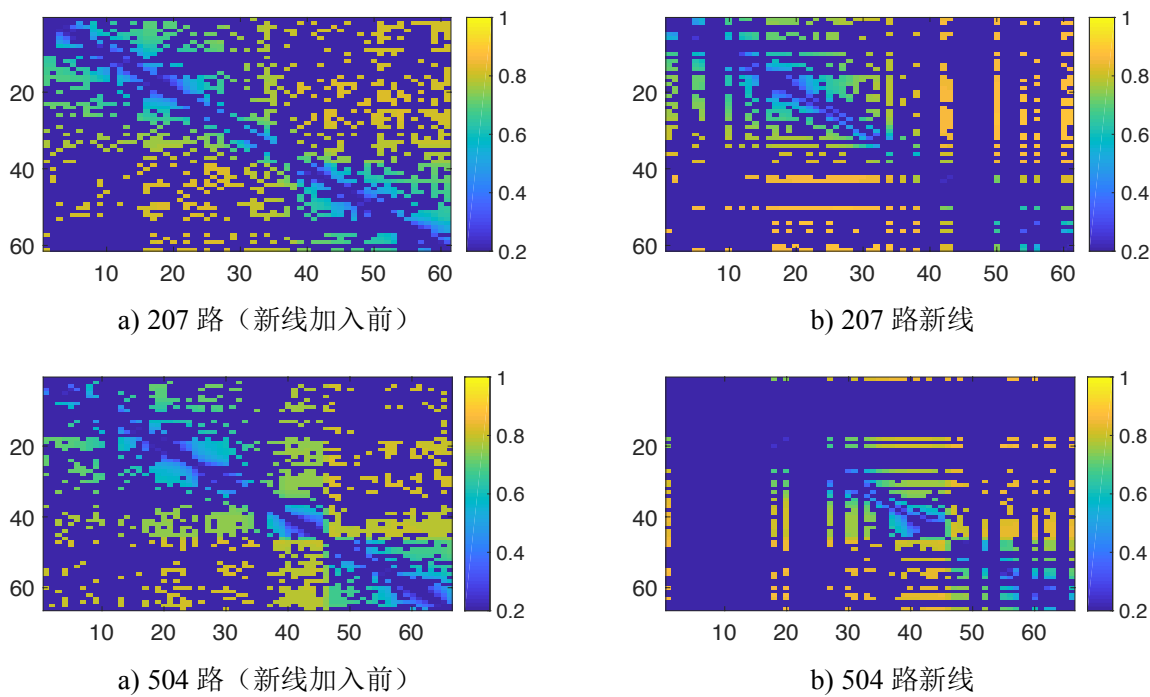

a) 207 路（新线加入前）    b) 207 路新线

a) 504 路（新线加入前）    b) 504 路新线

图 7-14 新线加入后出行效率分布变化

Fig. 7-14 Simulation of efficiency distribution before and after the establishment of new routes

207 路与 507 路新线的客流分布 OD 矩阵如图 7-13 所示，由图可知新线上的客流热点多远离 OD 矩阵的主对角线，且客流量取得峰值区域出现明显的条带状分布，部分乘客步行到达新线站点乘车。可以说明本文所提供的线路有效的连接了原 504 路沿线与 207 路沿线的客流热点区域，该方案得到的新线与旧线路可以共同满足乘客多样化的出行需求。

207 路与 504 路乘客在新线加入前出行效率在 OD 矩阵上的分布如图 7-14a，图 7-14b 所示，大量低效率出行分布在主对角线附近。因此整体出行效率偏低。相应按快速公交要求设计的新线出行效率分布如图 7-14c，图 7-14d 新线所覆盖站点形成的 OD 对（尤其是远离主对角线）的出行效率均有提升。即新线能初始服务里程转化率取得极大值同时为具有不同出行不同偏好乘客提供多样化的服务选项。

## 7.5 本章小结

本章基于乘客出行偏好分布模型，提供了一套基于粒子群优化与客流分布推演的计快速公交线路设计方法与案例。具体工作内容如下：

（1）从乘客出行效率以及服务里程转化率、服务效益三个方面提出了城市公交系统的服务质量评价指标，同时，分析了目标城市公交系统的服务质量的变化规律与关联关系。

（2）提出了一种基于乘客个体出行偏好的客流重分配模型，以预测线路经停站点以及发车班次发生改变后的客流。

（3）基于客流数据推演与粒子群优化算法，提出了利用原有线路基础设施设计新的线的方法以及合理的目标函数和约束条件。

（3）评估了新线开通后对旧线路的影响以及新旧线路乘客出行效率变化，客流数据推演试验表明，所选方案提供的新线能有效提升沿线居民的出行效率，改善公共投入至乘客服务里程的转化率。

# 结论与展望

本研究提供了一套从公交多源数据，尤其是一票制公交运营系统中多源异构化的数据中挖掘宏观上的客流时空分布规律以及微观上公交乘客行为特征以及偏好特性的方案，并将挖掘得到的乘客出行需求时空分布规律与乘客出行线路选择偏好进行有机结合，以期提供一套更合理公交线路进行优化方案。具体工作包含以下部分：

（1）缺陷数据源背景下的数据预处理方法：根据现有的较为稳定的城市公交网络的数据采集方法，提出了乘客刷卡数据与车辆进站报站数据间的时间误差校正方法，且该方法无需对现有公交数据采集系统进行硬件更新，还能借助本文提供的信息压缩方案高效的完成时间误差校准；此外，提出了融合历史数据与乘客刷卡数据进行缺失的报站数据信息还原方案。**这部分研究的意义在于提供了一套高效修复数据源基础信息缺陷的方案。**

（2）乘客出行轨迹与出行阶段还原方法。对传统的下车站点推断方法进行改进，使其能够在一定程度上容忍缺失数据带来的影响；还改进了传统的出行阶段（起点、换乘、终点）识别算法，使其能够识别乘客短时活动；通过跟车调查验证了乘客出行轨迹还原算法的有效性，并对目标城市的客流时空分布以及线网的满载率进行分析。**这部分研究的意义在于从孤立的刷卡事件中还原了乘客完整的出行轨迹与出行阶段，为后续数据挖掘提供基础资源。**

（3）乘客闭合出行链挖掘方法。在乘客出行时空轨迹的基础上，提出了一种能够融合多天的 OTD 数据的乘客出行拓扑关系图构建准则，以及在此基础上的乘客闭合出行链搜索、关联算法；此外，还对目标城市乘客的闭合出行链进行了分析，从个体乘客出行分布的基础上挖掘了城市的公交客流集散通道。**这部分研究意义在于将乘客在公交系统中碎片化的轨迹映射到闭合出行链空间内，为更准确的刻画乘客个体出行特征提供了一种新的更合理选择。**

（4）乘客线路选择偏好挖掘。在乘客个体出行链以及 OTD 数据的基础上，提取了每个乘客 8 大关键特征，并借助 MNL 模型拟合乘客的线路选择偏好，以客流重分配推演验证了个体偏好模型的有效性。并分析了具有不同极大偏好选项的乘客在工作日以及双休日在公交客流中的比重时变规律。**这部分研究的意义在于提供了一套不依赖调查问卷即可从公交大数据挖掘乘客出行模式的方法与案例，还发现了具有不同偏好的乘客的活动特点。**

（5）基于乘客选择偏好的公交线路优化方法研究。提出了一种基于乘客个体出行偏好以及粒子群优化算法的公交线优化设计方法，该方法借助信息编码，使得粒子群能够同时从站点部署、发车班次两方面搜索新线路最优方案。此外，本研究还以珠海市207路与504路为案例，探讨了不同的优化目标函数所得到的最优解特点，最终选出了最佳方案，为城市公交系统的运营调度管理供决策依据和理论支撑。**这部分研究的意义在于将大数据中挖掘出的乘客个体出行模式知识库运用到了线网优化实践中，还给出了一套更科学合理的优化目标策略。**

尽管本文取得了一定的研究成果，但仍存在一些不足，展望未来，本课题还需要从以下几个方面进行深入的研究：

（1）尽管我们得到了公交线网中乘客的时空轨迹以及其它一系列出行信息，但还未能对乘客未来的活动轨迹或未来的出行需求做出准确预测。

（2）虽然我们成功拟合了大部分乘客的出行线路选择偏好，但是我们对于乘客的线路选择偏好现阶段仍以线性模型进行描述，难以刻画乘客行为中的非线性部分，因此，有待进一步探索更精准、高效的乘客出行方案选择偏好模型。

（3）本研究未考虑从 IC 卡及 GPS 数据中挖掘乘客的社会、经济属性，因此，与传统交通规划方法间欠缺可比性，后续应结合交通调查与数据挖掘，进一步完善乘客偏好数据推演算法。

（4）本研究虽然对基于乘客个体出行偏好的公交线路优化方法进行了研究，但未能考虑将同时协调多种具有不同属性的线路以使得线网的综合服务效益最大化。

（5）在本研究开展过程中，还发现，公共交通领域的乘客轨迹数据对乘客行为的采样以及刻画程度仍然有所欠缺，无法实时的跟踪任何一个个体，且空间定位精度仅仅在上下车活动发生瞬间足够精确；另一方面，公交运营大数据在目前大多数城市的应用中难以与其它数据源做到微观层面上的知识共享，现阶段本文挖掘得到的乘客线路选择偏好数据暂时无法得到乘客社会经济属性数据的进一步支撑；因此，我国目前的公交系统仍有待进一步进行数据资源整合以打破信息壁垒。

# 参考文献

[1] 马晓磊, 刘从从, 刘剑锋, 等. 基于公交 IC 卡数据的上车站点推算研究[J]. 交通运输系统工程与 信息, 2015, 15(4):78-84.

[2] 胡继华, 邓俊, 黄泽. 一种基于乘客出行轨迹的公交断面客流估算方法[J]. 计算机应用研究, 2014, 31(5):1399-1402.

[3] 胡继华, 邓俊, 黄泽. 结合出行链的公交 IC 卡乘客下车站点判断概率模型[J]. 交通运输系统工程 与信息, 2014, 14(2):62-67.

[4] 徐瑢. 不完全信息下的公交客流 OD 推算方法的研究[D]. 北京: 北京交通大学, 2011.

[5] 王超. 基于 IC 卡信息的公交客流 OD 推算方法研究[D]. 北京: 北京交通大学, 2012.

[6] 高联雄. 智能公交系统数据挖掘研究与应用[D]. 北京: 北京邮电大学, 2011.

[7] 章玉. 基于数据挖掘的动态公交客流 OD 获取方法研究[D]. 北京: 北京交通大学, 2010.

[8] 王月玥. 基于多源数据的公共交通通勤出行特征提取方法研究[D]. 北京: 北京工业大学, 2014.

[9] 陈君, 杨东援. 融合智能调度数据的公交 IC 卡乘客换乘站点判断方法[J]. 长安大学学报: 自然科学版, 2013, 33(4):92-98.

[10] Wang W., Attanucci J., Wilson N. Bus passenger origin-destination estimation and related analyses using automated data collection systems[J]. 2011.

[11] Munizaga MA., Palma C. Estimation of a disaggregate multimodal public transport Origin-Destination matrix from passive smartcard data from Santiago, Chile[J]. Transportation Research Part C: Emerging Technologies, 2012, 24:9-18.

[12] Munizaga M., Devillaine F., Navarrete C., et al. Validating travel behavior estimated from smartcard data[J]. Transportation Research Part C: Emerging Technologies, 2014, 44:70-79.

[13] Gaudette P., Chapleau R., Spurr T. Bus Network Microsimulation with General Transit Feed Specification and Tap-in-Only Smart Card Data[J]. Transportation Research Record: Journal of the Transportation Research Board, 2016(2544):71-80.

[14] Alsger A., Assemi B., Mesbah M., etal. Validating and improving public transport origin-destination estimation algorithm using smart card fare data[J]. Transportation Research Part C: Emerging Technologies, 2016, 68:490-506.

[15] Liu Y., Weng X., Wan J., et al. Exploring Data Validity in Transportation Systems for Smart Cities[J]. IEEE Communications Magazine, 2017, 55(5):26-33.

[16] Robinson S., Narayanan B., Toh N., et al. Methods for pre-processing smartcard data to improve data quality[J]. Transportation Research Part C: Emerging Technologies, 2014, 49:43-58.

[17] Ma X., Wang Y. H., Chen F., et al. Transit smart card data mining for passenger origin information extraction[J]. Frontiers of Information Technology and Electronic Engineering, 2012, 13(10):750-760.

[18] Zhang F., Yuan N. J., Wang Y., et al. Reconstructing individual mobility from smart card transactions: a collaborative space alignment approach[J]. Knowledge and Information Systems, 2015, 44(2):299-323.

[19] Kusakabe T., Asakura Y. Behavioural data mining of transit smart card data: A data fusion approach[J]. Transportation Research Part C: Emerging Technologies, 2014, 46:179-191.

[20] WHITE P., et al. Use of public transport smart card data for understanding travel behaviour[A]. In: PROCEEDINGS OF THE EUROPEAN TRANSPORT CONFERENCE (ETC) 2003 HELD 8-10 OCTOBER 2003, STRASBOURG, FRANCE[C], 2003.

[21] Seaborn C., Attanucci J., Wilson N. Analyzing multimodal public transport journeys in London with smart card fare payment data[J]. Transportation Research Record: Journal of the Transportation Research Board, 2009(2121):55-62.

[22] Morency C., Trépanier M., Agard B. Measuring transit use variability with smart-cardd ata[J].Transport Policy, 2007, 14(3):193-203.

[23] Jang W. Travel time and transfer analysis using transit smart card data[J]. Transportation Research Record: Journal of the Transportation Research Board, 2010(2144):142-149.

[24] Foell S., Phithakkitnukoon S., Veloso M., et al. Regularity of Public Transport Usage: A Case Study of Bus Rides in Lisbon, Portugal[J]. Journal of Public Transportation, 2016, 19(4):10.

[25] Li G., Yu L., Ng W. S., et al. Predicting home and work locations using public transport smart card data by spectral analysis[A]. In: Intelligent Transportation Systems (ITSC), 2015 IEEE 18th International Conference on[C], 2015. 2788-2793.

[26] 龙瀛, 张宇, 崔承印. 利用公交刷卡数据分析北京职住关系和通勤出行[J]. 地理学报, 2012, 67(10):1339-1352.

[27] 周家中, 张殿业. 多模式交通网络下的城市交通出行链行为模型[J]. 华南理工大学学报 (自然科学版), 2014(2):125-131.

[28] Golob T. F. A simultaneous model of household activity participation and trip chain generation[J]. Transportation Research Part B: Methodological, 2000, 34(5):355-376.

[29] Morency C., Valiquette F. Trip Chaining and Its Impact on Travel Behaviour. Activity-Based Analysis and Modeling[A]. In: 12th World Conference on Transport Research, Lisbon, Portugal[C], 2010.

[30] Zhenru L., Xuemei L. Review of trip-chain-based travel activity study of residents[A]. In: Logistics Systems and Intelligent Management, 2010 International Conference on[C], 2010. 3:1527-1531.

[31] Xianyu J. An exploration of the interdependencies between trip chaining behavior and travel mode choice[J]. Procedia-Social and Behavioral Sciences, 2013, 96:1967-1975.

[32] Lee Y., Hickman M., Washington S. Household type and structure, time-use pattern, and trip-chaining behavior[J]. Transportation Research Part A: Policy and Practice, 2007, 41(10):1004-1020.

[33] Golob T. F., Hensher D. A. The trip chaining activity of Sydney residents: a cross-section assessment by age group with a focus on seniors[J]. Journal of Transport Geography, 2007, 15(4):298-312.

[34] Ye X., Pendyala R., Gottardi G. An exploration of the relationship between mode choice and complexity of trip chaining patterns[J]. Transportation Research Part B: Methodological, 2007, 41(1):96-113.

[35] Walle S.V., Steenberghen T. Space and time related determinants of public transport use in trip chains[J]. Transportation Research Part A: Policy and Practice, 2006, 40(2):151-162.

[36] Ma X., Wu Y.J., Wang Y., et al. Mining smartcard data for transit riders' travel patterns[J]. Transportation Research Part C: Emerging Technologies, 2013, 36:1-12.

[37] Jian-chuan X., Zhi-cai J. Research on the Interdependencies between Trip Chaining Behavior and Travel Mode[J]. Journal of Shanghai Jiaotong University, 2010, 6:017.

[38] 李妲. 基于活动的节假日出行分析方法[D]. 北京: 北京交通大学, 2008.

[39] 褚浩然, 郑猛, 杨晓光, 等. 出行链特征指标的提出及应用研究[J]. 城市交通, 2006, 4(2):64-67.

[40] 宗芳, 隽志才. 基于活动的出行方式选择模型与交通需求管理策略[J]. 吉林大学学报 (工), 2007, 37(1):48-53.

[41] 宗芳. 基于活动的出行时间与方式选择模型研究[D]. 吉林: 吉林大学, 2005.

[42] 李萌. 基于活动链特征分析的出行方式选择模型研究[D]. 南京: 东南大学, 2008.

[43] 王迎. 基于活动的城市居民出行方式选择模型研究[D]. 西安: 长安大学, 2007.

[44] 杨柳. 城市通勤出行链模式选择行为研究[J]. 科技资讯, 2012(7):226-229.

[45] 石心怡, 郭英, 王元庆. 基于出行链的公共交通走廊探讨[J]. 城市轨道交通研究, 2010, 13(5): 59-63.

[46] 马飞, 刘飞, 孙启鹏, 等. 复杂通勤出行链脆弱性感知结构模型[J]. 长安大学学报(自然科学版), 2017(6):99-104.

[47] 何流, 陈大伟, 卢静, 等. 基于出行链的居民出行次数建模与仿真[J]. 深圳大学学报 (理工版), 2012, 29(3):264-269.

[48] 吕婷婷. 基于大都市郊区空间演化的居民通勤特征变化研究[D]. 南京: 南京 林业大学, 2013.

[49] Zhao J., Wang J., Deng W. Exploring bikesharing travel time and trip chain by gender and day of the week[J]. Transportation Research Part C: Emerging Technologies, 2015, 58:251-264.

[50] Hui Y., Ding M., Zheng K., et al. Research on trip chain characteristics of round-trip car-sharing users in China: A case study in Hangzhou City[A]. In: Proceedings of the Transportation Research Board 96th Annual Meeting, Washington, DC, USA[C], 2017. 8-12.

[51] 王俊兵. 基于出行链的公交乘客出行特征分析[D]. 北京: 北京交通大学, 2017.

[52] 王周全. 基于IC卡数据与GPS数据的公交客流时空分布研究[D]. 重庆: 西南交通大学, 2016.

[53] 聂瑶, 王军芝, 匡凯. 基于可达性的公交网络分层优化[J]. 交通科学与工程, 2017(4):70-76.

[54] 罗孝羚, 蒋阳升. 中小城市公交线网及发车频率同步优化[J]. 工业工程, 2017, 20(4):25-30.

[55] 胡继华, 高立晓, 梁嘉贤, 等. 一种多目标的公交线网规划模型[J]. 重庆交通大学学报 (自然科 学版), 2017, 36(12):102-109.

[56] 蒋阳升, 罗孝羚, 刘媛, 等. 公交线网空间可达性优化研究[J]. 公路交通科技, 2016, 33(4): 102-107.

[57] 徐茜, 俞礼军. 基于"多站点"的城市常规公交线网优化[J]. 公路与汽运,

2016(4):32-35.

[58] 齐振涛, 李文勇, 余子威, 等. 基于直达客流量的公交路径优化模型及求解算法[J]. 桂林电子科技大学学报, 2016, 36(5):412-415.

[59] 蒋阳升, 罗孝羚. 考虑首末站约束和站间客流强度的公交线网优化[J]. 长安大学学报 (自然科学版), 2017, 37(1):106-111.

[60] 郭戎格. 基于 IC 卡数据的定制公交线路优化[D]. 北京: 北京交通大学, 2017.

[61] 宋子杭. 基于公交 GPS 和 IC 卡数据的公交线网优化方法[D]. 北京: 北京交通大学, 2017.

[62] Cevallos F., Zhao F. Minimizing transfer times in public transit network with genetic algorithm[J]. Transportation Research Record: Journal of the Transportation Research Board, 2006(1971):74-79.

[63] Hadas Y., Ceder A.A. Optimal coordination of public-transit vehicles using operational tactics examined by simulation[J]. Transportation Research Part C: Emerging Technologies, 2010, 18(6):879-895.

[64] Schröder M., Solchenbach I. Optimization of transfer quality in regional public transit[J]. 2006.

[65] Shafahi Y., Khani A. A practical model for transfer optimization in a transit network: Model formulations and solutions[J]. Transportation Research Part A: Policy and Practice, 2010, 44(6):377-389.

[66] Cortés C.E., Sáez D., Milla F., et al. Hybrid predictive control for real-time optimization of public transport systems'operations based on evolutionary multi-objective optimization[J]. Transportation Research Part C: Emerging Technologies, 2010, 18(5):757-769.

[67] Fu L., Liu Q., Calamai P. Real-time optimization model for dynamic scheduling of transit operations[J]. Transportation Research Record: Journal of the Transportation Research Board, 2003(1857):48-55.

[68] Chuanjiao S., Wei Z., Yuanqing W. Scheduling combination and headway optimization of bus rapid transit[J]. Journal of transportation systems engineering and information technology, 2008, 8(5):61-67.

[69] Zhao F.. Large-scale transit network optimization by minimizing user cost and transfers[J]. Journal of Public Transportation, 2006, 9(2):6.

[70] Nikolić M., Teodorović D. Transit network design by bee colony optimization[J]. Expert Systems with Applications, 2013, 40(15):5945-5955.

[71] Hadas Y., Shnaiderman M. Public-transit frequency setting using minimum-cost approach with stochastic demand and travel time[J]. Transportation Research Part B: Methodological, 2012, 46(8):1068-1084.

[72] Bagloee S.A., Ceder A.A. Transit-network design methodology for actual-size road networks[J]. Transportation Research Part B: Methodological, 2011, 45(10):1787-1804.

[73] Zhao F., Zeng X.. Simulated annealing-genetic algorithm for transit network optimization[J]. Journal of Computing in Civil Engineering, 2006, 20(1):57-68.

[74] Zhao F., Zeng X. Optimization of transit route network, vehicle headways and time tables for large-scale transit networks[J]. European Journal of Operational Research, 2008, 186(2):841-855.

[75] Yu B., Yang Z., Yao J. Genetic algorithm for bus frequency optimization[J]. Journal of Transportation Engineering, 2009, 136(6):576-583.

[76] Chen J. Validity research of judging boarding stop by cluster analysis smart card data[A]. In: International Conference on Remote Sensing, Environment and Transportation Engineering[C], 2011. 5625- 5628.

[77] Yuan N.J., Wang Y., Zhang F., et al. Reconstructing Individual Mobility from Smart Card Transactions: A Space Alignment Approach[A]. In: IEEE International Conference on Data Mining[C], 2014. 877-886.

[78] Wang W., Attanucci J., Wilson N. Bus Passenger Origin-Destination Estimation and Related Analyses Using Automated Data Collection Systems[J]. Journal of Public Transportation, 2010, 14(4):131-150.

[79] 于勇, 邓天民, 肖裕民. 一种新的公交乘客上车站点确定方法[J]. 重庆交通大学学报 (自然科学版), 2009, 28(1):121-125.

[80] 安冬冬. 基于数据挖掘技术的常规公交服务质量评价体系研究[D]. 重庆: 西南交通大学, 2015.

[81] Tucker W. Engineering Applications of Correlation and Spectral Analysis[J]. Technometrics, 1980, 24(1):79-80.

[82] Podobnik B., Stanley H.E. Detrended cross-correlation analysis: a new method for analyzing two non- stationary time series[J]. Physical Review Letters, 2008, 100(8):084102.

[83] Jerri A.J. The Shannon sampling theorem — Its various extensions and applications: A tutorial review[J]. Proceedings of the IEEE, 2005, 65(11):1565-1596.

[84] 张红. 基于公交 IC 卡信息的客流数据分析及静态调度研究[D]. 西安: 西安电子 科

技大学, 2012.

[85] 曹洁, 徐强, 李宇, 等. 利用数据挖掘技术对公交乘客特征状况的分析[J]. 计算机工程与设计, 2007, 28(17):4260-4262.

[86] Pelletier M.P., Trépanier M., Morency C. Smart card data use in public transit: A literature review[J]. Transportation Research Part C Emerging Technologies, 2011, 19(4):557-568.

[87] Jin K.E., Ji Y.S., Moon D.S. Analysis of public transit service performance using transit smart card data in Seoul[J]. Ksce Journal of Civil Engineering, 2015, 19(5):1530-1537.

[88] Ma X., Wang Y. Development of A Data-driven Platform for Transit Performance Measures Using Smart Card Data andGPSData[J]. Journal of Transportation Engineering, 2014, 140(12).

[89] Zhong C., Batty M., Manley E., et al. Variability in regularity: Mining temporal mobility patterns in london, singapore and beijing using smart-card data[J]. PloS one, 2016, 11(2):e0149222.

[90] 翁剑成, 王昌, 王月玥, 等. 基于个体出行数据的公共交通出行链提取方法[J]. 交通运输系统工程与信息, 2017, 17(3):67-73.

[91] 贾佃精. 基于出行链的公共交通出行需求预测研究[D]. 哈尔滨: 哈尔滨工业大学, 2015.

[92] 蒋家高. 公共交通出行链研究[D]. 昆明: 昆明理工大学, 2013.

[93] Tarjan R. Depth-first search and linear graph algorithms[A]. In: Switching and Automata Theory, 1971., 12th Annual Symposium on[C], 1971. 114-121.

[94] Arthur D., Vassilvitskii S. k-means++: The advantages of careful seeding[A]. In: Proceedings of the eighteenth annual ACM-SIAM symposium on Discrete algorithms[C], 2007. 1027-1035.

[95] Wold S., Esbensen K., Geladi P. Principal component analysis[J]. Chemometrics and intelligent labora- tory systems, 1987, 2(1-3):37-52.

[96] Ester M., Kriegel H.P., Sander J., et al. A density-based algorithm for discovering clusters in large spatial databases with noise.[A]. In: Kdd[C], 1996. 226-231.

[97] Goerigk M., Schmidt M. Line planning with user-optimal route choice[j]. European Journal of Operational Research. 2017 Jun 1;259(2):424-36.

[98] Anderson M.K., Nielsen O.A., Prato C.G. Multimodal route choice models of public transport passengers in the Greater Copenhagen Area[J]. EURO Journal on Transportation and Logistics. 2017 Sep 1;6(3):221-45.

# 附录 1 城市公交大数据挖掘与分析平台技术架构

本研究自主研发用以进行公交系统数据挖掘与可视化分析的系统平台架构如附图 1 所示，本系统硬件上由云服务集群中两个节点共同完成。其中，采集与存储节点负责原始数据抓取、预处理，接收运算服务节点的处理结果、并向授权用户提供数据集的网络访问接口；运算与图形渲染节点按用户请求进行数据处理、并提供可视化数据挖掘的接口，两节点间由高速通信链路相连。**本文所有地理信息数据可视化图均依托该数据分析平台生成。**

系统软件部分可分为数据抓取、数据处理与数据可视化三大子系统。其中，数据抓取子系统借助高德地图 API 提供的公交查询接口与城市地理兴趣点查询接口周期性抓取目标城市的公交线网结构轨迹数据；数据处理子系统为本系统核心模块，基于 QT 5.6.2 与 Python 混合框架开发，为数据分析提供高性能的函数库与完备的二次开发接口；数据可视化子系统则基于百度 MapV 与百度 E-Charts 实现空间数据可视化。

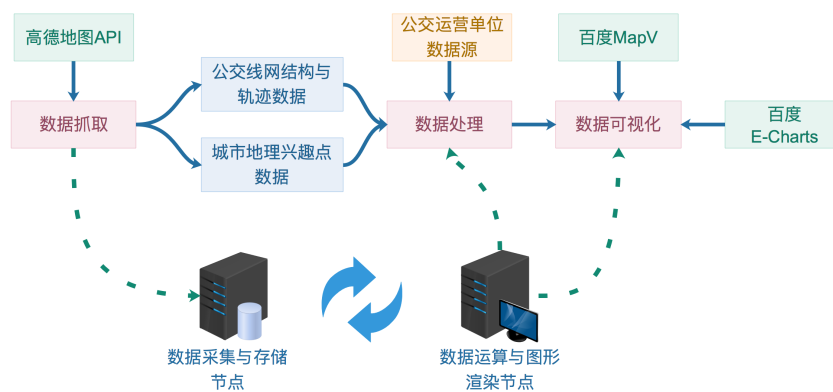

附图 1 数据分析平台硬件架构

本系统数据处理部分基于 C++实现，而数据可视化与线网数据抓取基于 Javascript 实现，两者不可直接调用，因此开发了 Web 交互组件，其技术架构如附图 1-2 所示，该组件采用了标准通信接口实现业务分离与数据共享，也便于未来在云端进行部署。

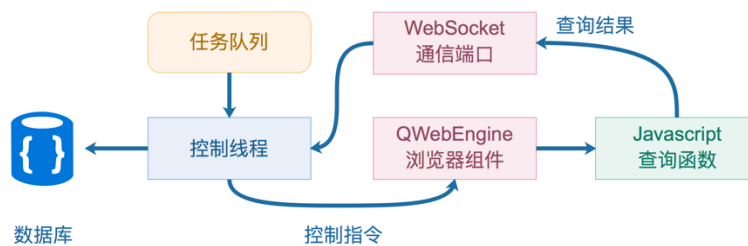

附图 2 Web 交互组件组件技术架构

该应用平台关键部分软件截图见图 6.3~图 6.5。

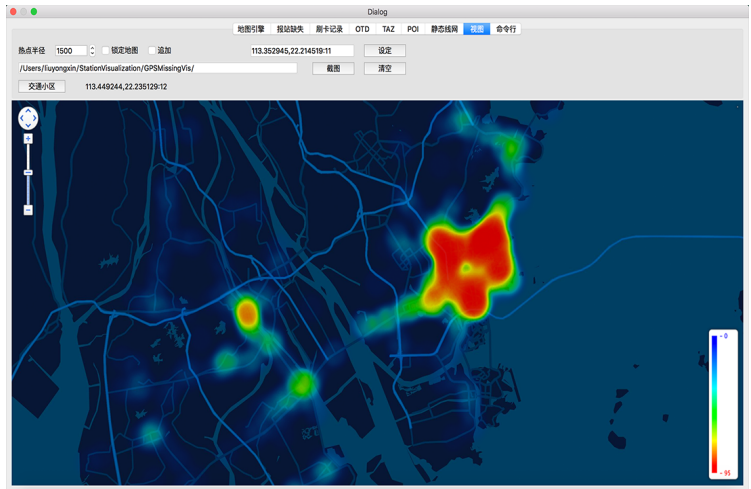

附图 3 客流热点动态分布可视化界面

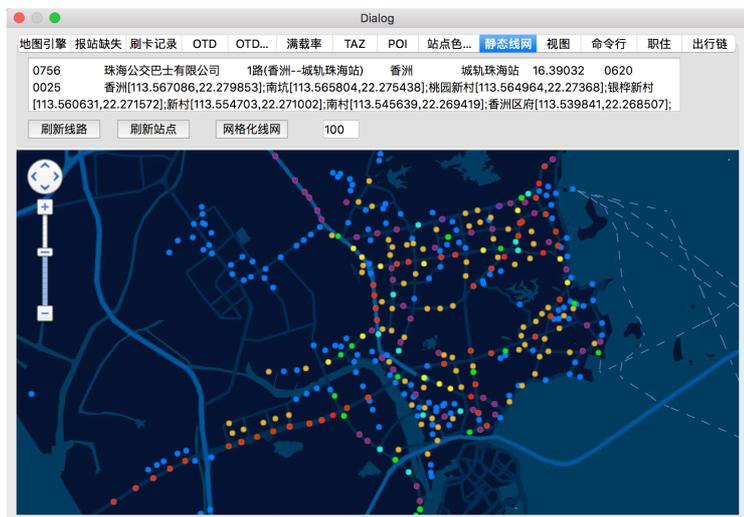

附图 4 精细化站点客流量分布可视化界面

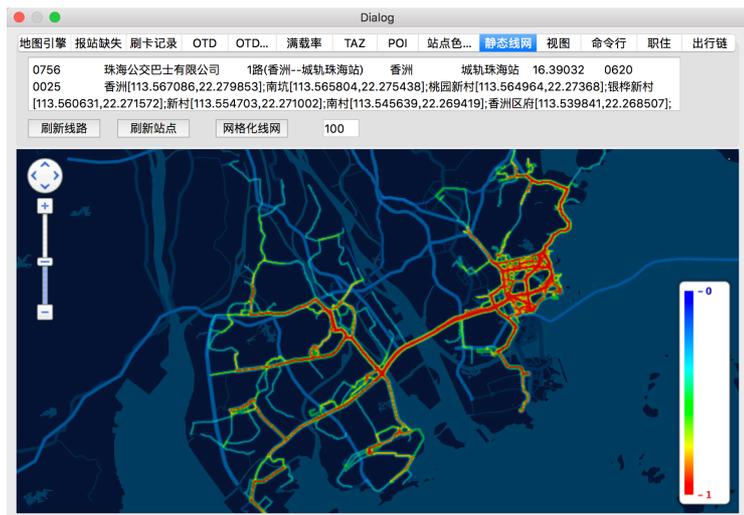

附图 5 线网客运力及满载率动态可视化界面

# 攻读博士学位期间取得的研究成果

一、已发表（包括已接受待发表）的论文，以及已投稿、或已成文打算投稿、或拟成文投稿的论文情况（只填写与学位论文内容相关的部分）：

| 序号 | 作者（全体作者，按顺序排列） | 题目 | 发表或投稿刊物名称、级别 | 发表的卷期、年月、页码 | 相当于学位论文的哪一部分（章、节） | 被索引收录情况 |
|---|---|---|---|---|---|---|
| 1 | **Y Liu**, X Weng, J Wan, X Yue, H Song, AV Vasilakos | Exploring data validity in transportation systems for smart cities | IEEE Communication Magazine (SCI) | 第46卷 2016年3月 pp.211-218 | 第三章 | SCI收录 |
| 2 | X Weng, **Y Liu**, H Song, S Yao, P Zhang | Mining urban passengers' travel patterns from incomplete data with use cases | Computer Networks (SCI) | 第10卷4期 2017年12月 | 第五章 | 待检索 |
| 3 | Xiaoxiong Weng, Yongxin Liu, Shushen Yao | Mining Residents' Usage of Urban Transit System from Big Data | IEEE Transactions on Big Data (SCI) | Major revision | 第四章 | |
| 4 | 翁小雄、**刘永鑫** | 基于数据挖掘的城市公交站点服务评价方法研究 | 现代电子技术（中文核心） | 2019年第4期 | 第四章 | 已录用 |
| 5 | 翁小雄、**刘永鑫** | 城市公交多源数据时间误差消除方法研究 | 现代电子技术（中文核心） | 2019年第11期 | 第三章 | 已录用 |
| 6 | 翁小雄、**刘永鑫** | 基于数据融合的城市公交系统缺失数据推断方法研究 | 现代电子技术（中文核心） | 2019年第9期 | 第三章 | 已录用 |

注：在"发表的卷期、年月、页码"栏：
1 如果论文已发表，请填写发表的卷期、年月、页码；
2 如果论文已被接受，填写将要发表的卷期、年月；
3 以上都不是，请据实填写"已投稿"，"拟投稿"。
不够请另加页。

二、与学位内容相关的其它成果（包括专利、著作、获奖项目等）

1. 翁小雄，刘永鑫，李莹，呙娟，姚树申. 一种公共交通乘客出行时空轨迹提取方法[P]. 公开号：CN106874432A. 2017-06-20.
2. 翁小雄，刘永鑫，呙娟，张腾月. 一种基于公共交通多源数据融合的 IC 卡刷卡站点匹配方法[P]. 公开号：CN105574137A. 2016-05-11．（发明专利授权）
3. 翁小雄，姚树申，刘永鑫. 一种基于极大概率估计的城市公交系统车载报站缺失数据修补方法[P]. 公开号：CN108230724A. 2018.01.31.
4. 翁小雄，张鹏飞，刘永鑫. 一种城市公交多源数据时间误差消除方法[P]. 公开号：CN108491932A. 2018.09.04.
5. 翁小雄，吕攀龙，刘永鑫. 一种基于不完全轨迹片段的公交乘客闭合出行链挖掘方法[P]. 公开号：CN108960684A. 2018.08.05
6. Liu Y, Song H. Levering mobile cloud computing for mobile big data analytics[M]//Mobile Big Data. Springer, Cham, 2018: 21-39.

# 致 谢

纸上得来终觉浅，绝知此事要躬行。博士四年如白驹过隙，转瞬即逝，回首四年感慨良多。在此，谨以感恩之心向你们致以最诚挚的谢意！

夫仁者，已欲立而立人，已欲达而达人。翁小雄教授学识渊博、治学严谨、思维敏捷而缜密。课题从选题、数据处理到具体的成果转化、结果分析直至文章撰写和论文修改都凝结了老师的心血和智慧。翁老师对科研孜孜追求的精神和对学术严谨的态度感染并激励着我，使我终身受益。在此，谨向教导、支持和鼓励我的老师致以深深的敬意。

作为一名从通信专业转到交通专业的学生，刚入学时所有的知识面只集中在物联网底层硬件方面，到今天，我已经学会通过数据了解城市交通系统各方面的特性，并建立数学优化模型，最后将成果以大数据可视化的形式发布，在此过程中导师翁小雄教授的指导、宽容与教诲，逐渐让我从只懂得电子工程的学生变成了一个复合型人才，也让我从只关心功能实现的初级工程师转变为学会不断深入思考事物内部关联关系的研究学者。对翁小雄教授表达深深的敬意！

还要感谢翁小雄教授在课题研究和论文修改方面给予的帮助和鞭策。尤其是感激翁老师在住院期间仍然不忘叮嘱我们按时完成论文并给予指导，翁老师敬业负责的工作态度令我们感激涕零！

感谢实验室全体同学四年来的陪伴与合作，感谢姚树申博士、宋明磊博士、呙娟硕士、张腾月硕士、李莹硕士、康擎彪硕士、彭新建硕士、汪周盼硕士、林龙硕士和张鹏飞同学在交通调查实验以及数据处理中对我的帮助与支持。

感谢华南理工大学为我提供了进行博士课题研究与成长环境，华南理工大学让我意识到了科学研究以及学习、生活的本质。

感谢硕士导师华南农业大学岳学军教授在硕士期间为我打下坚实的科研基础，感谢安普瑞德航天航空大学宋侯冰教授在发表论文过程中给予的无私帮助。感谢澳门大学杨丹硕士、华南农业大学王健硕士、王林慧博士在生活中给予的关心与帮助。

感谢我的良师益友和至爱的父母对我的鼓励、支持和帮助。

## Ⅳ - 2 答辩委员会对论文的评定意见

论文《基于多源数据融合的城市公交系统乘客出行模式挖掘及其应用研究》，论文针对公交系统乘客出行模式挖掘及其应用问题，从OTD复现、乘客出行模式识别、基于出行偏好的公交线路优化三个层面展开研究，选题切合当前智能交通数据挖掘研究的关键问题，选题合理，具有理论和实际意义。

论文针对从不完整、有缺陷的多源公交系统数据中挖掘乘客出行模式及其应用问题，首先提出一套在有缺陷数据环境下的一票制公交信息系统数据预处理方案；在此基础上提出了一种公交乘客完整公交出行信息提取方法；其次，提出一种基于出行拓扑关系图的新型数据融合方法，从轨迹片段中提取公交乘客的闭合出行链，并分析了乘客的活动特征，进而构建乘客的公交出行线路选择偏好模型；最后提出一种基于乘客个体出行偏好的公交线路优化方法。

综上所述，作者对国内外文献资料的归纳总结反映了作者具有智能交通数据挖掘领域前沿视角，能够把握本课题的发展动向；文中实际公交数据的收集分析平台搭建、论证方法和相关结论均合理，作者对公交系统乘客出行模式挖掘问题具有深刻见解，基础理论掌握较扎实，具备独立从事科学研究的能力；论文写作规范，结构层次清晰，文字表达通畅，论文基本达到博士学位的要求。经答辩委员会无记名投票表决，一致同意通过论文答辩，并建议授予工学博士学位。

论文答辩日期： 2018 年 6 月 2 日
答辩委员会委员共 5 人，到会委员 5 人
表决票数：优秀（ 1 ）票；良好（ 4 ）票；及格（ 0 ）票；不及格（ 0 ）票
表决结果（打"√"）：优秀（ ）；良好（ √ ）；及格（ ）；不及格（ ）
决议：同意授予博士学位（ √ ）    不同意授予博士学位（ ）

答辩委员会成员签名： ________（主席） ________ ________
________ ________ ________